\documentclass[twoside,12pt]{article}
\usepackage{epsfig}

\usepackage{amssymb}
\usepackage{amsfonts}
\usepackage{amsmath}
\usepackage{bm}
\usepackage{graphicx}
\usepackage{color}

\def\Journal#1#2#3#4{{#1} {#2} (#4) #3 }
\def\NIMA{{\em Nucl. Instrum. Meth.} A}

\def\NCA{{\em Nuovo Cimento} A}

\def\NPA{{\em Nucl. Phys.} A}

\def\NPB{{\em Nucl. Phys.} B}

\def\PLB{{\em Phys. Lett.} B}

\def\PRL{\em Phys. Rev. Lett.}
\def\PREV{\em Phys. Rev.}
\def\PREP{\em Phys. Rep.}

\def\PRD{{\em Phys. Rev.} D}
\def\PRC{{\em Phys. Rev.} C}

\def\ZPC{{\em Z. Phys.} C}

\def\ANNP{\em Ann. Phys. (N.Y.)}
\def\RMP{{\em Rev. Mod. Phys.}}

\def\INTA{{\em Int. J. Mod. Phys.} A}
\def\EPJC{{\em Eur. Phys. J.} C}
\def\JHEP{{\em JHEP}}
\def\SJNP{{\em Sov. J. Nucl. Phys.}}

\definecolor{nv}{rgb}{0.1,0.1,0.6}
\definecolor{pr}{rgb}{0.2,0.1,0.5}
\definecolor{mg}{rgb}{0.4,0.0,0.4}

\newcommand{\be}{\begin{equation}}
\newcommand{\ee}{\end{equation}}
\newcommand{\bea}{\begin{eqnarray}}
\newcommand{\eea}{\end{eqnarray}}
\newcommand{\nn}{\nonumber}

\newcommand{\beq}{\begin{equation}}
\newcommand{\eeq}{\end{equation}}
\newcommand{\beqy}{\begin{eqnarray}}
\newcommand{\eeqy}{\end{eqnarray}}
\newcommand{\beqyn}{\begin{eqnarray*}}
\newcommand{\eeqyn}{\end{eqnarray*}}

\newcommand{\bc}{\begin{center}}
\newcommand{\ec}{\end{center}}
\newcommand{\bmin}{\begin{minipage}}
\newcommand{\emin}{\end{minipage}}

\begin{document}

\title{ \vspace{1cm} Spin Structure of the Nucleon - Status and Recent Results}
\author{S.E.\ Kuhn,$^{1}$ J.-P.\ Chen,$^2$ E.\ Leader,$^3$\\ 
\\
$^1$Old Dominion University, Norfolk VA 23529, USA\\
$^2$Thomas Jefferson National Accelerator Facility, Newport News VA 23606, USA\\
$^3$Imperial College, London SW7 2AZ, U.K.}
\maketitle
\begin{abstract} After the initial discovery of the so-called 
``spin crisis in the parton model'' in the 1980's, a large set of polarization data in deep inelastic
lepton-nucleon scattering
 was collected at labs like SLAC, DESY and CERN. More recently, new high precision data at
large $x$ and in the resonance region have come from experiments at Jefferson Lab. These data, in
combination with the earlier ones, allow us to study in detail the polarized parton densities, the 
$Q^2$ dependence of various moments of spin structure functions, the duality between deep inelastic
and resonance data, and the nucleon structure in the valence quark region. Together with 
complementary data from HERMES,
RHIC and COMPASS, we can put new limits on the flavor decomposition and the
gluon contribution to the nucleon spin. In this report, we provide an overview of our present knowledge
of the nucleon spin structure and give an outlook on future experiments. We focus 
in particular on
the spin structure functions $g_1$ and $g_2$ of the nucleon and their moments.
\end{abstract}
\eject
\tableofcontents

\section{Introduction \label{sec:intro}}
The measurement of a spin dependent observable is generally a
daunting task, but with rich rewards, because spin seems to have a
scalpel-like ability to expose weaknesses and failures of
theories. Witness the switch from $S,T$ to $V-A$, which led to the
Weinberg-Salam electroweak theory, or the demise of Regge poles,
which had successfully described hadronic cross-sections and
shrinking diffraction peaks, or the spin crisis in the parton
model of deep inelastic scattering, supposedly resolved by an
anomalous gluon effect, now shown to be untenable, and, most
recently, the realization that some of the information, gathered for 40 years
 on the fundamental electromagnetic form factors of
the nucleon, is unreliable.

In this review we attempt to survey the tremendous experimental
and theoretical effort, mainly  involving studies of polarized
lepton-hadron scattering and polarized proton-proton reactions,
 that has led to  our present knowledge of the internal spin
structure of the nucleon,

 Deep inelastic lepton-hadron scattering (DIS) has played a seminal role
in the development of our present understanding of the
sub-structure of elementary particles. The discovery of Bjorken
scaling in the late nineteen-sixties \cite{Panofsky:1968pb}
provided the critical impetus for the idea that elementary
particles contain 
point-like constituents and for the
subsequent invention of the Parton Model. DIS continued to play an
essential role in the long period of consolidation that followed,
in the gradual linking of partons and quarks, in the discovery of
the existence of missing constituents, later identified as gluons,
and in the wonderful confluence of all the different parts of the
picture into a coherent dynamical theory of quarks and gluons --
Quantum Chromodynamics (QCD). Polarized DIS, involving the
collision of a longitudinally polarized lepton beam with a
longitudinally or transversely polarized target, provides
complementary and equally important insight into the structure of
the nucleon.

At first sight the theoretical treatment of the polarized case
seems to mimic the unpolarized case, with the structure functions
$F_{1,2}(x)$ replaced by the polarized structure functions
$g_{1,2}(x)$, and with parton densities $q(x)$ replaced by
polarized densities $\Delta q(x)$. But it turns out that the
polarized case is much more subtle: there is an anomalous gluon
contribution to $g_1(x)$, and $g_2(x)$ has no interpretation at
all in purely partonic language.

The latter insights were mainly inspired by the unexpected results
of the European Muon Collaboration (EMC) measurement of $g_1(x)$
in 1988 \cite{Ashman:1987hv}. The first excitement caused by this
experiment was its indication of the failure of a sum rule due to
Ellis and Jaffe based on the assumption that the contribution from
strange quarks to $g_1$ is negligible \cite{Ellis:1973kp}. It was
soon realized, however, that there were more profound
ramifications, which led to an intense scrutiny of the theory,
since they implied a ``spin crisis in the parton model''
\cite{Leader:1988vd}--- the spins of the quarks seemed to provide
only a tiny fraction of the spin of the nucleon, in contrast to
the situation in simple-minded constituent quark models of
hadrons, where the quark spins account for a very large fraction
of the proton spin.  The crisis was believed to be resolved via a
large polarized gluon contribution.

Of course the parton model predates QCD. In the more general
field-theoretic framework we know that Bjorken Scaling
\cite{Bjorken:1968dy}, i.e. the fact that structure functions and
parton densities depend only on $x$, cannot hold exactly, and
these functions have a $Q^2$-dependence which can be calculated
perturbatively in QCD, resulting in some of the most stringent
tests of the validity of the theory. Moreover, because of the
unfortunate need to renormalize the theory, the parton densities
 lose their simple physical meaning, and their actual functional
 form depends upon the renormalization scheme employed.

At present the situation, as will be discussed, is full of
interest.
\begin{itemize}
\item Measurements of the polarized gluon density suggest that it is
much too small to resolve the spin crisis
\cite{Ageev:2005pq,Leader:2005ci}. This almost certainly implies
that the partons possess orbital angular momentum, and it appears
possible to estimate this, at least for the quarks, via a study of
deeply virtual Compton scattering on protons \cite{Ji:2003qt}.
\item  More precise data  expected from the COMPASS
experiment at CERN and the $pp$ program at RHIC
 will allow further scrutiny of the  validity
of the above conclusion.
\item There are now significant measurements of $g_2(x)$ which can
be used to test the Wandzura-Wilczek approximation
\cite{Wandzura:1977qf}, the Burkhardt-Cottingham sum rule
\cite{Burkhardt:1970ti}, and the Efremov-Leader-Teryaev sum rule
\cite{Efremov:1996hd}.
\item Measurements at Jefferson Laboratory are probing the 
hitherto inaccessible region
of large $x$, where there are intriguing predictions about
the behavior of the ratios $\Delta q(x)/q(x)$, as well as the
region of low $Q^2$, where higher twist effects are important and
where the issue of ``duality" between the resonance and deep
inelastic regions can be studied.
\item By combining DIS data with the growing reservoir of data on
semi-inclusive DIS (SIDIS) it should become possible to learn
about the polarized sea densities $\Delta \bar{u}$ and $\Delta
\bar{d}$ and to resolve the present disagreement between DIS and
SIDIS  about the sign of the strange quark density $\Delta s(x) +
\Delta \bar{s}(x)$.
\end{itemize}

The aim of this review is phenomenological, i.e., it tries to strike
a reasonable balance between theory and experiment. The
theoretical treatment is thus conventional QCD and does not
address interesting, and sometimes profound, matters like the
Chern-Simons current and non-perturbative effects such as
instantons, axial ghosts, and the U(1) problem. These are discussed in
detail in the the review of Anselmino, Efremov and Leader
\cite{Anselmino:1994gn}, and in the review of Bass
\cite{Bass:2004xa}, who also examines the consequences of a fixed
pole in the virtual-photon Compton amplitude.

\subsection{\it Definitions and Formalism \label{subsec:formalism}}

\begin{figure}[htb!]
\begin{center}
\begin{minipage}[t]{9 cm}
\epsfig{file=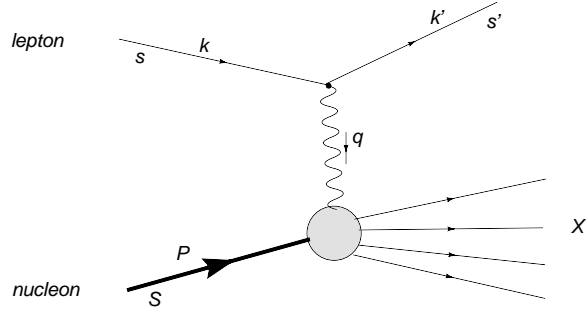,scale=0.7}
\end{minipage} 
\begin{minipage}[t]{16.5 cm}
\caption{Feynman diagram for deep inelastic lepton-hadron
scattering.} \label{fig1}
\end{minipage}
\end{center}
\end{figure}

Consider the inelastic inclusive scattering of polarized leptons on
polarized nucleons. We denote by $m$ the lepton mass,  $
k~(k^\prime$) the initial (final) lepton four-momentum and
$s~(s^\prime)$ its covariant spin four-vector, such that $s \cdot
k$ = 0 $(s^\prime \cdot  k^\prime = 0)$ and $s \cdot s = - 1$
$(s^\prime \cdot s^\prime = -1)$; the nucleon mass is $M$ and the
nucleon four-momentum and spin four-vector are, respectively, $P$
and $S$. Assuming, as is usually done, one photon exchange
 (see Fig.~\ref{fig1}), the
differential cross-section for detecting the final polarized
lepton in the solid angle $d\Omega$ and in the final energy range
$(E^\prime,~E^\prime + dE^\prime)$ in the laboratory frame, $P =
(M, \bm{0}), ~k = (E, \bm{k}), ~k^\prime = (E^\prime,
\bm{k}^\prime)$, can be written as

\begin{equation} \label{eq:DIS}
\frac{d^2 \sigma}{ d \Omega dE'} = \frac{\alpha^2}{ 2 M q^4}
\frac{E'}{E} L_{\mu \nu} W^{\mu \nu}
\end{equation}
where $q =  k- k'$ and $\alpha$ is the fine structure constant.

The leptonic tensor $L_{\mu\nu}$ (summed over the unobserved final lepton spin) is given by
\begin{eqnarray}\label{eq:L-tensor} \lefteqn{L_{\mu\nu} ( k, s; k^\prime )=}
\nn \\&& \sum_{s'} \,[\bar{u} (k^\prime, s^\prime)
~\gamma_{\mu}~u(k,s)]^\ast ~[\bar{u}(k^\prime,
s^\prime)~\gamma_{\nu}~u(k,s)]  \end{eqnarray}
 and can be split into
symmetric $(S)$ and antisymmetric $(A)$ parts under $\mu,\nu$
interchange: \begin{equation} \label{eq:L-split}
 L_{\mu\nu}(k, s; k^\prime)=  2 \{
L^{(S)}_{\mu\nu}~(k;k^\prime) + iL^{(A)}_{\mu\nu}~(k,s;k^\prime)\}
 \end{equation}
where
\begin{eqnarray} \label{eq:L-S,A}
L^{(S)}_{\mu\nu}~(k;k^\prime) &=& k_\mu k^\prime_\nu +
k^\prime_\mu k_\nu -
g_{\mu\nu}~(k\cdot k^\prime - m^2)  \nn \\
L^{(A)}_{\mu\nu}~(k,s;k^\prime) &=&
m~\varepsilon_{\mu\nu\alpha\beta}~s^\alpha~q^{\beta} .
\end{eqnarray}
The unknown hadronic tensor $W_{\mu\nu}$ describes the interaction
between the virtual photon and the nucleon and depends upon four
scalar structure functions, the unpolarized  functions $F_{1,2}$
and the spin-dependent  functions $ g_{1,2}$ (ignoring parity-violating
interactions). These must be
measured and can then be studied in theoretical models, in our
case in the QCD-modified parton model. They can only be functions
of the scalars $q^2$ and $ q\cdot P$. Usually people work with
\begin{equation} \label{eq:def-Q,xBj}
  Q^2 \equiv -q^2 \quad \textrm{and} \quad x_{Bj} \equiv Q^2/2q \cdot
  P= Q^2/2M\nu
\end{equation}
where $\nu=E-E'$ is the energy of the virtual photon in the Lab
frame. $x_{Bj}$ is known as ``$x$-Bjorken", and we shall simply
write it as $x$. We also refer to the invariant mass of the (unobserved)
final state, $W = \sqrt{(P+q)^2} = \sqrt{M^2 + 2M\nu - Q^2}$.

Analogous to Eq.~(\ref{eq:L-split}) one has  \be
\label{eq:Wmunu} W_{\mu\nu}(q; P,S) = W^{(S)}_{\mu\nu}(q;P) +
i~W_{\mu\nu}^{(A)}(q; P, S) .  \ee
The symmetric part, relevant to unpolarized DIS, is given by
\begin{eqnarray} \label{eq:WS}
 W^{(S)}_{\mu\nu}(q;P) &=&
  2 \left[ \frac{q_\mu q_\nu}{q^2} - g_{\mu \nu}
\right]\,F_1(x,Q^2) \nn \\
&& + \frac{2}{M \nu} \left[P_\mu -\frac{P\centerdot q}{q^2}q_\mu
\right]\left[P_\nu -\frac{P\centerdot q}{q^2}q_\nu
\right]\,F_2(x,Q^2) . \end{eqnarray}
The antisymmetric part, relevant for polarized DIS, can be written as
\begin{equation} \label{eq:WAG12}
 W^{(A)}_{\mu\nu }(q; P, s) = 2~\varepsilon_{\mu\nu\alpha\beta}~ q^\alpha \Biggl\{ M^2 S^\beta G_1
(\nu, Q^2) +   \left[M \nu S^\beta - (S\cdot q) \, P^\beta \right] G_2 (\nu, Q^2)\Biggl\} \end{equation}
or, in terms of the scaling functions $g_{1,2}$
\beq \label{eq:g12G12}
g_1(x,Q^2)=M^2 \nu G_1(\nu,Q^2), \qquad g_2(x,Q^2)= M \nu^2 G_2(\nu,Q^2) , \eeq
\begin{equation} \label{eq:WA}
 W^{(A)}_{\mu\nu }(q; P, s) = \frac{2M}{ P\cdot
q}~\varepsilon_{\mu\nu\alpha\beta}~ q^\alpha\Biggl\{ S^\beta g_1
(x, Q^2) +   \left[S^\beta - \frac{(S\cdot q) \, P^\beta}{
(P\cdot q)}\right] g_2 (x, Q^2)\Biggl\}\,. \end{equation}

Note that these expressions are electromagnetic gauge-invariant:
\begin{equation} q^\mu W_{\mu\nu} =0 . \end{equation}
In the Bjorken limit, or Deep Inelastic  regime,
\begin{equation}
\label{eq:Bjlim} -q^2 = Q^2 \to \infty\quad , \quad \nu = E -
E^\prime \to \infty \quad , \quad x \, \, \rm {fixed} \end{equation}
the structure functions $F_{1,2}$ and $g_{1,2}$
 are known to approximately scale, i.e., vary very slowly with $Q^2$
at fixed $x$ -- in the simple parton model they scale exactly; in
QCD their $Q^2$ evolution can be calculated perturbatively.

Differences of cross-sections with opposite target spins are given
by
\begin{eqnarray} \label{eq:Xsec-diff}
\lefteqn{\left[\frac{d^2\sigma}{d\Omega~dE^\prime}
 (k, s, P, - S; k^\prime) -\frac{d^2\sigma}{d\Omega
~dE^\prime}~(k, s, P, S; k^\prime)\right] } \nn \\ && =
\frac{\alpha^2}{2Mq^4}~\frac{E'}{
E}~4L_{\mu\nu}^{(A)}~W^{\mu\nu(A)}\phantom{BBBBBBBBBBBB}\,.
\end{eqnarray}

After some algebra (for a detailed derivation see
\cite{Anselmino:1994gn}),
 one obtains from Eqs.~(\ref{eq:Xsec-diff},\,\ref{eq:WA}) expressions for
 the following polarized cross-section differences (note that here, and in the following, we
 have included the lepton mass terms that are usually ignored) :
 \begin{itemize}
 \item For the lepton and  target
nucleon polarized longitudinally, i.e. along or opposite to the
direction of the lepton beam,  the
cross-section difference under reversal of the
nucleon's spin direction (indicated by the double arrow) is given by
\begin{equation} \label{eq:LongXsec}
\frac{d^2\sigma^{\begin{array}{c}\hspace*{-0.2cm}\to\vspace*{-0.2cm}\\
\hspace*{-0.2cm}\Leftarrow\end{array}}}{dx\,dy}-
\frac{d^2\sigma^{\begin{array}{c}\hspace*{-0.2cm}\to\vspace*{-0.2cm}\\
\hspace*{-0.2cm}\Rightarrow\end{array}}}{ dx\,dy}
 =
\frac{16\pi\alpha^2}{Q^2}\Biggl[\left(1-\frac{y}{2}-\frac{y^2(M^2 x^2 +
m^2) }{Q^2}\right)\,g_1 - \frac{2M^2x^2y}{Q^2}g_2\Biggl] \,.
\end{equation}
\item For nucleons   polarized transversely in the scattering plane,
 one finds
\begin{equation} \label{TransXsec}
\frac{d^2 \sigma^{\to\Uparrow}}{dx\,dy} - \frac{d^2
\sigma^{\to\Downarrow}}{dx\,dy} = - \frac{16
\alpha^2}{Q^2}~\left(\frac{2Mx}{Q}\right)\sqrt{1-y - \frac{M^2 x^2
y^2 }{Q^2}} ~\left[\frac{y}{2}\left( 1 + \frac{2 m^2y}{Q^2}
\right)g_1 + g_2\right]\,.
\end{equation}
\end{itemize}
Here we have used \be \label{eq:def-y} y\equiv \frac{\nu}{E} =
\frac{P\cdot q}{P\cdot k} . \ee

In principle, these two independent observables allow measurement
of both $g_1 $ and $g_2$
(as has been done at SLAC and in Jefferson Lab's Halls A and C),
but the transverse cross section difference is
generally smaller
because of kinematic factors and therefore more
difficult to measure. Only in the past few years has it been
possible to gather precise information on $g_2$, which turns out to be
usually smaller than $g_1$ in DIS.

Since experimental results are often presented in the form of
asymmetries, which are the ratios of cross-section differences to
the unpolarized cross-section, we comment here briefly on the
latter. The unpolarized cross-section is given by
 \begin{equation}
\label{eq:UnpolXsec}
 \frac{d^2 \sigma_{unpold}}{dx\,dy}=\frac{4 \pi  \alpha^2
 }{x\, y\, Q^2}\Biggl\{ xy^2 \left( 1 - \frac{2m^2}{Q^2}\right)F_1 +
 \left[1-y-\frac{M^2x^2 y^2}{Q^2}\right] F_2 \Biggl\} .
 \end{equation}

It is a non-trivial task to obtain separate information on $F_1$
 and $F_2$ and usually data are presented for $F_2$ and $R$ where
 \begin{equation} \label{eq:defR}
R\equiv [1 + \gamma^2] \,\Bigg(\frac{F_2}{2xF_1}\Bigg) - 1
\end{equation}
 and
 \begin{equation} \label{eq:gamma}
     \gamma^2    = \frac{4M^2x^2}{Q^2} .
\end{equation}
For a longitudinally polarized target the measured asymmetry is
\begin{equation} \label{eq:Apar} A_\| \equiv
\frac{d\sigma^{\begin{array}{c}\hspace*{-0.2cm}\to\vspace*{-0.2cm}\\
\hspace*{-0.2cm}\Leftarrow\end{array}}-
d\sigma^{\begin{array}{c}\hspace*{-0.2cm}\to\vspace*{-0.2cm}\\
\hspace*{-0.2cm}\Rightarrow\end{array}}}{2\,d\sigma_{unpold}}
\end{equation}
and for a transversely polarized target the measured asymmetry is
\begin{equation} \label{eq:Atrans} A_{\perp}\equiv
\frac{d\sigma^{\to\Uparrow} - d
\sigma^{\to\Downarrow}}{2d\sigma_{unpold}} . \end{equation}

It is customary to introduce the (virtual) photon-nucleon asymmetries
$A_{1,2}$,
\beq \label{eq:Aigi} A_1=\frac{g_1 - \gamma ^2 g_2}{F_1} \eeq
 and
 \beq \label{eq:A_2gi}
 A_2= \gamma \, \big[\frac{g_1 + g_2}{F_1}\big] .
 \eeq
 From Eqs.~(\ref{eq:LongXsec},\ref{TransXsec},\ref{eq:UnpolXsec}), it follows that
\begin{equation} \label{eq:AlongA12}
 A_\|= D\,(A_1 + \eta A_2) \end{equation}
 and
  \begin{equation} \label{eq:AperpA12}
A_{\perp}=d(A_2 - \xi A_1) \end{equation}
where
\begin{equation} \label{Dnew}
D= \frac{y[(1 + \gamma^2
y/2)(2-y)-2y^2m^2/Q^2]}{y^2(1-2m^2/Q^2)(1+\gamma^2) + 2
(1+R)(1-y-\gamma^2 y^2/4)} 
\end{equation}
\begin{equation} \label{dnew}
d= \left[\frac{[1+\gamma^2y/2(1 + 2m^2y/Q^2)]
\sqrt{1-y-\gamma^2y^2/4}}{(1-y/2)(1+\gamma^2y/2) -
y^2m^2/Q^2}\right]\,D
\end{equation}
\begin{equation} \label{etanew}
\eta=\gamma \,\frac{ [1-y-y^2(\gamma^2/4 + m^2/Q^2)
]}{(1-y/2)(1+\gamma^2y/2) - y^2m^2/Q^2}
\end{equation}
\begin{equation} \label{xinew}
\xi =\gamma \frac{1-y/2 -y^2m^2/Q^2}{1 + \gamma^2y/2(1
+2m^2y/Q^2)} .
\end{equation}

Measurements of both $A_{\perp}$ and $A_\|$ yield the values of
$g_1$ and $g_2$ directly (given a parametrization of $R$ and
$F_1$), but in practice the majority of
experiments in the past have only measured $A_\|$. These experiments
 have been
interpreted as measurements of $g_1$, neglecting or correcting for
$g_2$. Although it turns
out that $g_2$ is small, there is no \textit{a priori} reason for
this, so it is safer to neglect some quantity for which we know an
upper bound, e.g. we could utilize $|A_{\perp}|\leq 1$, but a
better approach is to utilize the (virtual) photon-nucleon asymmetries. The
point is that there exists a restrictive bound on $A_2$
(for the history of this see \cite{Soffer:1999zv,Artru:2008cp}):
 \begin{equation} \label{eq:A2bound}
 |A_2|\leq \sqrt{R\,(1+A_1)/2} . \end{equation}

 Replacing $g_2$ in terms of $g_1$ and $A_2$ yields
\begin{equation} \label{eq:Alongg1A2}
 \frac{A_\|}{D} = ( 1+ \gamma^2)
\bigg[\frac{g_1}{F_1}\bigg] + (\eta - \gamma )\, A_2 ,
\end{equation}
where $\eta - \gamma$ is typically very small.
Most experiments either use a parametrization for
$A_2$ or neglect the $A_2$ terms in Eqs.~(\ref{eq:AlongA12},
\ref{eq:Alongg1A2}) altogether, leading to the
approximations
\begin{equation} \label{eq:AlongA1approx}
 A_1 \approx \frac{A_\|}{D} \approx ( 1+ \gamma^2)
\bigg[\frac{g_1}{F_1}\bigg]  \end{equation}
to evaluate $g_1$ and then utilize the bound
Eq.~(\ref{eq:A2bound}) to estimate the error involved.

Strictly speaking, all of the above, from Eq.~(\ref{eq:WS})
onwards, applies only to a spin 1/2 target. For targets with
higher spin, like the deuteron, the hadronic tensor contains many
more structure functions. But in all the practical cases of
interest the target nucleus may be considered as a weakly bound
system of essentially independent nucleons, with a small
correction for binding effects, and the nuclear asymmetries  are
then expressed in terms of the asymmetries on protons and
neutrons. For a detailed analysis of this issue, and formulae
relating measured asymmetries on deuterium and on $^3$He to the
nucleon structure functions, see Section 2.1.4 of
\cite{Anselmino:1994gn}. For the more general theory of DIS on
spin 1 targets, see \cite{Hoodbhoy:1988am} and
\cite{Frankfurt:1983qs}.

Note that in the above we have kept terms of order $M^2/Q^2$ and
smaller. They are sometimes necessary in order to extract the
correct experimental values of the structure functions from the
measured asymmetries. However, very often, the QCD analysis of the
structure functions is carried out at \emph{leading twist} only,
ignoring higher twist terms, i.e. of order $M^2/Q^2$. This is clearly
inconsistent in those cases, for example the Jefferson Lab experiments,
where the above terms are important.

\subsection{\it Experiments \label{subsec:exp}}

When Bjorken proposed his famous sum rule~\cite{Bjorken:1968dy}
40 years ago, he considered it
``useless'' since the experimental technology needed to test it appeared far from
reach. It required substantial advances in both polarized target and polarized
beam technology before the first double-polarization lepton scattering 
experiment could begin~\cite{Alguard:1976bm}. The first round of
experiments at SLAC~\cite{Alguard:1976bm,Alguard:1978gf,Baum:1980mh,Baum:1983ha}
using polarized electrons impinging on a polarized proton target
seemed to confirm the expectations of the naive quark parton model (namely, that most of
the proton spin was carried by its quark helicities). However, the range in $x$ covered was
rather limited, and the data had large errors and were taken at fairly low $Q^2$. 
Nevertheless, these experiments started the exploration of the nucleon spin structure,
not only in the DIS region, but also in the region of the nucleon 
resonances~\cite{Baum:1980mh}.

The EMC collaboration~\cite{Ashman:1987hv}
used the (naturally polarized) muon beam  at CERN,
with much higher beam energy,
together with a large polarized proton target, to extend the covered range down to
significantly lower $x \approx 0.01$.
Together with the earlier SLAC data, their
result seemed to indicate that little to none of the proton spin was due to the
helicities of the quarks. This violated the Ellis-Jaffe sum rule~\cite{Ellis:1973kp}
within the simple quark-parton picture assuming that
SU(3) is a good symmetry and the strange (sea) quark
contribution to the nucleon spin could be ignored.
The final data from the EMC experiment are published in~\cite{Ashman:1989ig}.

\begin{figure}[htb!]
\begin{center}
\begin{minipage}[t]{8 cm}
\epsfig{file=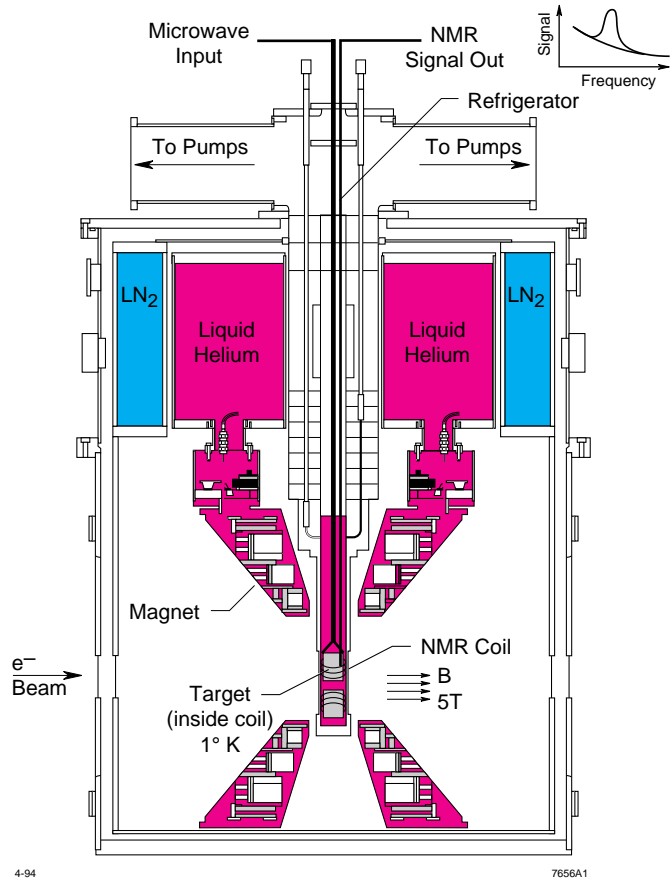,scale=0.5}
\end{minipage}
\begin{minipage}[t]{16.5 cm}
\vspace{-0.1in}
\caption{Typical solid state polarized proton or deuteron target
for electron scattering experiments. The cell containing the frozen
ammonia ($^{15}{\mathrm N\vec{\mathrm H}}_3$ or
 $^{15}{\mathrm N\vec{\mathrm D}}_3$) is at the center of
a Helmholtz-type magnet generating a homogeneous field of about 5 T.
A $^4$He evaporation refrigerator (a liquid helium bath in a low-pressure
environment) cools the target material to about 1 K. 140 GHz microwaves 
irradiate the target material to dynamically polarize the hydrogen
nuclei. The polarization is measured by a resonant NMR circuit (the
obtained signal vs. frequency is sketched in the top right).
Polarized targets for muon beams are typically much longer and
can be cooled to lower temperatures.
\label{poltarg1}}
\end{minipage}
\end{center}
\end{figure}

This puzzling result spurred the rapid development of several new
experiments that had the goal to verify the data on the proton with greater precision and,
most importantly at the time, test the Bjorken sum rule~\cite{Bjorken:1968dy}
by probing the spin structure of
the neutron, as well. This last goal required the use of targets containing
polarized neutrons, for which two very different technologies were developed.

One possibility is to use polarized deuterons as a target composed of equal
amounts of polarized protons and neutrons, from which neutron information
can be extracted by comparison with pure proton targets (taking nuclear
binding effects like the deuteron D-state into account). This method
follows closely the example of polarized proton target technology,
which typically involves chemical compounds (like alcohols or ammonia) rich in
hydrogen (either $^1$H for protons or $^2$H = D for deuterons)
 which are seeded with paramagnetic centers (with unpaired electrons), 
cooled to cryogenic temperatures
(1 K or lower) and immersed in a strong magnetic field (usually several T). The unpaired
electrons are polarized to nearly 100\% in the field, and this polarization is transferred
to the hydrogen nuclei via the process of Dynamic Nuclear Polarization (DNP),
by employing microwave radiation of the proper frequency to induce coupled electron-
nucleon spin transitions. This technique typically 
achieves a proton polarization of 80-90\%, and a deuteron polarization of 30-50\%,
both of which can be measured using a Nuclear Magnetic Resonance (NMR) system.
Figure~\ref{poltarg1} shows an example of such a target. In the case of deuteron
targets, one can also use $^6$LiD compounds, where both the Lithium nucleus
and the deuteron are polarized to increase the fraction of events from polarized
nucleons (the so-called dilution factor).

\begin{figure}[htb!]
\begin{center}
\epsfig{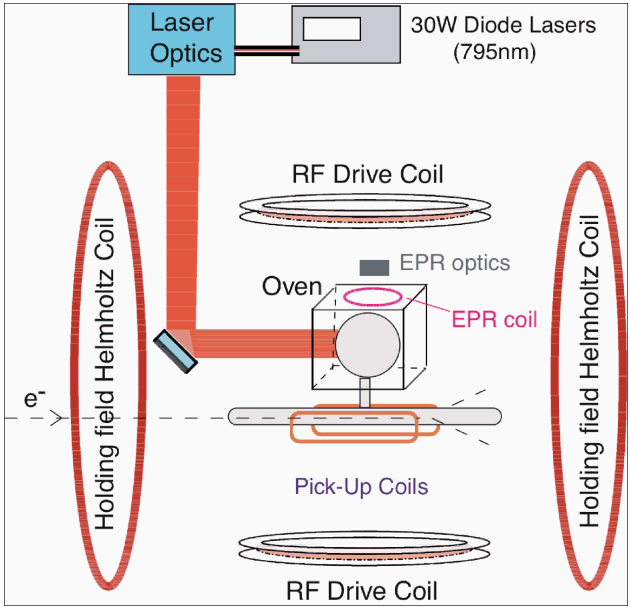}
\begin{minipage}[t]{16.5 cm}
\caption{Typical gaseous $^3$He target
for electron scattering experiments. The glass cell containing pressurized 
($\approx10$ atmospheres) $^3$He gas is at the center of
a Helmholtz magnet generating a homogeneous field of 25 gauss.
The cell is connected to a second, heated volume where 
the $^3$He is polarized by spin-exchange
collisions with alkali atoms, the latter being pumped by 
high-intensity lasers.
The polarization is measured either via a NMR circuit
(consisting of RF drive coils and NMR signal pickup coils)
 or using the EPR method.
\label{poltarg2}}
\end{minipage}
\end{center}
\end{figure}

A complementary approach uses polarized $^3$He nuclei to gain
information on the neutron, since to first order the $^3$He nucleus can be
considered as a bound state of a proton-proton pair (with their spins anti-aligned)
and an unpaired neutron carrying almost all of the nuclear spin. Again, nuclear corrections
to this picture are important, but the advantage is that one does not have to
subtract a large contribution from polarized protons to extract the neutron results.
The most widely used version of this target uses a glass vessel containing
pressurized $^3$He gas mixed
with a trace amount of an alkali atom vapor (typically Rubidium)
that can be electron-spin polarized via optical pumping with
circularly polarized laser light. The $^3$He nuclei are polarized through
spin exchange collisions with the alkali atoms. A typical example is shown in Fig.~\ref{poltarg2}. Polarizations of up to nearly 50\% have
been achieved in these targets, which can withstand much larger beam currents
than the solid-state DNP targets. As an example, the polarized $^3$He target
used in JLab Hall A~\cite{Alcorn:2004sb}, with a 15 $\mu$A electron beam on an 40 cm long,
10 atm
$^3$He target, achieved a luminosity of $10^{36}$ cm$^{-2}$s$^{-1}$.
Recently,  
a new development of a hybrid technique~\cite{hybrid}, using a mixture of Rubidium and 
Potassium atoms,
has significantly increased the spin-exchange efficiency and, in turn, 
increased the target polarization up to 60\% for the high density 
target. 
Further improvement to over 70\%
polarization has been achieved for this target by using recently available narrow-width
lasers.
The high luminosity and high polarization allow measurements
of the spin structure functions with very high statistical precision.
The target polarization is measured with both NMR and EPR 
(electron para-magnetic resonance) methods to a precision of 3\%.
Since the holding field is low, it is relatively easy to point the polarization
 in any direction in the horizontal plane 
using two pairs of Helmholtz coils. At Jefferson Lab, the addition of a new pair of vertical coils 
recently allowed the target also be polarized along the vertical direction, enabling
the study of transverse spin.

The successor collaboration to EMC at CERN, called SMC (Spin Muon Collaboration),
used large dynamically polarized cryogenic deuteron~\cite{Adeva:1993km}
and proton~\cite{Adams:1994zd} targets to
extract information on the neutron  and to
improve on the statistics and the kinematic reach of the EMC result. 
They also pioneered the use of semi-inclusive data, where a leading hadron
is identified in coincidence with the scattered lepton, to get more information
on the contribution of various quark flavors to the nucleon spin~\cite{Adeva:1995yi}.
The complete data set collected by the SMC resulted in precise
inclusive results both at the highest momentum transfer $Q^2$ ~\cite{Adeva:1998vv} 
and at the lowest quark momentum fraction
$x$ accessible to fixed target experiments~\cite{Adeva:1999pa}.

\begin{figure}[htb!]
\begin{center}
\epsfig{file=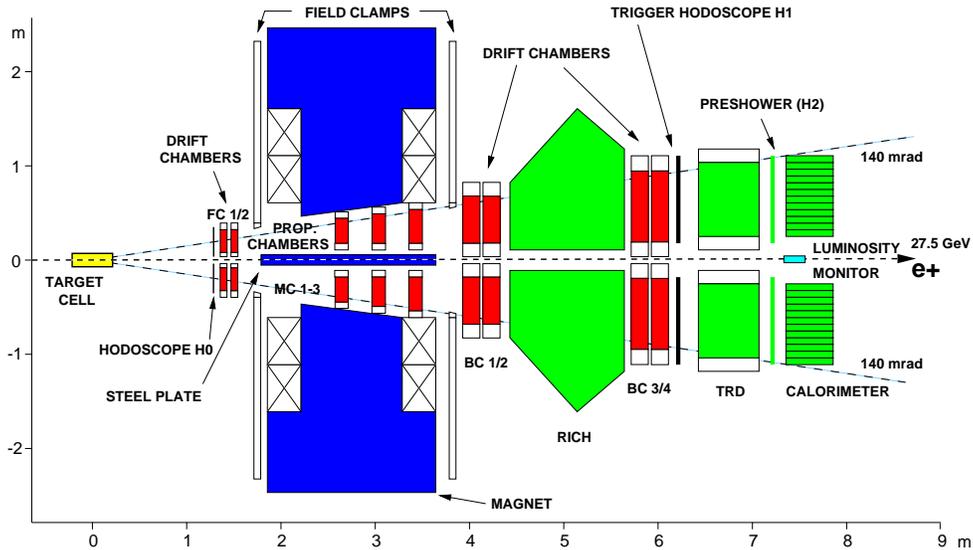,scale=0.6}
\begin{minipage}[t]{16.5 cm}
\caption{Layout of the HERMES experiment at HERA/Desy (Hamburg, Germany).
The stored electron or positron beam traverses an open storage cell fed by an atomic
beam source with polarized H, D or $^3$He. The scattered leptons (and leading hadrons)
are detected in a large acceptance spectrometer consisting of wire chambers, scintillator
hodoscopes, ring-imaging Cherenkov (RICH) and transition-radiation (TRD) detectors,
and an electromagnetic calorimeter.
\label{hermes}}
\end{minipage}
\end{center}
\end{figure}

The E142 collaboration at SLAC was the first to use a $^3$He gas target with high
luminosity to directly access the neutron spin structure 
functions $g_1^n$ and $g_2^n$~\cite{Anthony:1993uf,Anthony:1996mw}.
Together with the EMC and SMC experiments, the results showed that the Bjorken sum rule
including perturbative QCD (pQCD) corrections appeared to be valid. 
E142 was followed by a series of additional experiments at SLAC that used
all three nuclear targets, proton, deuteron and $^3$He, to accumulate
a highly precise data set on spin structure functions in the deep inelastic region.
Instrumental for achieving ever higher precision was a significant improvement in the polarization
(to over 80\%)
and intensity of available electron beams from strained GaAs cathodes irradiated
with circularly polarized laser light.
By using several electron beam energies and a set of up to 3 spectrometers,
the E143~\cite{Abe:1994cp,Abe:1995mt,Abe:1998wq} 
and the E155~\cite{Anthony:1999rm,Anthony:2000fn} collaborations collected data on the proton 
and the deuteron over a wide range of momentum transfer $Q^2$ at several values of
$x$, which were used to study scaling violations for polarized structure functions.
The E154 collaboration~\cite{Abe:1997cx,Abe:1997dp} 
added more neutron data at similar kinematics,
using a polarized $^3$He target. The E143 collaboration also published
the first precision results at lower $Q^2$ and in the nucleon resonance region
($W \leq 2$GeV)~\cite{Abe:1996ag}. The spin structure functions $g_2^{p,n,d}$
were measured with high precision by rotating the target polarization for
all 4 experiments from a longitudinal to a perpendicular orientation to the 
beam~\cite{Abe:1998wq,Abe:1997qk,Anthony:2002hy}.

The most innovative approach to measuring DIS structure functions came from the
HERMES collaboration (see Fig.~\ref{hermes})
which used positrons or electrons circulating in one of the HERA rings
at DESY together with internal low--density gas targets fed directly from atomic
beam sources~\cite{Ackerstaff:1997ws,Airapetian:1998wi}. The target atoms are
polarized using hyperfine transitions induced by radio frequency fields and 
Stern-Gerlach type separation with magnetic sextupoles. The atomic beam is
injected into a thin, windowless tube through which the beam circulates. This
method yields a pure polarized target without any dilution from unpolarized materials.
The polarization of the beam is accomplished by utilizing the Sokolov-Ternov effect
(the spontaneous vertical polarization through spin-dependent  synchrotron radiation
of leptons in a storage ring). Spin rotators turn the polarization into the longitudinal direction at
the target. The scattered electrons, as well as hadrons produced in coincidence,
were detected by a large acceptance forward spectrometer. This setup allowed the
HERMES collaboration not only to independently measure the inclusive spin
structure functions $g_1$ and $g_2$ (for final results see~\cite{Airapetian:2007mh}),
but also semi-inclusive structure functions for flavor--tagging~\cite{Airapetian:2004zf}.
In addition, they collected a large data set on related reactions of interest, from
Deeply Virtual Compton Scattering (DVCS)~\cite{Airapetian:2006zr} 
and transverse spin structure functions~\cite{Airapetian:2006rx}
to a first direct measurement of the gluon polarization~\cite{Airapetian:1999ib}.

\begin{figure}[htb!]
\begin{center}
\epsfig{file=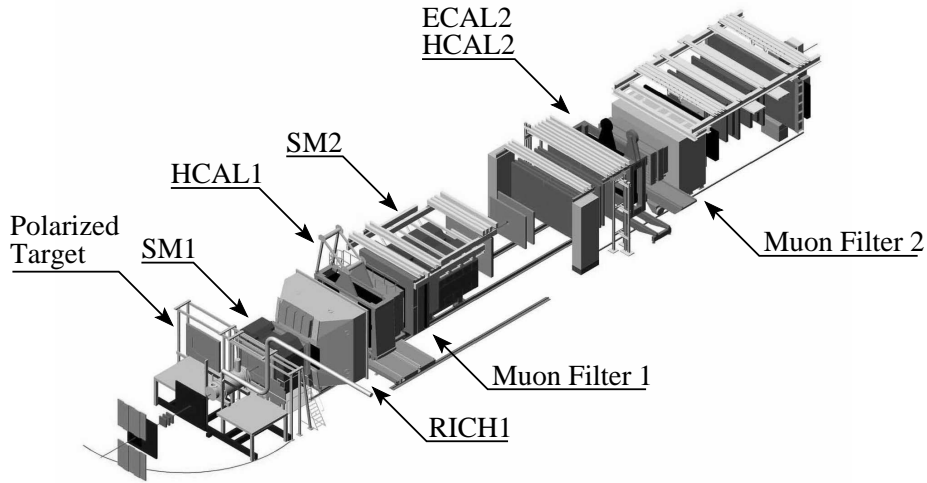,scale=0.6}
\begin{minipage}[t]{16.5 cm}
\vspace{-0.1in}
\caption{Layout of the COMPASS experiment at CERN (Geneva, Switzerland).
\label{compass}}
\end{minipage}
\end{center}
\end{figure}

After the shutdown of the HERA ring, there are now three laboratories left where experiments
studying the spin structure of the nucleon continue: CERN, with the SMC-successor
experiment ``COMPASS''; BNL (on Long Island, NY) with the polarized proton-proton
collision program at RHIC; and the Thomas Jefferson National
Accelerator Facility (``Jefferson Lab'' or JLab) in Newport News, Virginia, 
with an ongoing program of
electron scattering in all 3 experimental halls.

The COMPASS experiment (see Fig.~\ref{compass}) uses the secondary (naturally polarized)
muon beam at CERN together with large polarized deuteron and hydrogen
targets to extend the kinematic reach and precision of SMC, and, as its main
purpose, to extract information on the gluon polarization. This latter goal has been pursued
by measuring both the production of hadron pairs with high transverse 
momentum~\cite{Ageev:2005pq}
and by detecting charmed mesons in the final state (which are predominantly produced
via photon-gluon fusion). Indirect information on the gluon contribution to the nucleon
spin also comes from NLO analyses of inclusive DIS data, where the large kinematic
lever arm offered by COMPASS makes an important contribution. First results on the
deuteron have been 
published~\cite{Alexakhin:2006vx, Ageev:2007du,Alekseev:2007vi} and
the COMPASS experimental program will continue in the foreseeable future.

\begin{figure}[htb!]
\begin{center}
\epsfig{file=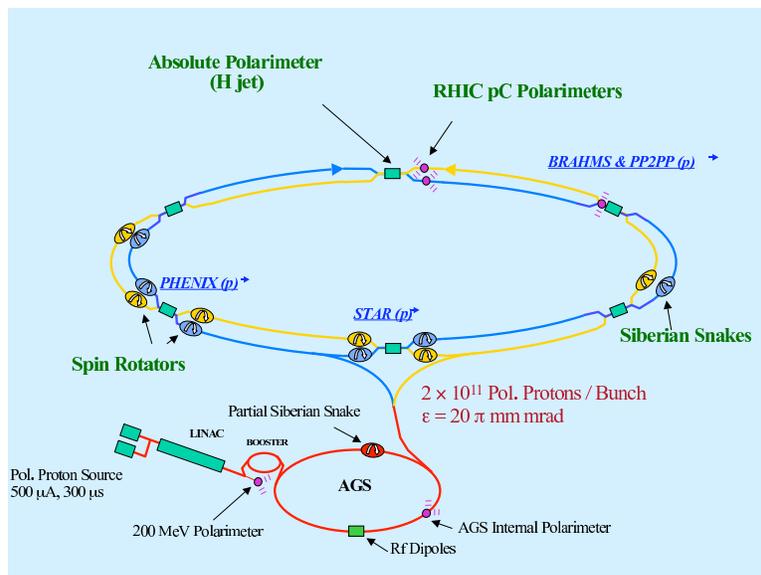,scale=0.4}
\begin{minipage}[t]{16.5 cm}
\caption{Layout of the RHIC accelerator complex at BNL (Long Island, NY).
\label{RHIC}}
\end{minipage}
\end{center}
\end{figure}

Another novel approach to studying the spin structure of the nucleon uses high-energy
collisions of counter-circulating proton beams in the Relativistic Heavy Ion Collider (RHIC)
at Brookhaven National Lab (BNL, Long Island, NY). Polarized protons are injected into a
series of accelerators that finally fill both RHIC rings, where energies up to 100 GeV 
(250 GeV in the future) can be reached (see Fig.~\ref{RHIC}). Siberian snakes 
rotate the proton spins to avoid depolarizing resonances, while spin rotators can
select  the desired spin direction at the interaction points.

At present, there are two large experiments (PHENIX and STAR) that use polarized
proton collisions to study the gluon helicity contribution $\Delta G$ to the nucleon spin.
The observables used so far include meson production with high transverse momentum
$p_T$ as well as jet production, both probing the gluons through 
quark-gluon and gluon-gluon
interactions in the initial state. First results from these experiments
have been 
published~\cite{Adler:2006bd,Abelev:2006uq,Adare:2007dg,Abelev:2007vt}
 and will be discussed in Section~\ref{subsec:PDFexp}.
By orienting the proton spins perpendicular to the beam direction,
both experiments can also study  reactions sensitive to transverse spins. 

Finally, for the last 10 years a large program using electron scattering to study the
spin structure of nucleons has been underway at Jefferson Lab (JLab).
This program has utilized the highest polarization electron beams (over 85\%) with
energies from 0.8 GeV to close to 6 GeV and all three species of polarized targets
(p, d, and $^3$He) to study spin-dependent structure functions both in the DIS
regime as well as in the nucleon resonance region (and even on nuclei).
This program is ongoing in all three experimental halls and will be
continued once the energy upgrade to 12 GeV of the JLab accelerator is
completed in 2014. In the following we give some of the experimental details
for all three halls; the results achieved so far will be discussed in the relevant
Sections~\ref{subsec:PDFexp}--\ref{sec:duality}.

The spin structure program in JLab's Hall A is focussed on the neutron, using 
a polarized $^3$He target (Fig.~\ref{poltarg2}) as an effective polarized neutron target. 
The target polarization direction can be oriented longitudinal or
transverse to the beam direction. Measurements of polarized
cross-sections and asymmetries in the two orthogonal directions
allow a direct extraction of $g_1$, $g_2$, $A_1$ and $A_2$. 
A series of high precision
experiments~\cite{Amarian:2002ar,Amarian:2003jy,Amarian:2004yf,Meziani:2004ne,
Chen:2005td,Zheng:2003un,Zheng:2004ce,Kramer:2005qe,Solvignon:2008hk,E97110} 
measured $g_1$ and
$g_2$ in a wide range of kinematics, from very low $Q^2$ ($\approx  0.01$~GeV$^2$) 
up to 5~GeV$^2$ and from the elastic peak to the DIS region ($W \approx  3$~GeV).
A pair of High Resolution Spectrometers (HRS) are used to detect the scattered
electrons. The HRS have angular acceptances of $\approx 6$ msr and momentum acceptances
of $\approx  9\%$. Their angular range is $12.5^\circ$ to $160^\circ$ and can reach as low
as $6^\circ$ with the addition of a septum magnet. The high luminosity of 
$10^{36}$ cm$^{-2}$s$^{-1}$ allowed for precision measurements at numerous HRS momentum and angular settings to cover a wide swath in the ($Q^2$,W)-plane. The electron
detector package consists of vertical drift chambers (for momentum analysis and vertex reconstruction), scintillation counters (for data acquisition trigger),
gas Cherenkov counters and lead-glass shower calorimeters (for particle 
identification). The HRS optical property and acceptance have been carefully
studied. Absolute cross sections are measured to a level of 2-3\% precision.
Asymmetries are measured to a level of 4-5\% precision, mostly due to the
uncertainties from the beam and target polarization measurements. 
The spin structure functions $g_1$ and $g_2$ are extracted using 
polarized cross section differences in which contributions from unpolarized 
materials, such as target windows and nitrogen,
cancel. Corrections for the two protons in $^3$He are still needed
since they are slightly polarized due to the D state (~8\%) and S$^\prime$  state
(1.5\%) of the $^3$He wave function~\cite{Friar:1990vx}. Corrections for binding
and Fermi motion are applied using state-of-the-art $^3$He 
calculations~\cite{CiofidegliAtti:1996cg,Bissey:2001cw}. Uncertainties due to the nuclear 
corrections have been studied~\cite{CiofidegliAtti:1996cg}. In the region of DIS and for the 
extraction of moments, the uncertainties are usually small, typically less 
than 5\%.

\begin{figure}[htb!]
\begin{center}
\epsfig{file=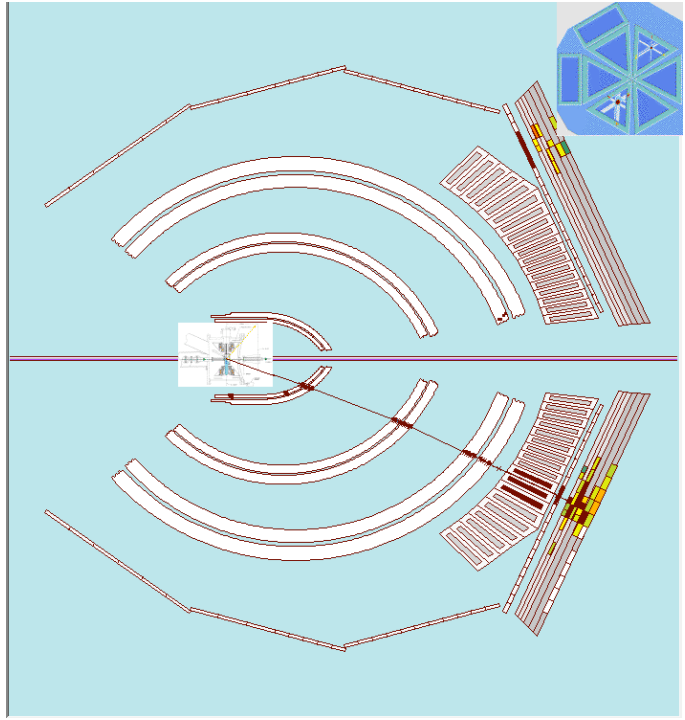,scale=0.75}
\begin{minipage}[t]{16.5 cm}
\caption{Schematic of the CEBAF Large Acceptance Spectrometer (CLAS)
with the polarized target. 
Charged particle tracks are measured with three
layers of drift chambers, while electrons can be identified using a set of
Cherenkov counters, time-of-flight scintillators and a electromagnetic 
calorimeter (the latter also shown in end-on view).
\label{CLAS}}
\end{minipage}
\end{center}
\end{figure}

The EG1--EG4 series of experiments
 in JLab's Hall B (see Fig.~\ref{CLAS}) has as its goal to
map out the asymmetry $A_1$ and the spin structure function $g_1$ of both 
nucleons over the largest, continuous kinematic range accessible. 
It makes use of the CEBAF Large Acceptance Spectrometer (CLAS) in Hall B
that covers an angular range of about 6 degrees to over 140 degrees in
polar angle and nearly $2 \pi$ in azimuth~\cite{Mecking:2003zu}. Because of the
open geometry and the toroidal magnetic field (maintained by 6 superconducting
coils evenly distributed in azimuth), one can simultaneously detect 
scattered electrons over a wide kinematic range, as well as
secondary produced hadrons (nucleons, pions and kaons) for semi-inclusive
or exclusive channels. By combining runs with several different beam energies
from 1 to 6 GeV, a continuous coverage in $Q^2$ from 0.015 to 5 GeV$^2$
and in final state mass $W$, from the  elastic peak  ($W = 0.94$ GeV) to
the  DIS region ($W \approx 3$ GeV),
has been achieved. Inclusive
results from the EG1 experiment have been
published~\cite{Yun:2002td,Fatemi:2003yh,Dharmawardane:2006zd,Bosted:2006gp}.

The relatively low luminosity limit of this open detector (about $10^{34}$ cm$^{-2}$s$^{-1}$)
is rather well matched to the luminosity limits typical for solid state, dynamically
polarized targets. EG1--EG4 took data on both hydrogen ($^{15}$NH$_3$) and
deuterium  ($^{15}$ND$_3$). So far, only targets polarized along the beam direction
have been utilized (because of the difficulty to combine a large transverse magnetic
field with the open geometry of CLAS), which necessitates (minor) corrections of 
the measured asymmetries for the unobserved contribution from 
$A_2$ (see Eq.~\ref{eq:Alongg1A2}). A fit
to the world data on $A_2$ and  on unpolarized structure functions $R$ and 
$F_1$~\cite{Bosted:2007xd,Christy:2007ve}
is used to extract the desired spin structure function information from the measured
asymmetries. In addition to the structure function $g_1(x,Q^2)$, the CLAS data
have also yielded new results on resonance excitation and decay (via exclusive
$\pi^+$, $\pi^0$ and $\pi^-$ channels)~\cite{DeVita:2001ue,Biselli:2008ug}, 
on deeply virtual Compton scattering~\cite{Chen:2006na}, and on
single and double spin asymmetries in semi-inclusive hadron 
production~\cite{Bosted:2009}.

The first experiment completed in Hall C used a standard DNP
ammonia  target ($^{15}$NH$_3$ and $^{15}$ND$_3$) and the existing
high momentum spectrometer (HMS) for a detailed look at the resonance region at
intermediate $Q^2 \approx 1.3$ GeV$^2$. This is the only experiment
at JLab on the proton and the deuteron where both longitudinal and transverse
double spin asymmetries were measured, allowing an unambiguous separation
of the structure functions $A_1$ and $A_2$ or $g_1$ and $g_2$ up to a final
state missing mass of $W \approx 2$ GeV. First results have been
published~\cite{Wesselmann:2006mw}.

\section{Spin-dependent Parton Density Functions}

\subsection{\it The Simple Parton Model \label{subsec:parton}}
\begin{figure}[htb!]
\begin{center}
\epsfig{file=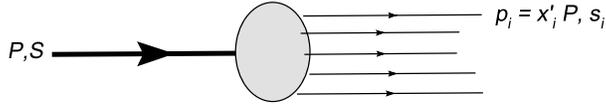,scale=0.8} 
\begin{minipage}[t]{16.5 cm}
\vspace{-0.1in}
\caption{The nucleon viewed as a collinear beam of partons.}
\label{fig2}
\end{minipage}
\end{center}
\end{figure}

The parton model was invented \cite{Feynman:1969ej} long before
QCD to explain the fact that the unpolarized
 structure functions $F_{1,2}$ apparently scale, i.e., do not decrease with $Q^2$, in contrast
 to the elastic form factors which decrease rapidly
\cite{Panofsky:1968pb}.  The partons were conceived of as
effectively massless, point-like constituents of the nucleon,
which interact electromagnetically like leptons, and it was argued
that a nucleon, in a frame where it is moving very fast, could be
viewed as a ``beam" of collinear partons, as shown in
Fig.~\ref{fig2}. The partons are characterized as having momentum
$\bm{p} = x' \bm{P} $ and covariant spin vector $s$.

\begin{figure}[htb!]
\hspace{6cm} \begin{center}
\epsfig{file=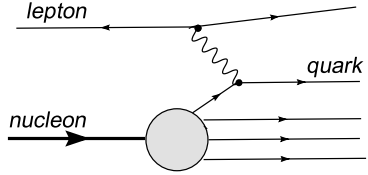,scale=0.8}
\begin{minipage}[t]{16.5 cm}
\vspace{-0.1in}
\caption{Parton model description of DIS.}
\label{fig3}
\end{minipage}
\end{center}
\end{figure}

The interaction with the hard photon is then visualized as in
Fig.~\ref{fig3}, in which the lepton-quark scattering is treated
analogously to elastic lepton-lepton scattering.
Requiring the final quark to be on mass shell,
i.e. $(p+q)^2=0$, selects the value $x'=x$.

For \textit{unpolarized} DIS one finds the scaling result
expressed in terms of the number density $q(x)$ of quarks and
 $\bar q(x)$ of antiquarks
\be \label{eq:F1partonmodel}
 F_1(x,Q^2) = \frac{1}{2}\, \sum_j \, e^2_j \,[\, q_j(x)
+ \bar q_j(x)] \ee where the sum is over flavors $j$,
$e_j$ is the charge of the quark, and the Callan-Gross
 relation~\cite{Callan:1968zza} yields
\be \label{eq:F2partonmodel}
 F_2(x)=2xF_1(x).
 \ee

\begin{figure}[htb!]
\begin{center}
 \epsfig{file=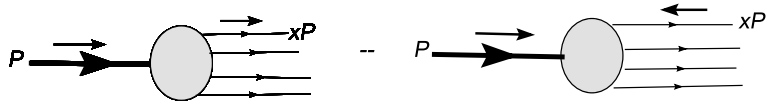,scale=0.8}
 \begin{minipage}[t]{16.5 cm}
 \vspace{-0.1in}
\caption{Visualization of the longitudinally polarized parton
density $\Delta q(x)$. The upper arrows show the spin
direction.} \label{fig5} \end{minipage} \end{center}
\end{figure}

For \textit{longitudinally polarized} DIS one obtains
\be \label{eq:g1partonmodel}
 g_1(x) = \frac{1}{2}\, \sum_j \,
e^2_j\,[ \triangle q_j(x) + \triangle \bar q_j(x)] \ee
with
\be \label{eq:deltaq}
 \triangle q(x) = q_{+}(x)- q_{-}(x) \ee
where $q_{\pm}(x)$ are the number densities of quarks whose spin
orientation is parallel or antiparallel to the 
longitudinal spin direction of
the proton (see Fig.~\ref{fig5}). In terms of these, the usual
(unpolarized) parton density is
\be \label{eq:qvsqplusminus} q(x) = q_{+}(x) + q_{-}(x) . \ee

Concerning the second spin structure function, there are many different, inconsistent results
for $g_2(x)$ in the literature. In fact there is no unambiguous
way to calculate $g_2$ in the simple parton model. For an
explanation of the problem see Section~3.4 of
\cite{Anselmino:1994gn}.

The only reliable result is the Wandzura-Wilczek relation
\cite{Wandzura:1977qf}

\be \label{eq:WW} g_2(x) \simeq -g_1(x) + \int_x^1
\,\frac{g_1(x')}{x'}\, dx' \ee

which was originally derived as an approximation in an operator
product expansion approach, but which has recently been shown to
be derivable directly in the simple parton model \cite{Cag07} .
It is discussed further in Sections~\ref{subsec:g2} and~\ref{sec:mom}.


Let us rewrite the expression for $g_1$,
Eq.~(\ref{eq:g1partonmodel}), in terms of linear combinations of
quark densities which have specific transformation properties
under the group of flavor transformations $SU(3)_F$:
\begin{equation}\label{eq:Deltaq3} \Delta q_3 =
(\Delta u + \Delta\overline{u}) - (\Delta d + \Delta\overline{d})
\end{equation}
\begin{equation}\label{eq:Deltaq8}
 \Delta q_8 = (\Delta u + \Delta\overline{u}) + (\Delta d + \Delta\overline{d}) -2(\Delta s + \Delta\overline{s})
\end{equation}
\begin{equation}\label{eq:Deltasigma}
 \Delta \Sigma = (\Delta u + \Delta\overline{u}) + (\Delta d + \Delta\overline{d}) + (\Delta s + \Delta\overline{s}) .
\end{equation}
These transform respectively as the third component of an isotopic
spin triplet, the eighth component of an $SU(3)_F$ octet, and a
flavor singlet. Then
\begin{equation}\label{eq:g1SU3}
  g_1(x) = \frac{1}{9}\left[ \frac{3}{4}\Delta q_3(x) +
  \frac{1}{4}\Delta q_8(x) +\Delta  \Sigma(x) \right] .
\end{equation}
Taking the first moment of this yields, for protons,
\begin{equation}\label{eq:Gamma1}
  \Gamma _1^p \equiv \int_0^1 g_1(x) dx = \frac{1}{9}\left[\frac{3}{4}a_3 +
  \frac{1}{4}a_8 + a_0 \right]
\end{equation}
where
\begin{eqnarray} \label{eq:a3,8,0}
a_3&=&\int^1_0 dx~\Delta q_3(x) \nn \\
a_8&=& \int^1_0 dx~\Delta q_8(x) \nn \\
a_0&= &\Delta \Sigma \equiv \int^1_0 dx~\Delta\Sigma (x) .
\end{eqnarray}

Now the hadronic tensor $W^{\mu\nu}$ is given by the Fourier
transform of the nucleon matrix elements of the commutator of
electromagnetic currents $J_\mu (x)$:
\begin{equation}\label{eq:Q}
W_{\mu\nu} (q;P,S) = \frac{1}{2\pi} \int d^4x ~e^{iq\cdot x}
\langle P,S|[J_\mu (x), J_\nu(0)]|P,S\rangle .
\end{equation}
In hard processes,  $x^2\simeq 0$ is important, so we can  use the
Wilson Operator Product Expansion (OPE). This
 gives moments of $g_{1,2}$ in terms of hadronic matrix elements
of certain operators multiplied by perturbatively calculable
Wilson coefficient functions. This is discussed in more detail in Section~\ref{subsec:mom}. The  $a_i$ in Eq.~(\ref{eq:a3,8,0})
turn out to be hadronic matrix elements of an octet of quark
$SU(3)_F$ axial-vector currents $J^j_{5\mu}\ (j =1,...,8)$ and a
flavor singlet axial current $J^0_{5\mu}$ 
(see Eqs.~\ref{eq:AVcurrent},\ref{eq:AVsinglet},\ref{eq:defain}). The  octet of currents
is precisely the one that controls the weak decays of the
neutron and of the octet hyperons, which implies that the
values of $a_3$ and $a_8$ are \textit{known} from
$\beta$-decay
measurements:

\be \label{eq:a38} a_3 \equiv g_A = 1.2670 \pm 0.0035 \qquad  a_8=0.585 \pm
0.025 . \ee
Hence a measurement of $\Gamma_1$, Eq.~(\ref{eq:Gamma1}), can be
considered as giving the value of the flavor singlet $a_0$.

It should be noted that the connection with hadronic matrix elements of local
operators is strictly only valid if the moments include the elastic contribution,
which appears as a delta-function at $x=1$. However in \emph{deep} inelastic scattering
the elastic contribution is completely negligible, so the experimentally tabulated moments
do not include an elastic term, as can be seen in the discussion of the extrapolation
to $x=1$ given in Sections~\ref{subsec:val} and \ref{subsec:delq}. However, in dealing with
 inelastic scattering at low $Q^2$,  especially in the resonance region, it is important to  distinguish between
 moments including the elastic contribution, for which the notation will remain $\Gamma$
 and those not including an elastic contribution which will be denoted $\bar{\Gamma}$.

Now we can put the  startling result of
 the European Muon Collaboration~\cite{Ashman:1987hv} into context:
Knowing the values of $a_3$ and $a_8$, the EMC measurement implied
\be \label{eq:a0EMC} a_0^{EMC} \simeq 0 . \ee
But in the naive parton model \be \label{eq:a0Deltasigma}
 a_0 =\Delta\Sigma = a_8 +3(\Delta s + \Delta\overline{s})
\ee
 where $\Delta\Sigma$ is given by Eq.~(\ref{eq:Deltasigma}).

 In 1974 Ellis and Jaffe \cite{Ellis:1973kp} had suggested that one could ignore the
 contribution from the strange quarks, i.e., from $\Delta s + \Delta \bar{s}$, 
 implying that
 \be \label{eq:a0EllisJaffe} a_0 \simeq a_8 \simeq 0.59 .  \ee
  Thus the EMC result Eq.~(\ref{eq:a0EMC}) is in gross contradiction with Ellis-Jaffe.
It was this contradiction which at first aroused interest in the
EMC result, but it was soon realized that their result had far
more serious consequences.

Consider the physical significance of $\Delta\Sigma(x)$. Since
$q_\pm (x)$ count the number of quarks of momentum fraction $x$
with spin component $\pm \frac{1}{2}$ along the direction of
motion of the proton (say the $z$-direction), the total
contribution to $J_z$ coming from the spin of a given flavor
quark is
\begin{eqnarray}\label{eq:Szfromquark}
\langle S_z \rangle&=& \int^1_0
dx\Biggl\{\Biggl(\frac{1}{2}\Biggl) q_+(x) +
\Biggl(\frac{-1}{2}\Biggl) q_-(x) \Biggl\} \nonumber\\
&=&\frac{1}{2} \int^1_0 dx~\Delta q(x) \,.
\end{eqnarray}
It follows that
\begin{equation}\label{eq:a0Sz}
a_0 = 2\langle S^{quarks}_z \rangle
\end{equation}
where $ \langle S^{quarks}_z \rangle $ is the contribution to
$J_z$ from the spin of all quarks and antiquarks.

Naively, in a non-relativistic constituent model one would have
expected all of the proton spin to be carried by the spin of its
quarks. In a more realistic relativistic model one expects 
$2\langle S^{quarks}_z \rangle \approx 0.6$, which, as first noted by Sehgal \cite{Sehgal:1974rz},
is actually close to the value obtained by neglecting the strange contribution in Eq.~(\ref{eq:a0Deltasigma}), but 
quite far from the EMC
value Eq.~(\ref{eq:a0EMC}) for $a_0$.

It was this discrepancy between 
 the contribution of the quark spins to the angular momentum of the proton, 
 as measured in DIS and as computed in both non-relativistic and relativistic
 constituent models of the proton, that was termed a ``spin crisis in the
parton model" \cite{Leader:1988vd}.
It should be noted, however, that Thomas and co-workers have developed a  constituent model,
 a more sophisticated version of the \emph{Cloudy Bag} model, 
 in which the nucleon is visualized as made up of valence quarks and a pion cloud,
 and the wave function includes the effect of single gluon exchange between the valence quarks. 
 The pion cloud and the strength of the one gluon exchange were treated phenomenologically, 
 determined from fits to various different static properties of the hadron octet and decuplet, 
 and it was not clear if , or to what extent, double counting was taking place. 
 However, lattice studies of the mass-splitting between the $\Delta$ and the nucleon \cite{Young:2002rx} 
 suggest that the pion cloud 
 contributes very little to the splitting and thus allows a fairly reliable estimate of the value of the 
 phenomenologically determined gluon coupling strength. It then turns out that the contribution
  of the quark spins to the angular momentum of the proton,
   in its rest frame, is $\Delta \Sigma \approx 0.35$, which is not too far from most recent experimental results.
Thus in this approach there is no significant ``spin crisis". For a review of this work see \cite{Thomas:2008bd}.

\subsection{\it The Parton Model in QCD \label{subsec:QCD}}
\begin{figure}[htb!*]
\begin{center}
\epsfig{file=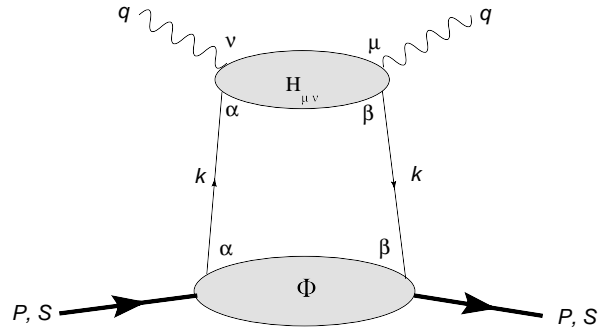,scale=0.75}
\begin{minipage}[t]{16.5 cm}
\caption{QCD generalization of parton model.} \label{fig6}
\end{minipage} \end{center}
\end{figure}

The parton model is a heuristic picture, much like the impulse
approximation in nuclear physics, and predates QCD. Once QCD is
accepted as the theory of strong interactions, with quark and
gluon fields as the fundamental fields, there will be interaction
dependent modifications of the simple parton model formulae for
DIS. The theory is invariant under color gauge transformations,
but the physical content of individual Feynman diagrams does
depend upon the gauge, and it turns out that a parton-like picture
emerges in the light-cone gauge $A_{+} =0$ where $A^\mu $ is the
gluon vector potential. The description of nucleon structure
becomes much more complicated, involving a set of twelve
functions, and the parton model number densities $q(x)$, $\Delta
q(x)$ (and the analogue for transversely spinning nucleons
$\Delta_Tq (x)$, which is not relevant for the longitudinal spin
structure) are only the principal, so called ``leading twist"
members of this set. In this review,
we shall only deal with $q(x)$ and $\Delta q(x)$. The
reaction is visualized as in Fig.~\ref{fig6}, where the top
``blob" involves a hard interaction (the photon is highly virtual)
and the bottom ``blob" involves non-perturbative soft
interactions.

The main impact of the QCD interactions will be twofold:\newline
 1) to introduce a mild, calculable logarithmic $Q^2$ dependence in the
 parton densities \newline
 2) to generate a contribution to $g_1$ arising from the
 polarization of the gluons in the nucleon \newline
We shall not go into the technical details, but simply indicate
 the physical source of these effects.

\begin{figure}[htb!*]
\begin{center}
\begin{minipage}[t]{8 cm}
\epsfig{file=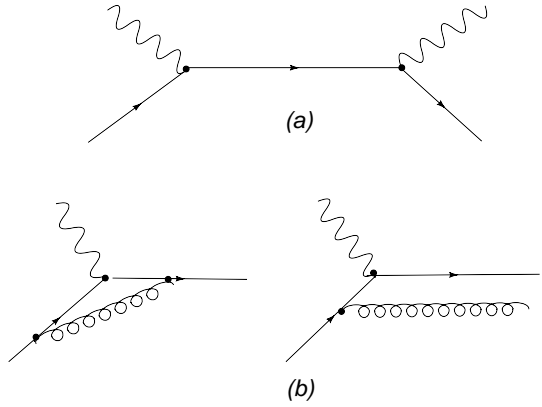,scale=0.8}%
\end{minipage} 
\end{center}
\begin{center}
\begin{minipage}[t]{16.5 cm}
\vspace{-0.1in}
\caption{Example of QCD correction terms (b), to the Born
approximation (a).} \label{fig7} \end{minipage} \end{center}
\end{figure}

1) \textit{QCD corrections and evolution}.\\
 In Fig.~\ref{fig7} we
show the Born term for the interaction of the virtual photon with
a quark (the hard blob), and the simplest correction terms, a
vertex correction and a diagram where a gluon is radiated from the
active quark before it interacts with the photon.

Unfortunately these correction terms are infinite. The infinity is caused by
collinear divergences which occur because of the masslessness
 of the quarks and which are removed by a process known as
 factorization.
 In this process the reaction is factorized (separated) into a hard and soft
 part and the infinity is absorbed into the  soft part, which in any case cannot be
  calculated and has to be parametrized and studied experimentally. 

  The point at which this separation is made is referred to as the
  \textit{factorization scale} $\mu^2$.  Schematically, one finds terms of the form
$\alpha_s \, ln\,\frac{Q^2}{m_q^2}$ which one splits as follows
\be \label{eq:factorization}
 \alpha_s \, ln\,\frac{Q^2}{m_q^2}=\alpha_s \,
 ln\,\frac{Q^2}{\mu^2}+ \alpha_s \, ln\,\frac{\mu^2}{m_q^2}
 \ee
and one then absorbs the first term on the right hand side into the hard part
 and the second into the soft part. $\mu^2$ is an arbitrary number, like
  the renormalization scale, and, in an exact calculation,  physical results cannot
  depend on it. However it does mean that what we call the parton
  density has an extra label  $\mu^2$ specifying our choice. Moreover,
  since we never calculate to all orders in perturbation theory, it
  can make a difference what value we choose. It turns out that an
  optimal choice is $\mu^2 = Q^2$, so the parton densities now
  depend on both $x$ and $Q^2$ i.e. we have $q(x,Q^2)$ and $\Delta q(x,Q^2)$,
  and perfect Bjorken scaling is
  broken. But the variation with $Q^2$ is gentle (logarithmic), and
  can be calculated via what are called the \textit{evolution
  equations} which will be discussed later.

 It turns out to be crucial in handling these divergences to use the
 technique of \textit{dimensional regularization}, which is
 straightforward in the unpolarized case, but which runs into
 ambiguities in the polarized case. As a consequence there are
 several different factorization schemes in use and it is crucial,
 when presenting results on the parton densities to specify
 which scheme is being utilized.

 At present there are three schemes in use:

i) The Vogelsang, Mertig, van Neerven scheme
\cite{Vogelsang:1995vh,Mertig:1995ny}, $\overline{MS} - MNV$
(usually just abbreviated as $\overline{MS}$). In this
scheme $a_3$ and
$a_8$ are independent of $Q^2$.

ii) The $AB$ scheme of Ball, Forte and Ridolfi \cite{Ball:1995td},
which, in addition, has the first moment

\be  \label{delsigma} \Delta \Sigma = \int^1_0 \,dx \, \Delta
\Sigma (x, Q^2) \ee

independent of $Q^2$.

iii) The JET scheme of Carlitz, Collins and Mueller
\cite{Carlitz:1988ab}, Anselmino, Efremov and Leader
\cite{Anselmino:1994gn} and Teryaev and M\"{u}ller
\cite{Mueller:1997zu}, and which is identical to the Chiral
Invariant scheme of Cheng \cite{Cheng:1997xf}. In this scheme
$a_3$ and $a_8$ are independent of $Q^2$ as is $\Delta \Sigma $,
but it can be argued that the JET scheme is superior to the others
in that \textit{all} hard effects are included in $H_{\mu \nu}$ (see Fig.~\ref{fig6}).

Of course, if one could work to all orders in perturbation theory
it would make no difference which scheme one used, but given that
we work to leading order (LO), next to leading order (NLO), and in
some cases to NNLO, the choice of scheme can be of importance.

For the polarized densities the evolution equations are
\bea \label{eq:deltaqevol} \frac{d}{d\,lnQ^2}\, \Delta q(x,Q^2)
&=& \frac{\alpha_s(Q^2)}{2 \pi }\,\int^1_x \,\frac{dy}{y}\lbrace
\Delta P_{qq}(x/y)\, \Delta
q(y,Q^2) \nn \\
& + & \Delta P_{qG}(x/y) \, \Delta G(y,Q^2)\rbrace \eea
\bea \label{eq:deltaGevol} \frac{d}{d\,lnQ^2}\, \Delta G(x,Q^2)
&=& \frac{\alpha_s(Q^2}{2 \pi }\,\int^1_x \,\frac{dy}{y}\lbrace
\Delta P_{Gq}(x/y)\, \Delta
q(y,Q^2) \nn \\
& + & \Delta P_{GG}(x/y) \, \Delta G(y,Q^2)\rbrace , \eea
 where
$\Delta G(x)$ is analogous to $\Delta q(x)$
\begin{equation}\label{eq:delG}
  \Delta G(x) = G_{+}(x)  -  G_{-}(x) .
\end{equation}

The $\Delta P$ are the polarized \textit{splitting functions} and
are calculated perturbatively
\be  \label{eq:deltaP}   \Delta P(x) = \Delta P^{(0)}(x) +
\frac{\alpha_s}{2 \pi}\, \Delta P^{(1)}(x) \ee
where the superscripts $(0)$ and $(1)$ refer to LO and NLO
contributions. For details about these the reader is referred 
to~\cite{Vogelsang:1995vh}. A pedagogical introduction to
scheme dependence can be found in Section 11.8 of
\cite{Leader:2001gr}. For a more complete set of relations between
the densities and Wilson coefficients in the various schemes, see
\cite{Leader:1998nh}.

\begin{figure}[htb!*]
\begin{center} \begin{minipage}[t]{8 cm} 
\epsfig{file=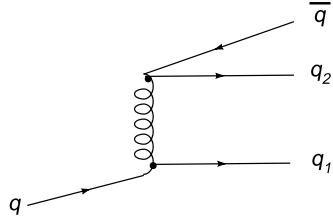,scale=0.8}
\end{minipage}
 \begin{minipage}[t]{16.5 cm}
 \vspace{-0.1in}
\caption{NLO amplitude for the $q\rightarrow \bar q$ transition.}
\label{fig8}
\end{minipage} \end{center}
\end{figure}

\begin{figure}[htb!*]
\begin{center} \begin{minipage}[t]{8 cm}
\epsfig{file=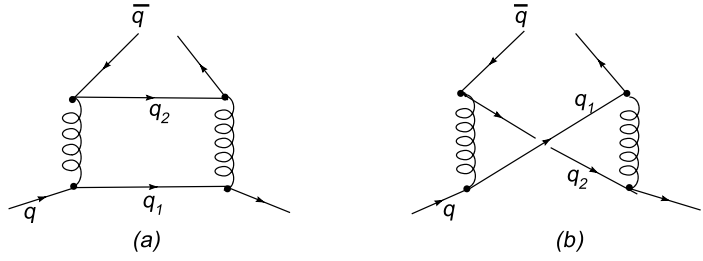,scale=0.8}
\end{minipage}  \begin{minipage}[t]{16.5 cm}
\vspace{-0.1in}
\caption{NLO contributions to the splitting function for a
$q\rightarrow \bar q$ transition.} \label{fig9}
\end{minipage} \end{center}
\end{figure}

Note that in LO flavor combinations like $q_f - q_{f'} $  (e.g.,
$u(x) -d(x)$) and valence combinations like $ q_f - \bar q_f$ (e.g.,
$ u(x) -\bar u(x)$)  are \textit{non-singlet} and evolve in the
same way, without the $\Delta G$ term in
Eq.~(\ref{eq:deltaqevol}). (There is no splitting in LO from a $q$
to a $\bar q$, nor from, say, a $u$ to a $d$.) However, in NLO
flavor non-singlets like $u(x) -d(x)$ and charge-conjugation
non-singlets like  $u(x) -\bar u(x)$ evolve differently. The
origin of this difference can be seen in Figs.~\ref{fig8} and
\ref{fig9}. Fig.~\ref{fig8} shows an NLO amplitude for a quark to
split into a $ \bar q$ .

Fig.~\ref{fig9} shows two possible contributions to $\Delta P_{q
\bar q}$ from taking the modulus squared of this amplitude. In (a)
the contribution is pure flavor singlet and involves only gluon
exchange, whereas in (b) the contribution is non-singlet. However,
if we try to do something similar for a flavor changing splitting
function e.g. $\Delta P_{du}$ we find that we cannot construct the
non-singlet diagram.

The expression for $g_1(x,Q^2) $ now becomes
\bea \label{eq:g1NLO}
 g_1(x,Q^2)&=& \frac{1}{2}\sum_{flavors}\, e_q^2 \, \Big\lbrace  \Delta
 q(x,Q^2) +  \Delta \bar {q}(x,Q^2)  \nn \\
 & +& \frac{\alpha_s(Q^2)}{2 \pi}\int_x^1 \, \frac{dy}{y}
 \{ \Delta C_q(x/y) \, [\Delta
 q(y,Q^2) +  \Delta \bar {q}(y,Q^2) ] \nn \\
 &+& \Delta C_G(x/y)\, \Delta G(y,Q^2) \} \Big\rbrace
 \eea
 where $\Delta C_G$ and $\Delta C_q $ are Wilson coefficients
 evaluated from the hard part calculated beyond the Born
 approximation.
 Note that very often the evolution equations are written  using
 the \textit{convolution notation}, for example,
 \be \label{eq:convolution}
\Delta C_q \otimes \Delta q \equiv \int_x^1 \frac{dy}{y}\, \Delta
C_q(x/y) \,\Delta q(y) . \ee

\begin{figure}[htb!*]
\begin{center} \begin{minipage}[t]{10 cm}
\epsfig{file=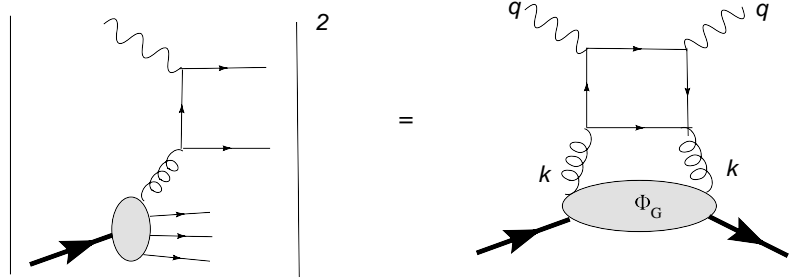,scale=0.8}
\end{minipage} 
\begin{minipage}[t]{16.5 cm}
\vspace{-0.05in}
\caption{QCD diagram leading to the anomalous gluon contribution.}
\label{fig10}
\end{minipage} \end{center}
\end{figure}

2) \textit{The gluon contribution to $g_1$.} \\
Fig.~\ref{fig10}
demonstrates a NLO gluon-initiated contribution to DIS.

\begin{figure}[htb!*]
\begin{center} \begin{minipage}[t]{6 cm}
\epsfig{file=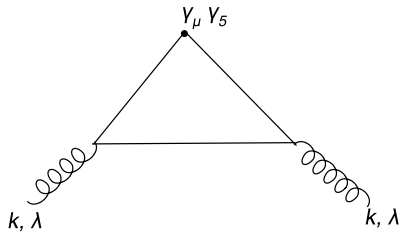,scale=0.8}
\end{minipage}
 \begin{minipage}[t]{16.5 cm}
 \vspace{-0.1in}
\caption{Feynman diagram responsible for the anomaly.}
\label{fig11}
\end{minipage} \end{center}
\end{figure}

In the Bjorken limit, for the longitudinal polarized case, this
turns out to involve the gluonic version of the famous Adler
\cite{Adler:1969gk}, and Bell and Jackiw \cite{Bell:1969ts}
anomalous triangle diagram shown in Fig.~\ref{fig11}.
The net result is that there is an anomalous gluonic contribution
to the  flavor singlet 
$a_0$ \cite{Carlitz:1988ab, Efremov:1988zh,Altarelli:1988nr,Leader:1988iq}
\bea \label{eq:a0gluon}
a^{gluons}_0(Q^2)&=&-3\,\frac{\alpha_s(Q^2)}{2\pi}~\int^1_0 dx~\Delta
G(x,Q^2)\nonumber \\
&\equiv&-3\,\frac{\alpha_s(Q^2)}{2\pi}\,\Delta G(Q^2) . \eea

Note that the factor $3$ corresponds to the number of
\textit{light} flavors i.e. $u,d,s$. Heavy flavors do not
contribute.

So, there exists potentially a \textit{gluonic} contribution to the first moment
of $g_1$:
\begin{equation}\label{eq:g1gluon}
\Gamma^{gluons}_{1}(Q^2) = -\frac{1}{3}\,
\frac{\alpha_s(Q^2)}{2\pi}\,\Delta G(Q^2)\, !
\end{equation}
\textit{This result is of fundamental importance}.
It implies that the  simple parton model formula
Eq.~(\ref{eq:a3,8,0}) for $a_0$ (and hence for $\Gamma^p_1$) in terms
of the $\Delta q_f$ could be considered as incomplete.
Instead,
\begin{equation}\label{eq:a0NLO}
a_0 = \Delta \Sigma -3 \, \frac{\alpha_s}{2\pi}~\Delta G .
\end{equation}

However, it should be noted that there is a subtlety concerning
Eq.~(\ref{eq:a0NLO}). It turns out that the result actually
depends on the factorization scheme utilized. Eq.~(\ref{eq:a0NLO})
is correct in the AB and JET schemes, but the gluon contribution
to $a_0$ is zero in the $\overline{MS}$ scheme. Now the spin
crisis emerged  from comparing the size of $\Delta \Sigma $ with
what would be expected if the quark spins dominated the nucleon's
angular momentum. But in NLO, in the $\overline{MS} $ scheme,
$\Delta \Sigma $ varies with $Q^2$, so one could argue
that it  cannot be directly 
interpreted as the spin contribution of the quarks. It is the
invariant AB or JET value of $\Delta \Sigma $ which should be
interpreted directly as a spin.

The fundamental conclusion is that the small measured value of
$a_0$ does not necessarily imply that the physically meaningful,
invariant $\Delta \Sigma$ is small. This discovery was hailed as a
resolution of the spin crisis. We shall see later that this seems
to be a false hope.

\subsection{\it Experimental Determination of Polarized Parton Densities \label{subsec:PDFexp}}
\begin{figure}[htb!]
\begin{center}
\parbox[t]{3.2in}{\epsfig{file=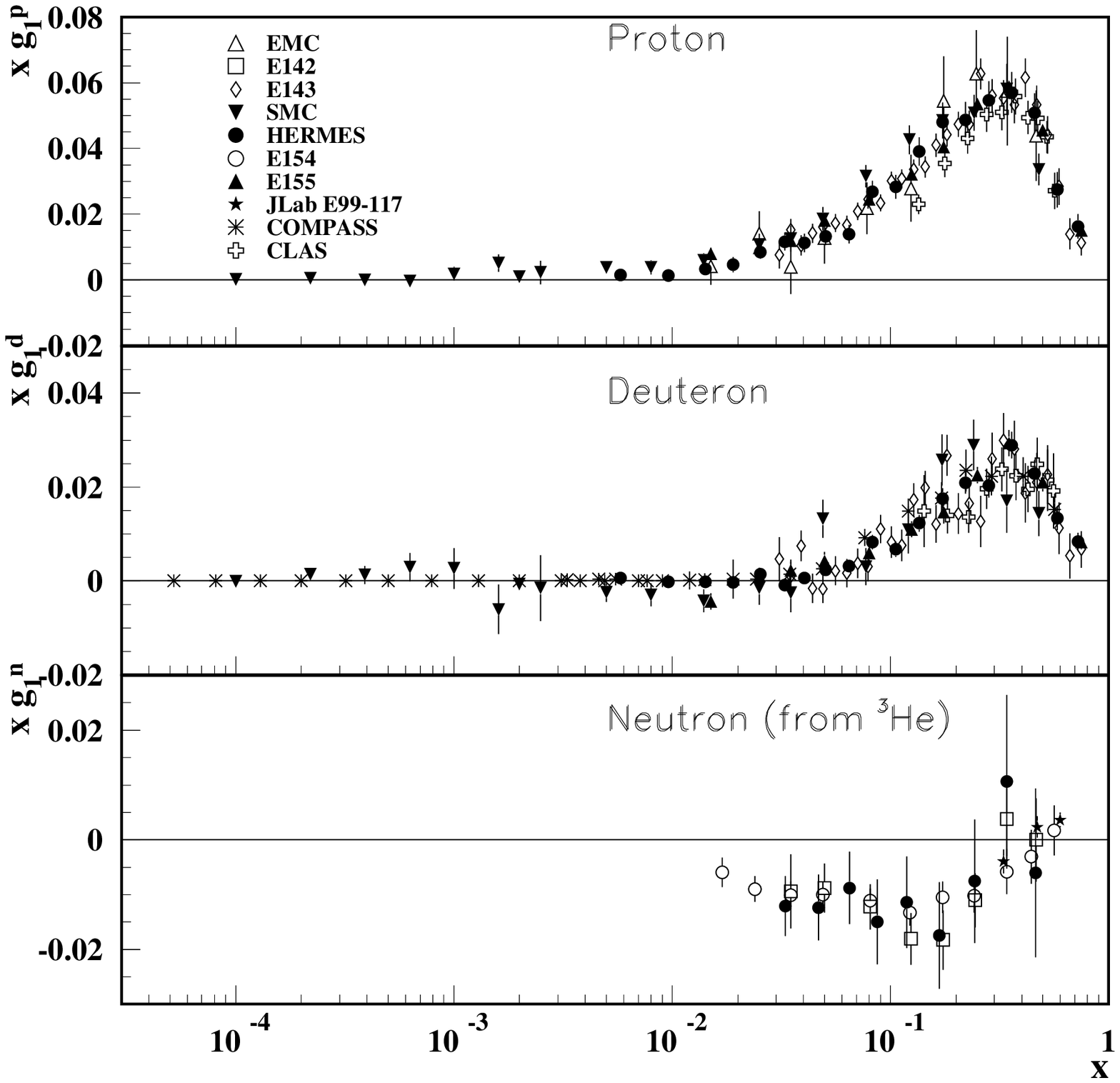,scale=0.4}}
\parbox[t]{3.2in}{\epsfig{file=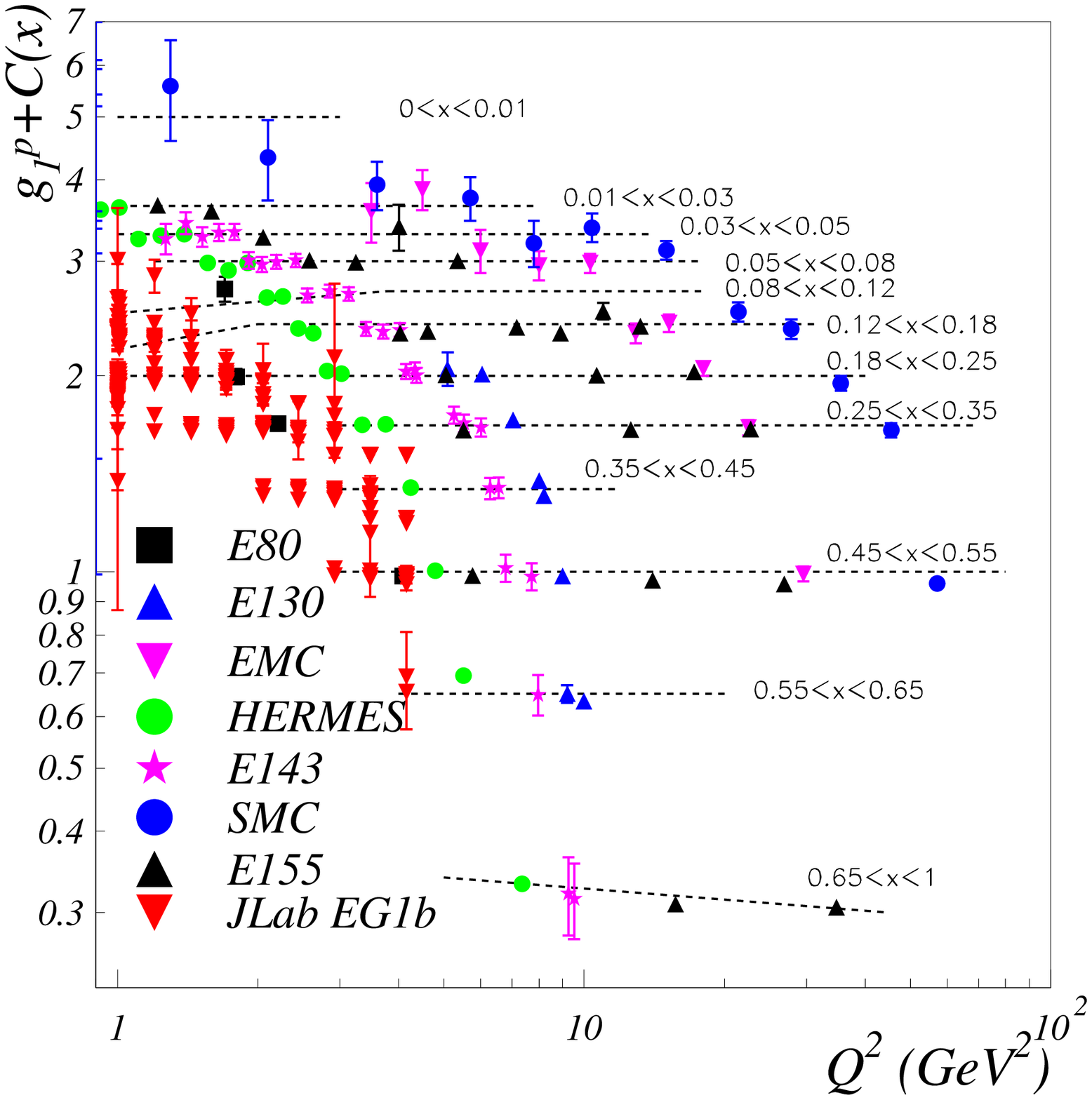,scale=0.4}}
\begin{minipage}[t]{16.5 cm}
\vspace{-0.2in}
\caption{World data on the polarized structure function $x g_1(x)$ 
for the proton, deuteron and neutron
in the DIS region ($W > 2$~GeV), taken by
different experiments, at several different values of $Q^2$, as compiled by the
Particle Data Group~\protect{\cite{Amsler:2008zz}} (left panel). 
The proton data are also shown versus $Q^2$, for several bins in $x$~\protect{\cite{Deur:2008}}
(right panel; the dashed lines just connect the data to guide the eye).
Note that much of the data
at very small $x$ is for $Q^2 < 1$~GeV$^2$.
\label{g1world}}
\end{minipage}
\end{center}
\end{figure}

From the discussion in the previous two sections, 
it is clear that the bulk of information on polarized parton densities in the nucleon
comes from charged lepton scattering experiments in the deep inelastic (DIS) region,
with final state masses larger than $W = 2$ GeV and momentum transfers
in excess of $Q^2 = 1$ GeV$^2$. As described in Section~\ref{subsec:exp},
a vast amount of data on the inclusive spin structure function $g_1(x,Q^2)$
has been accumulated by lepton scattering experiments at SLAC, CERN, DESY
and Jefferson Lab. The Particle Data Group's compilation~\cite{Amsler:2008zz}
 of these data plotted vs. $x$
are shown in Fig.~\ref{g1world} on the left side, while
the right side displays the $Q^2$-dependence of $g_1^p(x,Q^2)$ for several fixed $x$ bins.
The observed mild violation of scaling 
is sensitive to the polarized gluon density (see Sections~\ref{subsec:QCD}
and~\ref{subsec:delq}). 
In Section~\ref{subsec:val} we highlight in particular
the new data taken at high values of $x$, where valence quarks dominate the
measured spin structure functions. In that region, valence parton densities can be
uniquely determined from DIS data on the proton and the neutron.

\begin{figure}[htb!]
\begin{center}
\epsfig{file=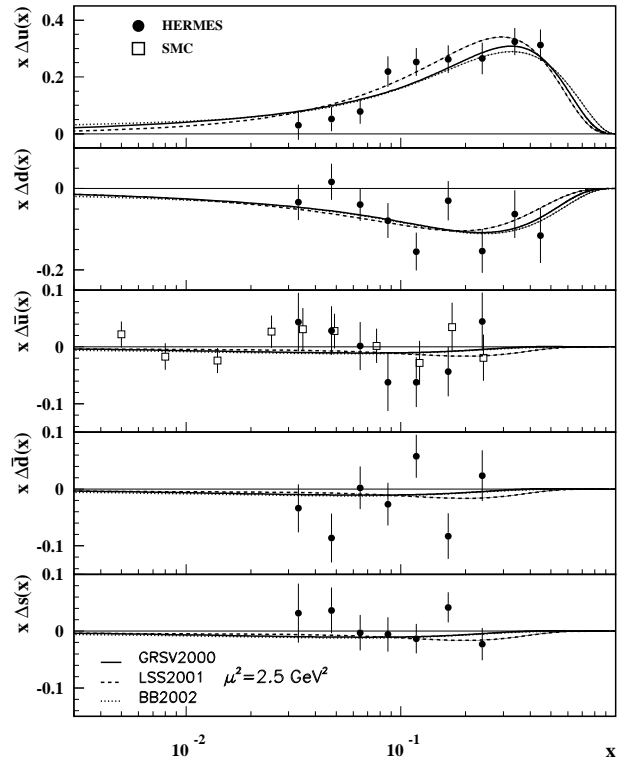,scale=0.45}
\begin{minipage}[t]{16.5 cm}
\vspace{-0.2in}
\caption{World data on polarized parton densities $\Delta q(x)$
extracted from semi-inclusive DIS data, as compiled by the
Particle Data Group~\protect{\cite{Amsler:2008zz}}.
\label{SIDISworld}}
\end{minipage}
\end{center}
\end{figure}

Inclusive data alone, however, are not able to distinguish the contributions from 
quark and antiquark densities. One approach to gather additional information
has been to use semi-inclusive lepton scattering (SIDIS), where in addition to the scattered
lepton one detects a leading hadron (typically a pion or kaon) in the final state.
Struck quarks of different flavors have varying probabilities to produce 
those final state hadrons, as expressed through the various fragmentation functions. 
These are principally extracted from $e^+ e^-$ collision experiments, 
and still have considerable uncertainties (in particular for kaons).
By comparing the asymmetries of 
semi-inclusive $\pi^+, \pi^-$ and $K^+, K^-$ production from protons
and neutrons, one can try to separate out the contributions from all quark
flavors. This approach was pioneered by the SMC collaboration~\cite{Adeva:1995yi}
and the most detailed data set stems from the HERMES 
collaboration~\cite{Airapetian:2004zf,Airapetian:2008qf}, as shown in Fig.~\ref{SIDISworld}.
The COMPASS collaboration has also measured semi-inclusive
channels~\cite{Alekseev:2007vi}.

While the virtual photon in deep inelastic lepton scattering couples directly to the
quarks (via their electric charges), gluons can only be probed indirectly in this reaction.
Their main influence is in the gentle (logarithmic) evolution of the quark densities with $Q^2$
(see r.h.s. of Fig.~\ref{g1world}),
as outlined in Section~\ref{subsec:QCD}. 
The resulting violation of Bjorken-scaling can be exploited to extract the contribution of gluons to the 
nucleon momentum and spin. This has been spectacularly successful in the case of unpolarized
DIS, where (thanks to the existence of the lepton-nucleon collider HERA) the range of $Q^2$ that has
been accessed experimentally exceeds 5 orders of magnitude for some (fixed) values of $x$.
The situation is much more difficult in the case of polarized structure functions; so far, only
fixed target experiments have measured double spin observables in lepton-nucleon
scattering. The highest $Q^2$ available are of order 100 GeV$^2$ (using the muon beam at CERN).
Recent data from COMPASS~\cite{Alexakhin:2006vx} 
have dramatically improved the precision of $g_1$ at relatively high $Q^2$
 for the deuteron which is an isoscalar target and therefore more sensitive
to gluon contributions.
For the maximum lever-arm in $Q^2$, it is important to push for high precision at both extremes
of the $Q^2$ range. In this context, new deuteron
results from Jefferson Lab~\cite{Dharmawardane:2006zd}
play an important role. 

\begin{figure}[htb!]
\begin{center}
\epsfig{file=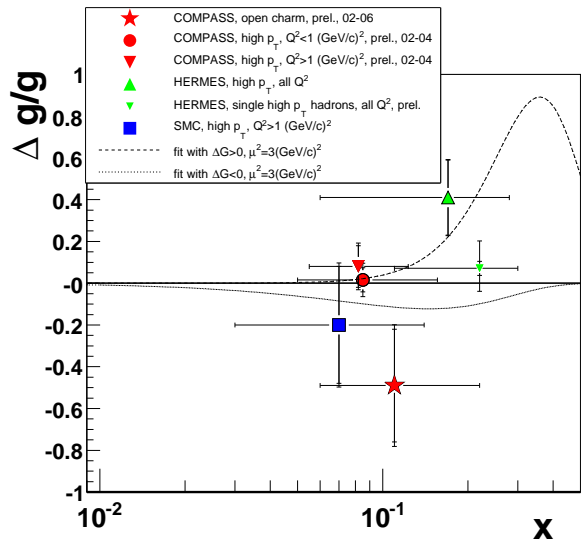,scale=0.4}
\begin{minipage}[t]{16.5 cm}
\vspace{-0.2in}
\caption{Results (some preliminary) on the gluon polarization $\Delta G/G$ extracted
(at leading order) from photon-gluon fusion processes. (Data compiled by the
COMPASS collaboration~\protect{\cite{Robinet:2008}}).
\label{dGComHerm}}
\end{minipage}
\end{center}
\end{figure}

For a direct approach, it is highly desirable to measure observables that are more closely
linked to the polarized gluon density in the nucleon, e.g. 
virtual or real photon-gluon fusion (see Fig.~\ref{fig10}) in semi-inclusive
lepton scattering or photon absorption, with two quarks in the final state stemming
from the splitting of a gluon. In order to distinguish this process from the more copious
direct quark diagrams, one has to select either events in which the final state
hadrons or jets carry large transverse momentum $p_T$ or where the final state quark
has a heavy flavor (e.g., charm) unlikely to stem from the nucleon itself. This
approach has been used by HERMES and COMPASS 
and was to be employed by the (unfortunately terminated) SLAC experiment E161.

The existing results (including preliminary data)
are shown in Fig.~\ref{dGComHerm}. The 
new  Hermes data point (inverted triangle)
 is extracted from high-$p_T$ virtual photon production of a single hadron, while the COMPASS data points come from high-$p_T$ hadron pairs or jets and from
open charm production data
($D^0$ and $D^*$; indicated by the star symbol in Fig.~\ref{dGComHerm}).
These data put some restrictions on the magnitude of
the polarized gluon density $\Delta G(x)$ in the moderate $x$--region
to which they are sensitive
but still have rather large statistical errors. Also, at present no NLO analyses of these
data exist, and backgrounds from ``ordinary'' SIDIS processes as well as ``resolved photon''
contributions have to be taken into account.

An alternate route has been employed in the polarized proton collision program
at BNL, using the Relativistic Heavy Ion Collider (RHIC). One selects final state
signatures like hadrons, jets or even direct photons with large transverse momentum that indicate
an underlying hard interaction between two constituents from the two colliding
protons. 
The cross section for the production of this final state
can be expressed as a convolution of the parton densities of the two
initial protons with the hard parton-parton scattering cross section and the final
state fragmentation function (see, e.g.,~\cite{deFlorian:2008mr}). 
There is good agreement between the NLO pQCD predictions and data for the unpolarized
cross section~\cite{Adare:2007dg},
suggesting that pQCD can be applied reliably to such reactions.

\begin{figure}[hbt!]
\begin{center}
\parbox[t]{2.8in}{\epsfig{file=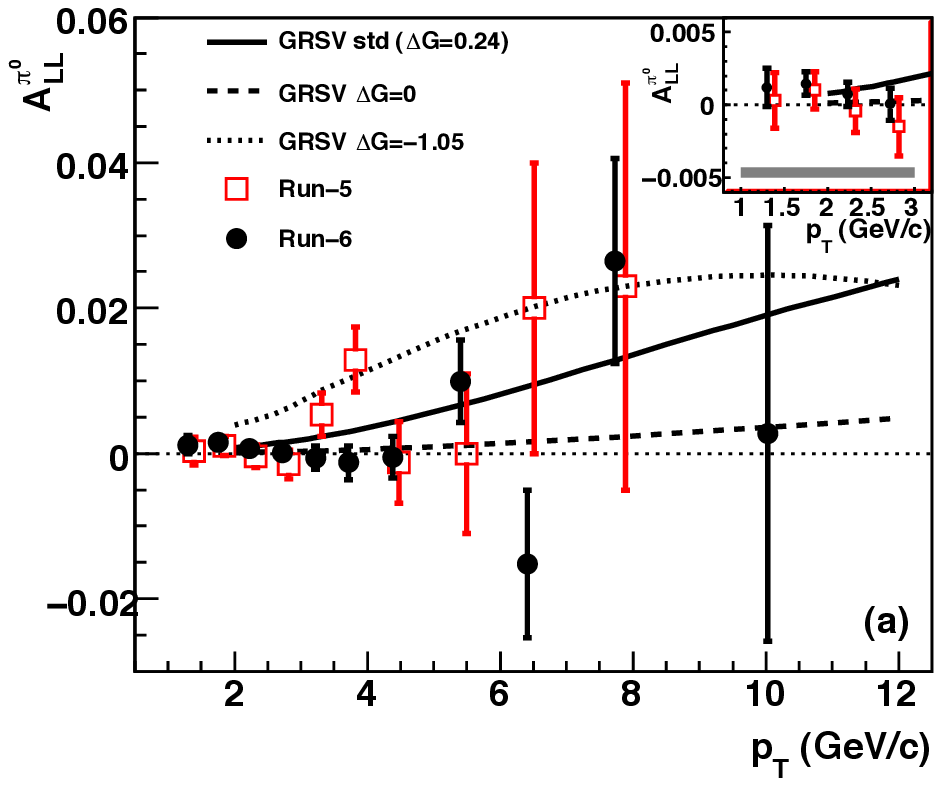,scale=0.7}}
\parbox[t]{3.5in}{\epsfig{file=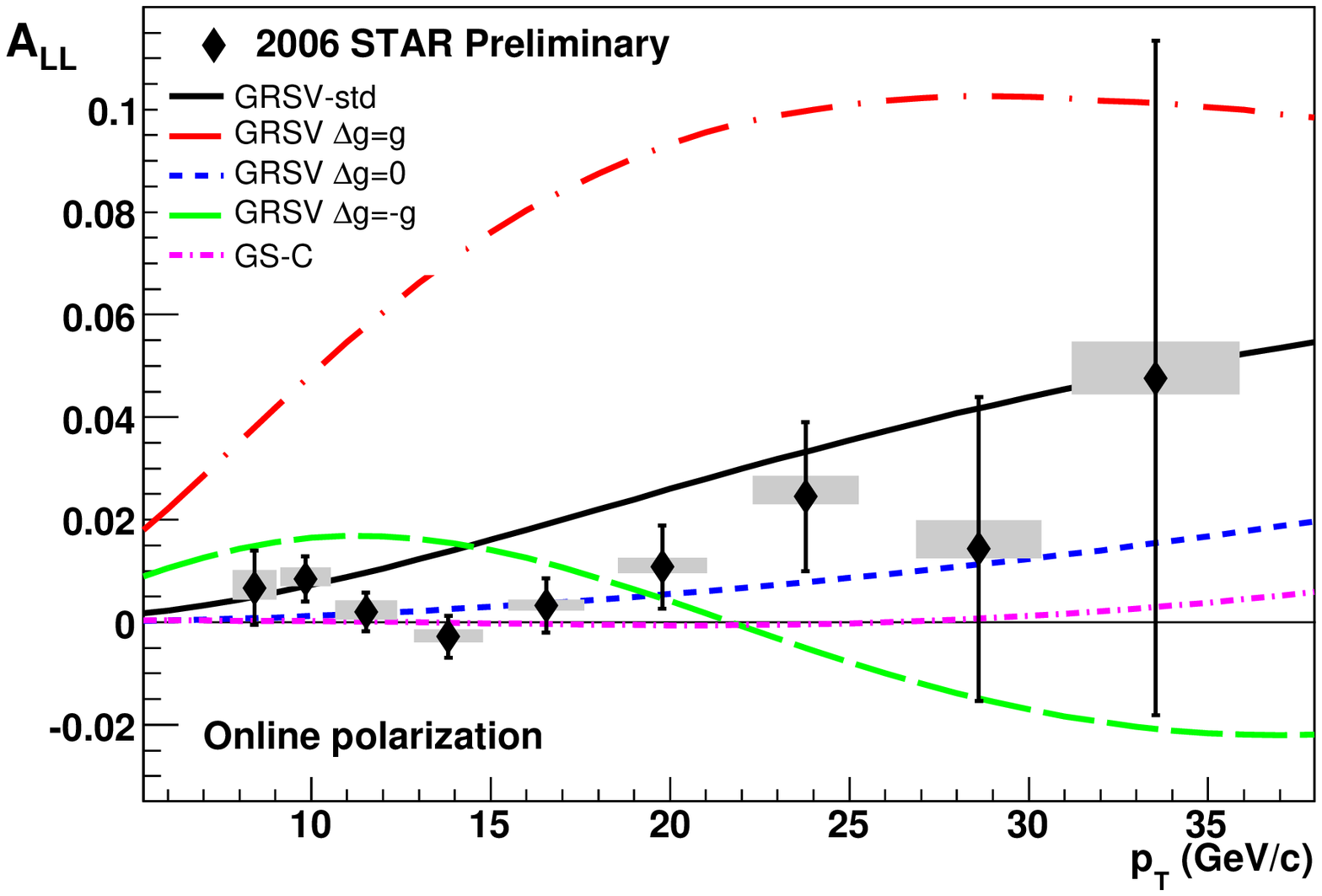,scale=0.45}}
\begin{minipage}[t]{16.5 cm}
\vspace{-5mm}
\caption {Double spin asymmetries in high $p_T$ $\pi^0$ production 
(PHENIX~\protect{\cite{Adare:2008px}}, left panel)
and jet-production 
(STAR preliminary~\protect{\cite{Staszak:2008xp}}, right panel) 
from proton proton collisions at 200 GeV
center-of-mass energy at mid-rapidity. The various curves shown are the NLO pQCD 
predictions from several versions of the polarized gluon density parametrizations
from Vogelsang et al.~\protect{\cite{Jager:2002xm}}.}
\label{RHICglue}
\end{minipage}
\end{center}
\end{figure}

For the double spin asymmetry, one divides the cross section difference for opposite proton
spins by the sum. The result depends on products of polarized parton densities; depending
on the kinematics one is mostly sensitive to quark-quark, quark-gluon or gluon-gluon
subprocesses. While this quadratic dependence does not allow a ``naive'' direct extraction
of polarized parton distributions (not even in LO), the observed asymmetry can be
compared to predictions from parametrizations of quark and gluon densities.
Figure~\ref{RHICglue} shows the preliminary
results from PHENIX ($\pi^0$ production,~\cite{Adare:2008px}) 
and STAR (jet production,~\cite{Staszak:2008xp}) 
for the double spin
asymmetry $A_{LL}$ versus transverse momentum $p_T$, compared with several 
model calculations based on the GRSV parametrization~\cite{Gluck:2000dy}
 of polarized parton densities. 
One clearly can see a preference for the curves labeled ``$\Delta G = 0$'' while some of the
more extreme possibilities ($\Delta G = \pm G$) can be
excluded. 
However, depending on the particular
overall shape of the polarized gluon distribution with $x$, non-zero values of the overall
first moment of $\Delta G$ are not excluded.

The most comprehensive analysis of all DIS, SIDIS and pp data so far has been 
conducted by de Florian, Sassot, Stratmann and Vogelsang~\cite{deFlorian:2008mr}.
The authors find good agreement between all data sets, which together yield a preference
for the integral over $\Delta G(x)$ from $x=0.05$ to $x=0.2$ to be close to zero.
For a definite answer on the question ``what is the contribution $\Delta G$ to
the nucleon spin'' we need additional data that can better constrain the higher and lower
$x$ region, to pin down both the shape and overall magnitude of $\Delta G(x)$. These
data will be forthcoming from future experiments at JLab (precision measurements
of DIS on the deuteron, both at 6 GeV and at 12 GeV beam energy) and the
continuation of the COMPASS experiment at CERN, as well as higher precision data
for different center-of-mass energies and different final state channels from RHIC.

In Section~\ref{subsec:delq}, we survey
the status of the polarized parton densities
extracted from all the data described above.

\subsection{\it Measurements of Spin Structure in the Valence Region \label{subsec:val}}

The properties of nucleon structure functions in the
high-$x$ region are of special interest, because this is where the valence 
quark contributions are expected to dominate.
With minimal sea quark and explicit gluon contributions, 
it is a clean region to test our understanding of nucleon
structure. Relativistic constituent quark models~\cite{Isgur:1998yb,Thomas:2008bd}
should be applicable in this region
and perturbative QCD~\cite{Brodsky:1994kg} can be used to make predictions in the large
$x$ limit. 

To first approximation, the constituent quarks in the nucleon are
described by SU(6) wave functions with orbital angular momentum of zero.
SU(6) symmetry leads to the following predictions~\cite{Close:1974ux}: 
\begin{equation}
d(x)/u(x)=1/2;\ 
\Delta u(x)/u(x)=2/3; \ 
\Delta d(x)/d(x)=-1/3; \ 
A_1^p(x)=5/9;\ {\rm and} \ A_1^n(x)=0.
\label{eq:SU6}
\end{equation}

Relativistic Constituent Quark Models (RCQM) with broken SU(6) symmetry, e.g., 
the hyperfine 
interaction model~\cite{Isgur:1998yb}, lead to a dominance of `quark-diquark' 
configurations 
with the spectator-diquark spin $S=0$ at high $x$. This implies that as $x\rightarrow1$:
\begin{equation}
 d/u\rightarrow 0;\ 
 \Delta u/u \rightarrow 1;\ 
 \Delta d/d \rightarrow -1/3;\ 
A_1^p\rightarrow 1;\ {\rm and} \ 
   A_1^n\rightarrow 1.
\label{eq:rnpqcd}
\end{equation}
\noindent In these RCQM models, relativistic effects lead to a 
non-zero quark orbital angular momentum and reduce the valence quark 
contributions to the nucleon spin from 1 to $0.6 - 0.75$.
 
Another approach is leading-order pQCD~\cite{Brodsky:1994kg}, which assumes the 
quark orbital angular momentum to be negligible and leads to hadron helicity 
conservation. 
For a nucleon with helicity $+1/2$
  \beq \label{Brodskyxto1}
  q_{\pm}(x) \rightarrow (1-x)^{2n-1+(1\mp 1)} \eeq
  as $x \rightarrow 1$,
  where $n=2$ is the number of spectators. The exponent for $q_+$ is therefore
  3, while the strength of $q_-$ is suppressed by an extra factor of $(1-x)^2$ at large $x$.
  This, plus SU(6), leads to:
\begin{equation}
d/u\rightarrow 1/5;\ 
\Delta u/u \rightarrow 1;\ 
\Delta d/d \rightarrow 1;\ 
A_1^p\rightarrow 1;\ {\rm and} \ 
 A_1^n\rightarrow 1.
\label{eq:rnppqcd}
\end{equation}
in agreement with \cite{Farrar:1975yb}. 
\noindent
 Eq.~(\ref{Brodskyxto1})
cannot hold at all $Q^2$, since evolution causes the power of $(1-x)$ to grow
like $loglog\,Q^2$ at large $Q^2$, but Eq.~(\ref{eq:rnppqcd}) should hold at 
all $Q^2$. 

Not only are the limiting values at $x=1$ important, but also
the behavior as $x$ approaches 1, which is sensitive to
the dynamics in the valence quark region. 
A new approach~\cite{Avakian:2007xa} in pQCD, including quark orbital angular momentum 
(and therefore not requiring hadron helicity conservation), 
shows a different behavior at large $x$ region while keeping the
same limiting values at $x=1$. In particular, the approach of $\Delta d/d$ towards
unity only  sets in at a significant larger $x$ than predicted by hadron helicity conservation 
(see below).

\begin{figure}[!hbt!]
\begin{center}
\begin{minipage}[t]{8 cm}
\epsfig{file=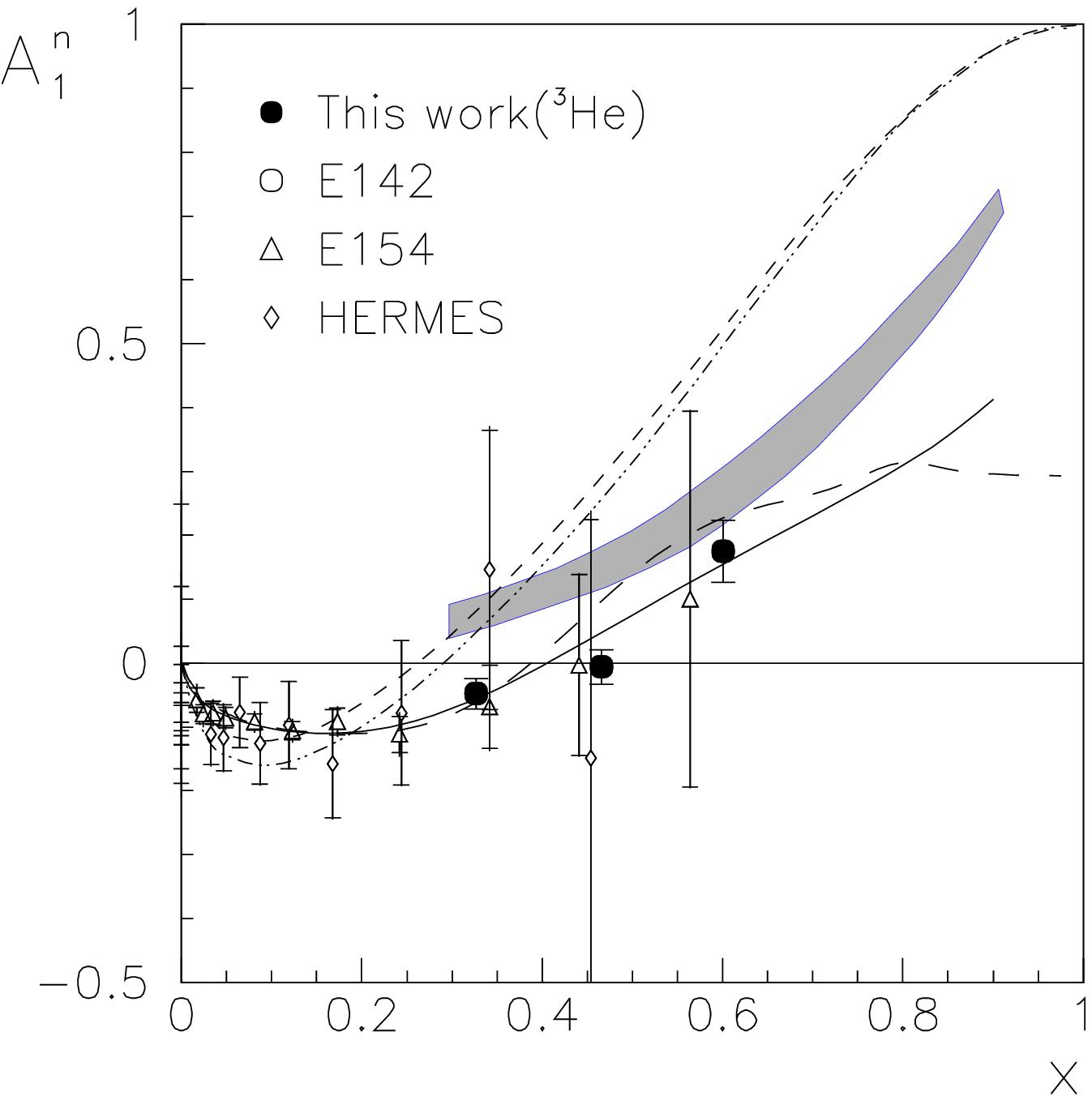,scale=0.55}
\end{minipage}
\begin{minipage}[t]{8 cm}
\epsfig{file=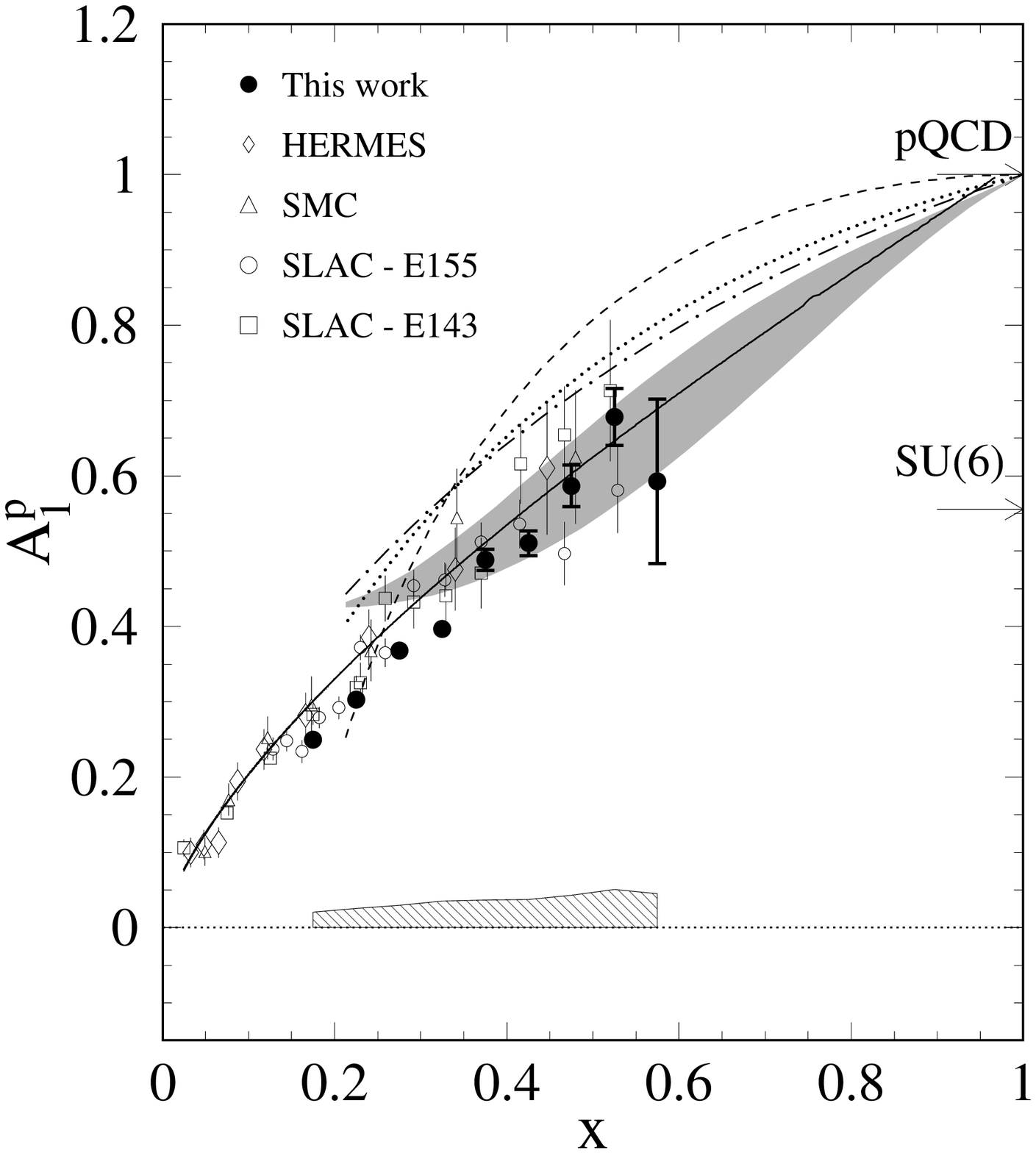,scale=0.4}
\end{minipage}
\begin{minipage}[t]{16.5 cm}
\vspace{-0.2in}
\caption {A$_1^n$ (left-panel) and A$_1^p$ (right-panel)
results from JLab Hall A 
E99-117~\protect\cite{Zheng:2003un,Zheng:2004ce} and CLAS 
EG1b~\protect\cite{Dharmawardane:2006zd} experiments
(filled circles), compared with 
the world data
and theoretical predictions (see text for details).}
\label{highxfig}
\end{minipage}
\end{center}
\end{figure}



Experimentally, it is difficult to access the high-$x$ region because most
high-energy experiments so far lacked the necessary luminosity and
fine-enough resolution in $x$. This situation has dramatically changed
with the advent of the high-current, moderate energy
 continuous electron beam at Jefferson Lab.
 
JLab Hall A experiment E99-117~\cite{Zheng:2003un,Zheng:2004ce} measured
the neutron asymmetry  $A_1^n$ with 
high precision in the $x$ region from 0.33 to 0.61
($Q^2$ from 2.7 to 4.8 GeV$^2$). 
Asymmetries from inclusive scattering of 
a highly polarized 5.7 GeV electron beam 
on a high pressure ($>10$ atm) (both longitudinally and
transversely) polarized $^3$He target were measured. 
Parallel and perpendicular asymmetries
were extracted. After taking into account the beam and target 
polarizations and the dilution factor,
they were combined to form $A_1^{^3He}$. Using a recent 
model~\cite{Bissey:2001cw}, nuclear
corrections were applied to extract $A_1^n$. The results on $A_1^n$
are shown in the left panel of Fig.~\ref{highxfig}. 
The experiment greatly improved the precision
of data in the high-$x$ region, providing the first evidence that 
$A_1^n$ becomes positive at large $x$, a clear evidence for SU(6) symmetry 
breaking. The results are in good agreement with the LSS 2001 pQCD
fit to previous world data~\cite{Leader:2001kh} (solid curve) and 
the statistical model~\cite{Bourrely:2001du} (long-dashed curve).
The trend of the data is consistent with the RCQM~\cite{Isgur:1998yb} predictions
(the shaded band). The data disagree with the predictions from the 
leading-order pQCD models~\cite{Brodsky:1994kg} (short-dashed and dash-dotted curves).
These data provide crucial input for the global fits to the world data to 
extract the 
polarized parton densities and the extractions of 
higher-twist effects. 

\begin{figure}[!hbt!]
\begin{center}
\begin{minipage}[t]{6 cm}
\epsfig{file=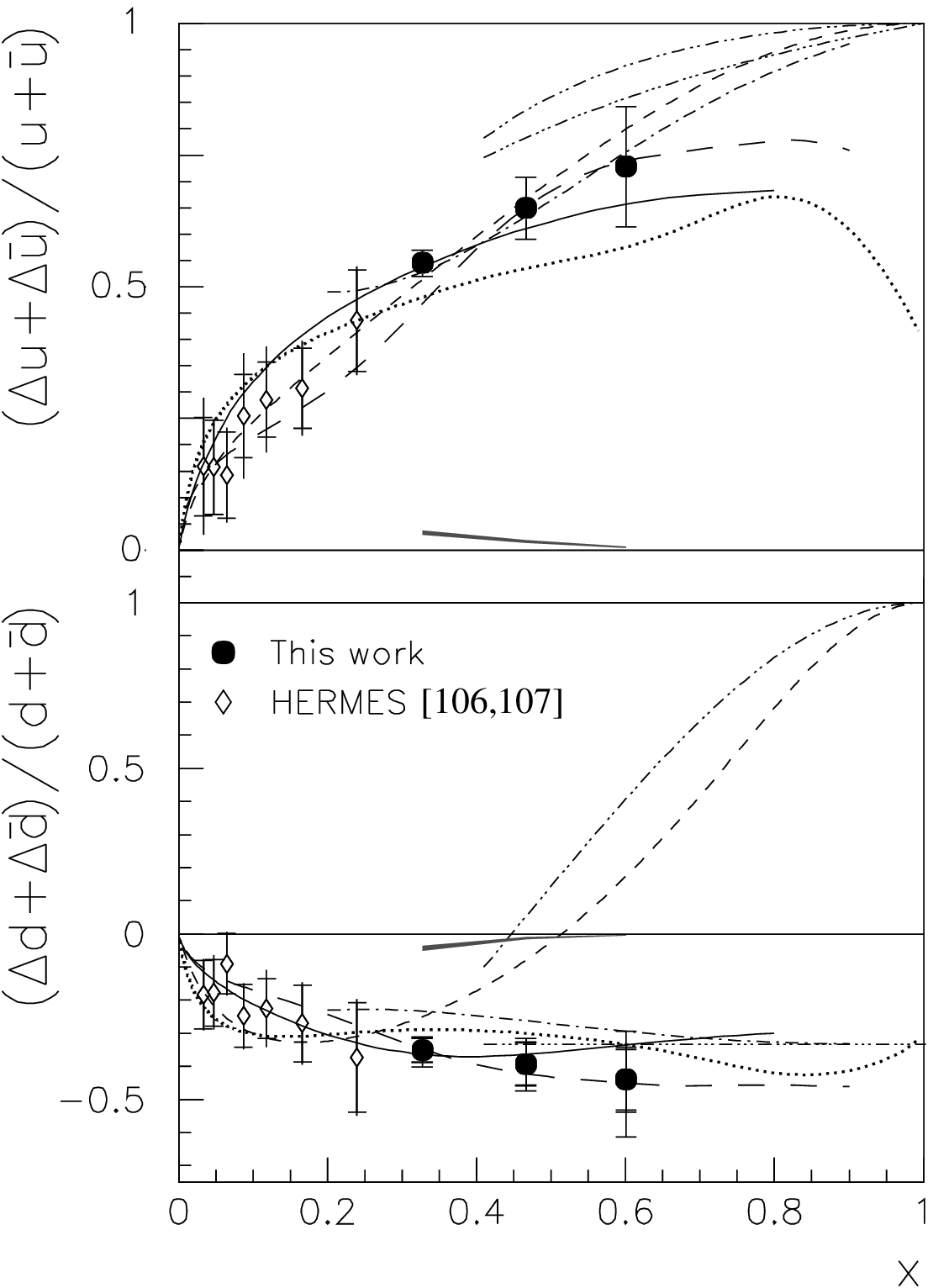,scale=0.5}
\end{minipage}
\begin{minipage}[t]{10 cm}
\epsfig{file=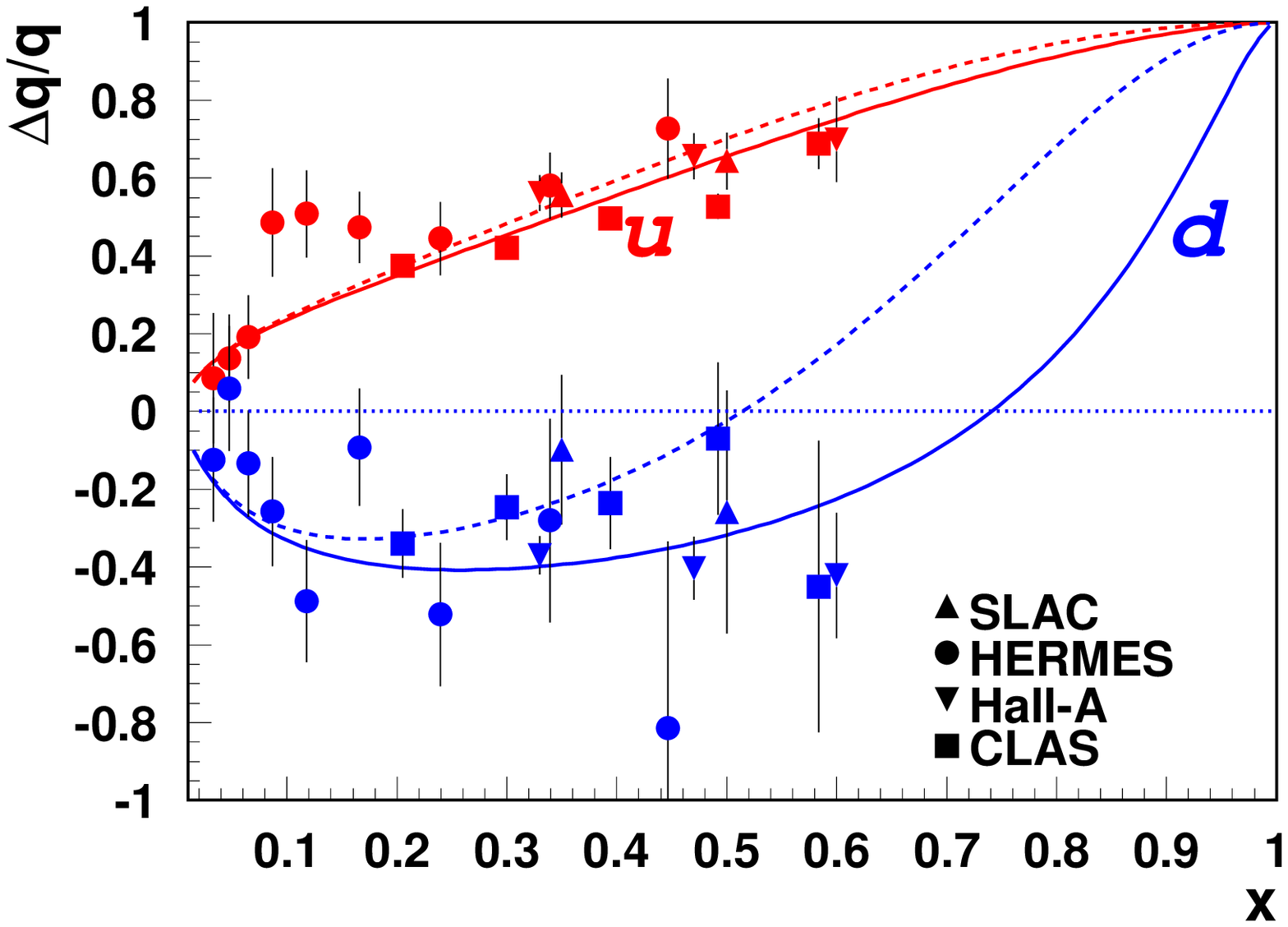,scale=0.5}
\end{minipage}
\begin{minipage}[t]{16.5 cm}
\vspace{-0.2in}
\caption {$\Delta u/u$ (upper side of both panels) 
and $\Delta d/d$ (lower side of both panels) 
results from JLab Hall A 
E99-117~\protect\cite{Zheng:2003un,Zheng:2004ce} 
(left panel)
and  from CLAS 
EG1b~\protect\cite{Dharmawardane:2006zd} (right panel), 
compared with 
the world data
and theoretical predictions (see text for details).}
\label{valencepdf}
\end{minipage}
\end{center}
\end{figure}

New results for $A_1^p$ and $A_1^d$ from the Hall B EG1b
 experiment~\cite{Dharmawardane:2006zd} 
are also available. 
The data cover the $Q^2$ range of 1.4 to 4.5 
GeV$^2$ for $x$ 
from 0.2 to 0.6 with an invariant mass larger than 2 GeV.
 The results on $A_1^p$
are shown in the right panel of Fig.~\ref{highxfig}.
The precision of the data
improved significantly over that of the existing world data.  
Similar data also exist on the deuteron and once again exhibit
a trend to exceed the asymmetries predicted by SU(6) 
(Eq.~\ref{eq:SU6}) at large $x$.
In addition to the prediction by the RCQM model~\cite{Isgur:1998yb} (again
indicated by the shaded band), several curves based on
different scenarios of SU(6) symmetry
 breaking as presented in the paper by Close and Melnitchouk~\cite{Close:2003wz}
 are also shown
in the right panel of Fig.~\ref{highxfig}. These scenarios are based on duality
arguments and include final state helicity-1/2 dominance (dashed curve),
spin-1/2 dominance (dotted) and symmetric wave function suppression (dash-dotted),
which is closest to the data.

The polarized quark distribution functions $\Delta u/u$ and $\Delta d/d$ in the high-$x$ region were 
first extracted, in the leading-order approximation, 
from the Hall A neutron data and the world proton data
(left panel of Fig.~\ref{valencepdf})~\cite{Zheng:2003un,Zheng:2004ce}.
A recent leading-order extraction used the CLAS EG1b proton and deuteron 
data~\cite{Dharmawardane:2006zd} as well. 
The results are shown
in the right panel of Fig.~\ref{valencepdf}, 
along with predictions from 
leading-order pQCD~\cite{Brodsky:1994kg} (dashed curves) 
and a pQCD fit (solid line) including quark orbital angular momentum 
contributions~\cite{Avakian:2007xa}.  The results of $\Delta d/d$ are in 
significant disagreement with the predictions from the
leading-order pQCD model assuming hadron helicity conservation.
Data agree better with the fit including quark-orbital angular momentum contributions, 
suggesting that the quark orbital
angular momentum may play an important role in this kinematic region.

\subsection{\it Status of Polarized Parton Densities \label{subsec:delq}}
 
From measurements of $g_1(x,Q^2)$ at various values of $x$ and
$Q^2$, one can, in principle, via Eq.~(\ref{eq:g1NLO}), extract
the quark and gluon polarized parton densities. We shall first consider general
issues related to the determination of the parton densities and then discuss
what is known about the densities themselves.
 \\

1) \textit{General properties of the parton densities}

\begin{itemize}
  \item In DIS, one can only obtain information on the combinations $
\Delta q(x,Q^2) +  \Delta \bar {q}(x,Q^2) $, i.e., there is no
information at all about the $\Delta \bar{u}$ and $\Delta \bar{d}$
sea quarks, though $ \Delta s + \Delta \bar{s} $ is determined.
Nonetheless, in presenting results, some authors use conventions
such as $ \Delta \bar{u} = \Delta \bar{d}=\Delta \bar{s}$, or
variants thereof, in order to show results for the valence quarks.
It is important to bear in mind that these results are not
physical and are entirely convention dependent.

\item To determine $\Delta \bar{u}$ and $\Delta \bar{d}$ one
 needs to study semi-inclusive deep inelastic scattering or semi-inclusive
 hadron-hadron reactions, where,
however, one also needs information about fragmentation functions.

 \item Because the gluon has zero electric charge it does not
 couple directly to the photon, so it contributes in
 Eq.~(\ref{eq:g1NLO}) only in NLO. However, as shown in
 Eq.~(\ref{eq:deltaqevol}), it also plays a role in the evolution of the parton densities,
  but since the evolution is only logarithmic and the lever arm in
  $Q^2$ of present day experiments is relatively small (\emph{much} smaller than
  in the unpolarized case), there is considerable uncertainty about the form and magnitude of
  $\Delta G(x,Q^2)$.

  \item The parametrizations assumed for the $\Delta q(x) $ at some value of $Q^2$
  follow general intuition based on the parton model.
   In particular, if the $q_{+,-}(x)$ are regarded as positive number
   densities then, via Eqs.~(\ref{eq:deltaq}, \ref{eq:qvsqplusminus})
   one imposes the \emph{positivity} condition
   \beq \label{eq:positivity}
   |\Delta q(x)| \leq q(x) .
   \eeq
 It can be shown that if positivity holds at some $Q_0^2$, then it is
   preserved under evolution to larger values of $Q^2$ \cite{Bourrely:1997nc,Artru:2008cp}, but care must be taken since there are fits
   to the \emph{unpolarized} data with
     parton densities which are
   \emph{negative} at $Q^2=1GeV/c^2$ \cite{Watt:2008hi}! This is possible
   because in the unpolarized case one has the luxury of using
   only data typically with $Q^2\geq 4 GeV/c^2 $, by which point the
   density has evolved into a positive one.

   \item There is some guidance as to the behavior as
   $x\rightarrow 1$. Perturbative QCD arguments \cite{Farrar:1975yb} suggest
  that quark densities vanish like a power of
  $(1-x)$ which depends on the helicity; see the discussion in
  Section~\ref{subsec:val}.
%
%
%
%
 In practice the exponents of $(1-x)$ are left as  free
parameters to be determined in the fit to the data, and tend to
come out close to the expected values.

\item What the behavior should be as $x\rightarrow 0$ is not very clear. Based on the
 correspondence between $x\rightarrow 0$ in the structure
functions, and energy $\nu \rightarrow \infty $ in virtual-photon
Compton scattering, one would expect the power of $x$ to be
controlled by the leading Regge intercept i.e
\beq \label{xto0Regge}
 g_1(x)\sim x^{- \alpha }
 \eeq
 where  $\alpha$ is the intercept of the $a_1(1260)$ meson Regge trajectory in the isovector channel, and
 the $f_1(1285)$ meson trajectory in the isoscalar channel \cite{Heimann:1973hq}. These
intercepts are not well known (see discussion in
\cite{Bass:2004xa} ). Roughly
 \bc  $ -0.4 \, \leq \alpha_{a_1} \approx \alpha_{f_1}\, \leq -0.18 $ . \ec

However, we know that for unpolarized DIS the effective power
$\alpha$ increases with $Q^2$ for the range of $x$ explored thus
far, contrary to the Regge picture, and this is explained as due
to  QCD evolution (soft Pomeron vs. hard Pomeron). Bass \cite{Bass:2006dq}
suggests that something
similar might be expected for the polarized case.

Study of the small $x$ behavior of the DGLAP evolution equations \cite{Ball:1995ye}
yields parton densities that grow more rapidly than any power of $ln(x_0/x)$ (where $x_0$ is the upper
limit where the treatment is valid), and the growth rate increases with $Q^2$. On the other hand
summation of double logarithmic terms at small $x$, which are not included in DGLAP, leads to the
expectation $C_s\,x^{-a_s}$ and $C_{ns}\,x^{-a_{ns}}$ for the
singlet and non-singlet contributions to $g_1$ respectively
(The coefficients C are sensitive to
the structure of the parton densities). This was originally studied  in
 \cite{Bartels:1995iu,Bartels:1996wc}
using a constant, non-running value of $\alpha_s$ and later by Ermolaev,
 Greco and Troyan \cite{Ermolaev:2003zx}
using a running coupling. Both approaches give the same structure
with $a_s > a_{ns}$, but with somewhat
 different numerical values of the powers of $x$.  In \cite{Ermolaev:2003zx} the intercepts are \bc
$a_s \approx 2 \, a_{ns} \approx 0.86 $,  \ec
implying that the singlet flavor combination ($g_1^p + g_1^n$) should diverge more rapidly
than the non-singlet combination ($g_1^p - g_1^n$) as $x$ goes to zero.
Thus one would expect
\beq  \label{ smallxineq} |g_1^p + g_1^n | \geq |g_1^p - g_1^n |
\qquad \textrm{as} \qquad x\rightarrow 0 . \eeq The only data on both flavor combinations at
extremely small $x$ values \emph{and} $Q^2\geq 1 GeV^2$ are from
the  SMC experiment \cite{Adeva:1998vw} shown in Fig.~\ref{g1p +/-
g1n}, where precisely the opposite trend is visible.
COMPASS data on the deuteron further confirm the smallness of the flavor singlet
(see Fig.~\ref{g1world}).
Further information will become available
when COMPASS publish their new data on $g_1^p$. However, as
experiments probe smaller and smaller $x$, there ought to be a
dramatic change in the trend of the data, or else the theoretical
arguments are incomplete.
 Indeed Blumlein and Vogt \cite{Blumlein:1996hb} suggest that corrections for sub-leading terms 
 could significantly change these results.
 (A more detailed discussion can be found in Section 11.8 of \cite{Leader:2001gr}; 
 for possible connections to the infrared renormalon model see the work of
 Kataev \cite{Kataev:2003jv}).
 
\begin{figure}[htb!*]
\hspace{4cm} \begin{minipage}[t]{8 cm} \epsfig{file=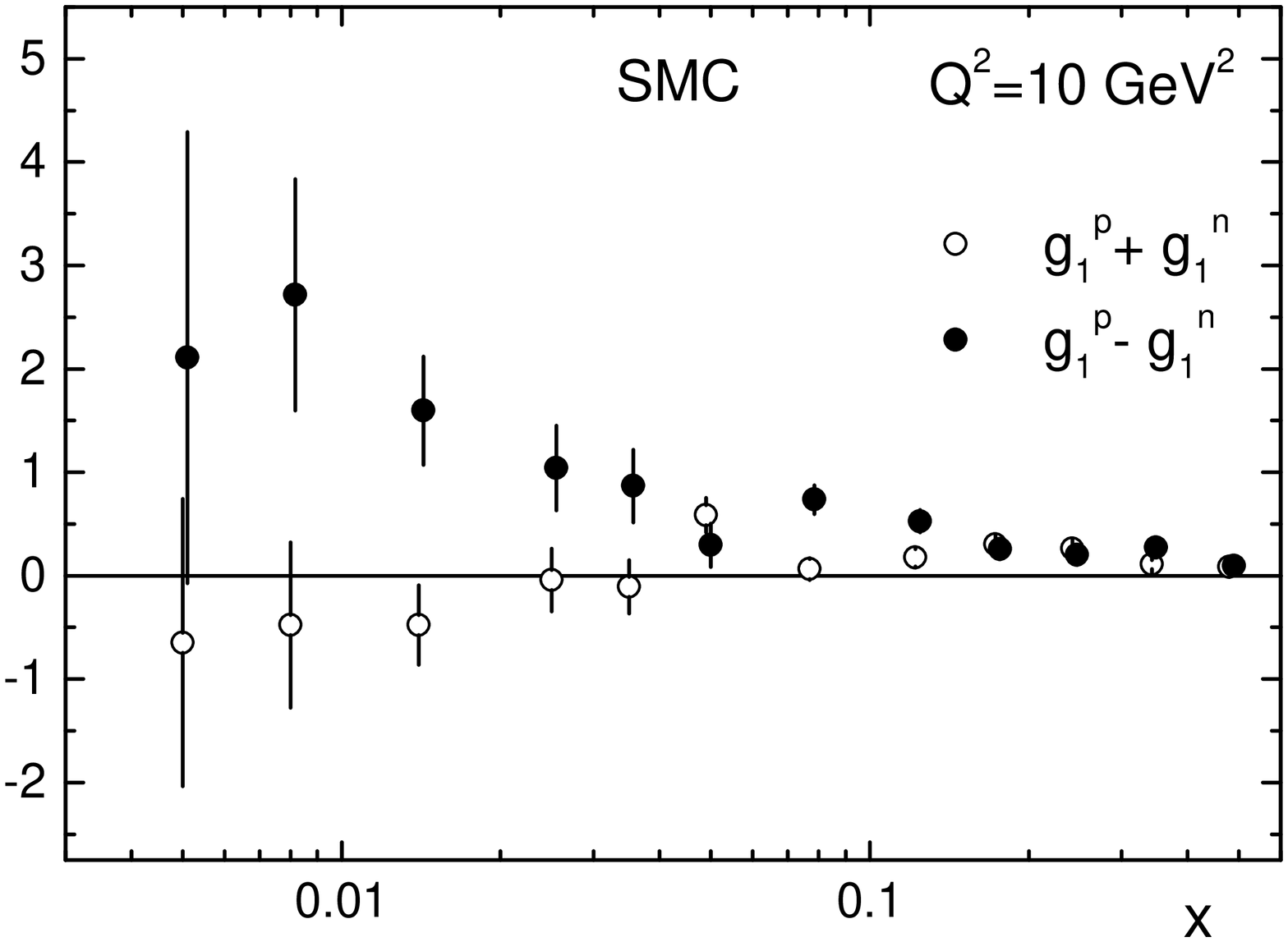,
scale=0.35, angle=-90}
\end{minipage} \begin{center}
\begin{minipage}[t]{16.5 cm}
\vspace{-0.3in}
\caption{Comparison of $g_1^p \pm g_1^n$ .} \label{g1p +/- g1n}
\end{minipage}
\end{center}
\end{figure}

It is also argued \cite{Brodsky:1994kg} that
\beq \label{smallxdG} \frac{\Delta G(x)}{G(x)}\propto x \qquad
\textrm{as}\qquad x\rightarrow 0  .\eeq
Again, in practice, the power of $x$ is taken as a parameter to be
determined by the fit to the data.

\item As explained in Section 2.2 there is, alas, no absolute physical meaning
  to parton densities in QCD. One has to specify in what factorization
  scheme they are determined. Although the differences between
  densities in different schemes are proportional to $\alpha _s(Q^2)$,
  these differences can be surprisingly large for the smaller densities
  e.g. $\Delta s(x)|_{\overline{MS}} \approx 2 \, \Delta s(x)|_{JET}$
  at  $Q^2=1 GeV/c^2$~\cite{Leader:1998nh}.

   \item Although there has been a tremendous increase in the
  amount and quality
  of polarized data in the past few years, it is still
  usual to impose, rather than test, the various sum rules
  (Bjorken, $SU(3)_{flavour}$) which should be satisfied by the
  quark densities. For a discussion of sum rule tests,
  see Section~\ref{subsec:bjork}.

 \item It is important to appreciate that the QCD formula for
  $g_1(x,Q^2)$, Eq.~(\ref{eq:g1NLO}), is a \emph{leading
  twist}(LT)
  expression, i.e it neglects terms of order $M^2/Q^2$. For
  experiments at relatively low values of $Q^2$ the expression Eq.~(\ref{eq:g1NLO})
  must be supplemented by \emph{higher twist} (HT) terms. There are two types of HT terms: target mass corrections,
  which are purely kinematical, and dynamical corrections. The target mass corrections
  can be calculated exactly and
  were first given in a closed form for $g_1$ by Piccione and Ridolfi \cite{Piccione:1997zh}.
  For a new approach, see also~\cite{Accardi:2008pc}.
   The details
  and $Q^2$ evolution of the dynamical HT terms are not known, so they are simply
  parametrized in the form $h(x)/Q^2$, with $h(x)$ determined, after applying the target mass corrections, from
  experiment at a few values of $x$. For access to the literature see \cite{Leader:2006xc}.
  
  \item It is well known that parton densities extracted in LO and
  NLO sometimes differ significantly, and one may therefore worry
about the effect of NNLO etc. corrections. Some indication of the
reliability of the NLO approximation can be obtained via the LSS
Transformation Test \cite{Leader:1998nh}. In each order of
perturbative QCD there exist transformation formulae relating
parton densities in different factorization schemes. Let us
indicate this symbolically for two schemes, $A$ and $B$:
\beq \label{eq:transform} \Delta q(x)|_B = T_{B\leftarrow
A}\,\Delta q(x)|_A . \eeq
Suppose now that $ T_{B\leftarrow A}$ is known to NLO accuracy,
and the parton densities are extracted from the data,
\emph{independently}, in NLO, using schemes $A$ and $B$, with
results $\Delta q(x)|_{A,B}^{data} \,$, respectively. If the
densities can be extracted  reliably in NLO i.e NNLO effects are
unimportant, then one should find
\beq \label{eq:tranformdata} \Delta q(x)|_B^{data} =
T_{B\leftarrow A}\,\Delta q(x)|_A^{data} . \eeq
Any failure of this equality is a measure of the importance of
NNLO effects. Thus the ratio
\bc $ \frac{ \Delta q(x)|_B^{data}\, - \, T_{B\leftarrow
A}\,\Delta q(x)|_A^{data}}{ \Delta q(x)|_B^{data} \, + \,
T_{B\leftarrow A}\,\Delta q(x)|_A^{data} } $ \ec
 gives some  indication of the reliability of the parton
 densities.

\end{itemize}

A detailed  discussion of many of the above  issues can be found
in \cite{Leader:1998qv}.
 \\

2) \textit{Extraction of parton densities from DIS}
 \\
 
Following upon the remarkable EMC experiment in 1988, there have
been  many new data on polarized DIS and many QCD analyses. We
shall only be concerned with the  recent analyses which have
included all or most of the present world data, namely, Leader,
Sidorov and Stamenov (LSS'05 and LSS'06)
\cite{Leader:2005ci,Leader:2006xc}, Alexakhin et al (COMPASS'06)
\cite{Alexakhin:2006vx},  Hirai, Kumano and Saito (AAC'03)
\cite{Hirai:2003pm}, Bl\"{u}mlein and B\"{o}ttcher (BB)
\cite{Bluemlein:2002be} and Gl\"{u}ck, Reya, Stratmann and
Vogelsang (GRSV2000) \cite{Gluck:2000dy}. We shall also comment on
the ground-breaking paper of de Florian, Sassot, Stratmann and
Vogelsang (DSSV) \cite{deFlorian:2008mr} which analyzed
simultaneously data on DIS, SIDIS and pion production in polarized
$pp$ collisions at  RHIC, and on the work of Hirai and Kumano (AAC'08) \cite{Hirai:2008aj} who analyzed data on DIS and on pion production at RHIC. Details of the parametrizations etc. can be found in the papers cited and also on the Durham HEP
Databases website~\cite{Durham}.

\begin{figure}[htb!]
\begin{center}
\parbox[t]{3.2in}{\epsfig{file=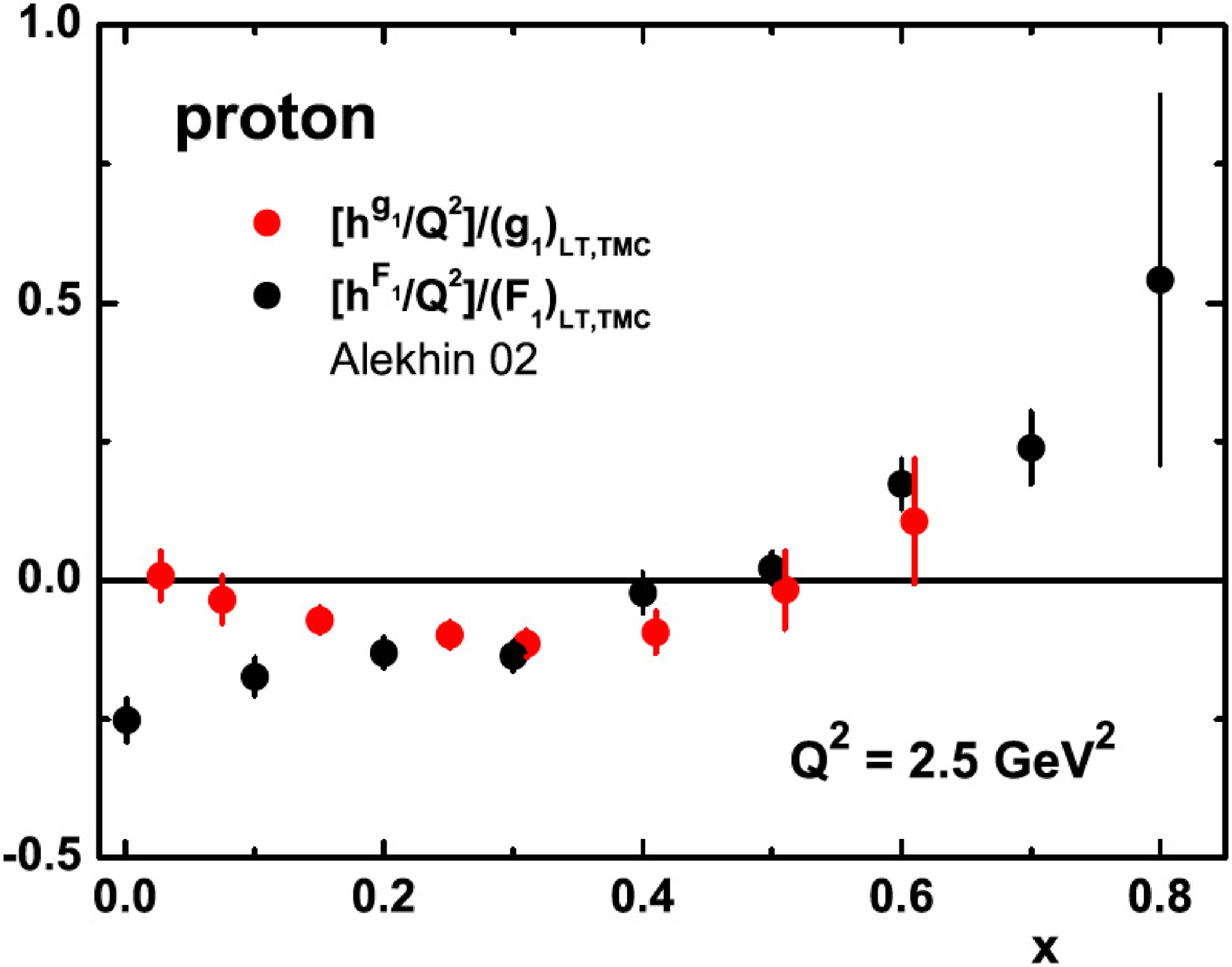,scale=0.25}}
\parbox[t]{3.2in}{\epsfig{file=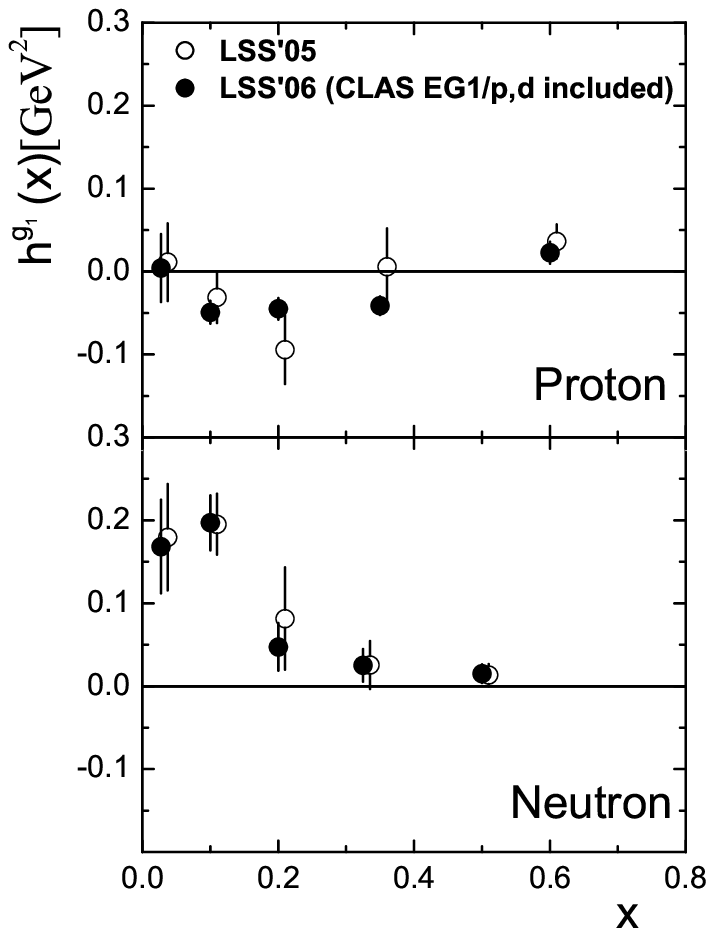,scale=0.8}}
\end{center}
\begin{center}
\begin{minipage}[t]{16.5 cm}
\vspace{-0.2in}
\caption{Comparison of HT terms in $g_1$ and $F_1$
(left panel) and higher twist terms for protons and neutrons
extracted from data (right panel).
\label{HT}}
\end{minipage}
\end{center}
\end{figure}

One of the problems in any analysis is to decide whether to use the lowest twist (LT) expression
Eq.~(\ref{eq:g1NLO})
or  to include higher twist (HT) terms.
This is not just a question of the kinematic region in a given experiment, but also of
what quantity one fits. Thus long ago we discovered empirically that HT terms cancelled out in the ratio
$\frac{g_1}{F_1}$.
Put
 \beq \label{eq:g_1/F_1HT}
 g_1^{EXP} = g_1^{LT} + g_1^{HT} \qquad F_1^{EXP} = F_1^{LT} + F_1^{HT}  \nn \eeq
Then
\beqy  \big[\frac{g_1}{F_1}\big]^{EXP} & \approx &\frac{g_1^{LT}}{F_1^{LT}} \big[ 1 + \frac{g_1^{HT}}{g_1^{LT}} - \frac{F_1^{HT}}{F_1^{LT}} \big] \nn \\
& \approx & \frac{g_1^{LT}}{F_1^{LT}}
\eeqy
provided there is a cancellation between 
$\frac{g_1^{HT}}{g_1^{LT}} $ and $\frac{F_1^{HT}}{F_1^{LT}} $.


The left panel of Fig.~\ref{HT} which compares HT terms for $g_1$
\cite{Leader:2006xc}  and for $F_1$ \cite{Alekhin:2002nv}
 demonstrates the validity of this for $x\geq 0.15$, but clearly indicates that ignoring HT terms in the ratio
 $\frac{g_1}{F_1}$ below $x=0.15$ is incorrect.
The HT terms for $g_1$ determined in \cite{Leader:2006xc} are
shown in the right panel of Fig.~\ref{HT} using 5 bins in $x$. (The results change
negligibly if 7 bins are utilized.)


A more subtle problem relates to the kinematic relations between
various quantities. The exact relations between $A_\| $, $A_1$ and
 $g_1$ include HT terms involving $\gamma^2$, both in
Eq.~(\ref{eq:AlongA1approx}) and in the expression for $D$,
Eq.~(\ref{Dnew}). If it is safe to ignore HT terms in $g_1$, $F_1$
 is it safe to ignore them in these kinematic relations as
well? DSSV assume it is, whereas LSS insist that if $\gamma^2$ is
significant such terms must be kept in the kinematic relations.
The validity of the latter viewpoint is borne out by the empirical
fact (accidental?) that HT terms can be ignored in
$\frac{g_1}{F_1}$ in the region $x\geq 0.15$, where, in some
cases, $\gamma^2$ is non-negligible. Thus the DSSV analysis, which
treats $\frac{g_1}{F_1}$ in LT, but ignores $\gamma^2$, may be unreliable in those regions
where $\gamma^2$ is important. This issue is particularly important for
data from  HERMES and JLab.

\begin{itemize}
\item \textit{The light quark densities.}
There is broad agreement between the various analyses for the
$\Delta u(x) + \Delta \bar{u}(x)$ and $\Delta d(x) + \Delta
\bar{d}(x)$ parton densities. Generally the constraint
Eq.~(\ref{eq:rnppqcd}) has not been enforced in the
parametrizations used for the $d$ quark . Early  data demanded
negative values of $\Delta d(x) + \Delta \bar{d}(x)$ and continued
to do so even when the measured region was extended to $x=0.6$ at
Jefferson Laboratory \cite{Zheng:2003un,Dharmawardane:2006zd} .
These data are discussed in detail in Section~\ref{subsec:val}.
 Note that a fit \cite{Leader:1997kw} using the Brodsky,
Burkardt and Schmidt type parametrization \cite{Brodsky:1994kg},
which respects Eq.~(\ref{eq:rnppqcd}), led to a $\Delta d(x) +
\Delta \bar{d}(x)$ which became positive just beyond $x=0.6$. With
the $12 \,GeV$ upgrade at Jefferson it should be possible to explore
out to $x=0.8$ and to settle the matter. Fig.~\ref{uubar} shows a comparison between
the LSS'06,  DSSV and AAC'08 $\Delta u + \Delta \bar{u}$ and $\Delta d +\Delta \bar{d}$ densities.
 It is seen that the agreement is impressive,
 although the d-quark polarization  $\Delta d(x)/d(x)$ is still rather uncertain at high $x$
 (see also Section~\ref{subsec:val}).

\begin{figure}[htb!*]
 \hfill
\begin{minipage}[t]{.45\textwidth}
\begin{center}
\epsfig{file=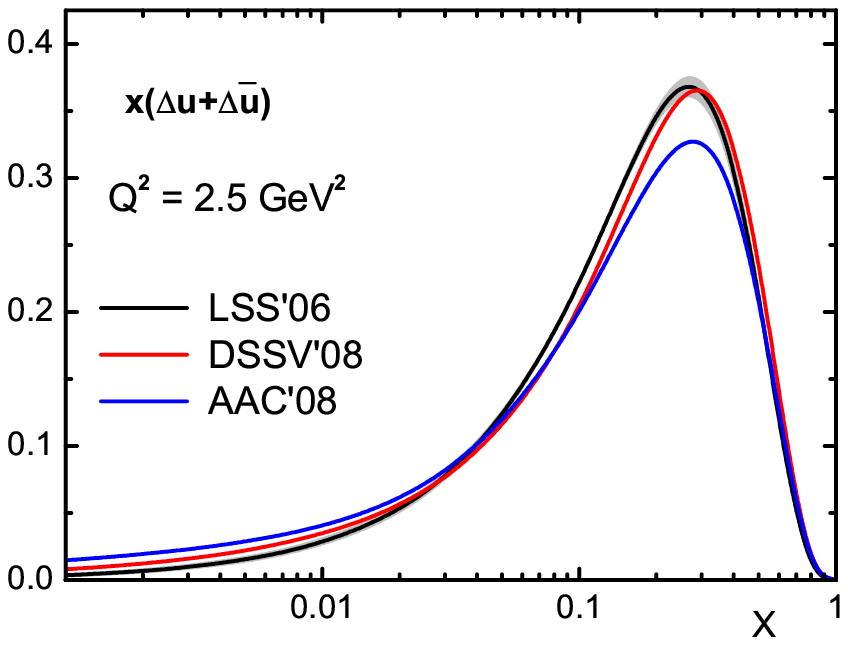, width=\textwidth}
\end{center}
\end{minipage}
\hfill
\begin{minipage}[t]{.45\textwidth}
\begin{center}
\epsfig{file=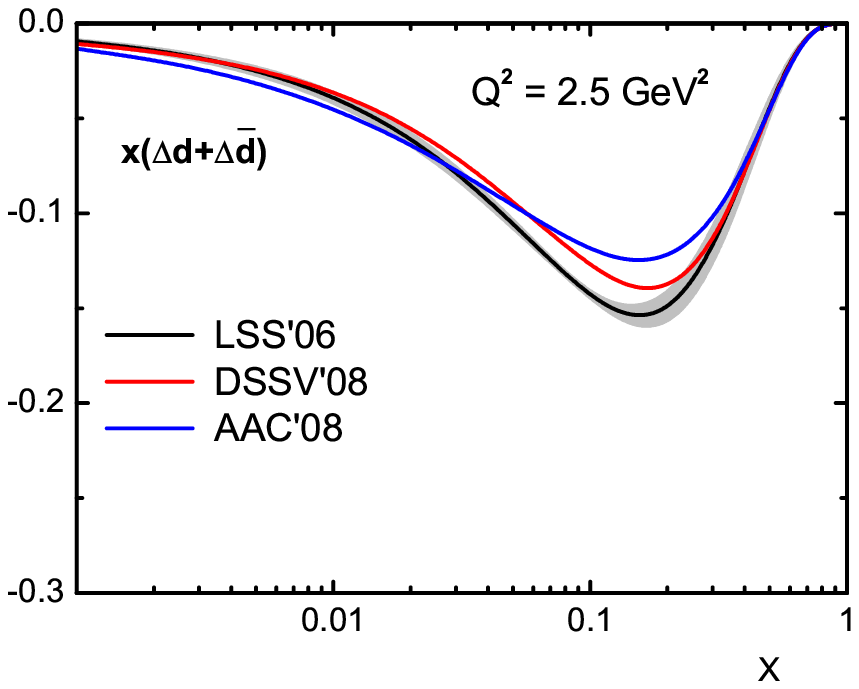, width=\textwidth}
\end{center}
\end{minipage}
\hfill
\vspace{-1.5cm}
\begin{center}
\begin{minipage}[t]{16.5 cm}
\caption{Comparison of LSS'06 (with shaded error band),
 DSSV  and AAC'08 $(\overline{MS})\, \,x(\Delta u+\Delta \bar{u})$ and $x(\Delta d +\Delta \bar{d})$ densities. }\label{uubar}
\end{minipage}\end{center}
\end{figure}


In Fig.~\ref{err-u}  we show the impact of the CLAS'06 and COMPASS'06 data on the uncertainties in
$x(\Delta u +\Delta \bar{u})$ and $x(\Delta d +\Delta \bar{d})$  which had been
found in the LSS'05 analysis. The uncertainties are estimated by the usual Hessian method, utilizing
$\Delta \chi^2 =1$.
It should be borne in mind that these uncertainties, and all those to follow,
must be treated with caution. They are calculated relative to a chosen analytical
form for the parametrization of the parton density
and do not at all reflect possible uncertainties in the form of parametrization.

\begin{figure}[htb!]
 \hfill
\begin{minipage}[t]{.45\textwidth}
\begin{center}
\epsfig{file=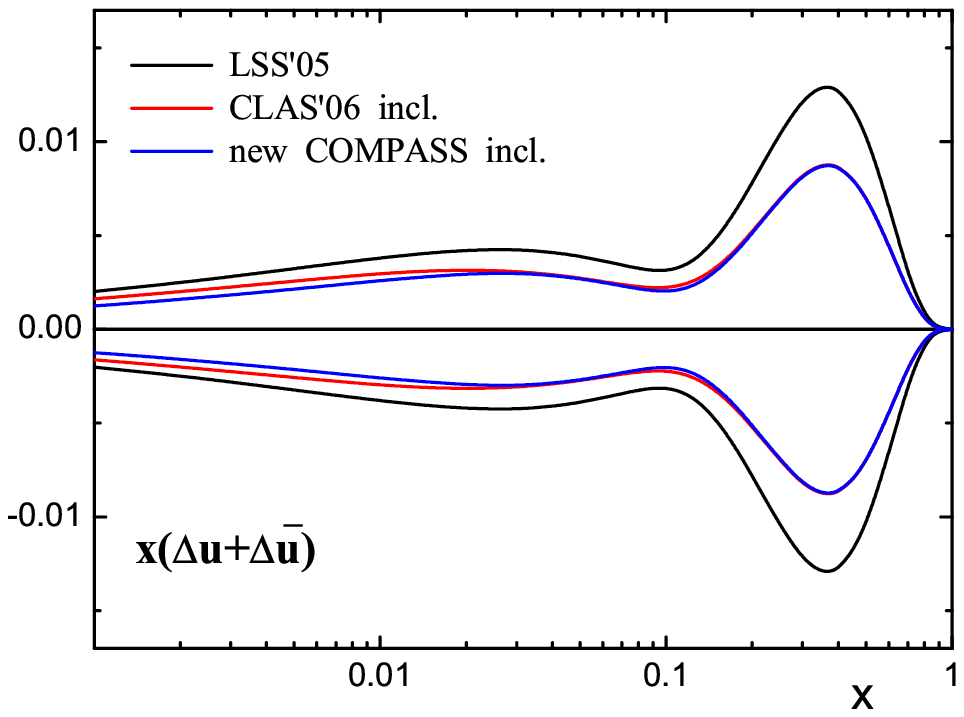, width=\textwidth}
\end{center}
\end{minipage}
\hfill
\begin{minipage}[t]{.45\textwidth}
\begin{center}
\epsfig{file=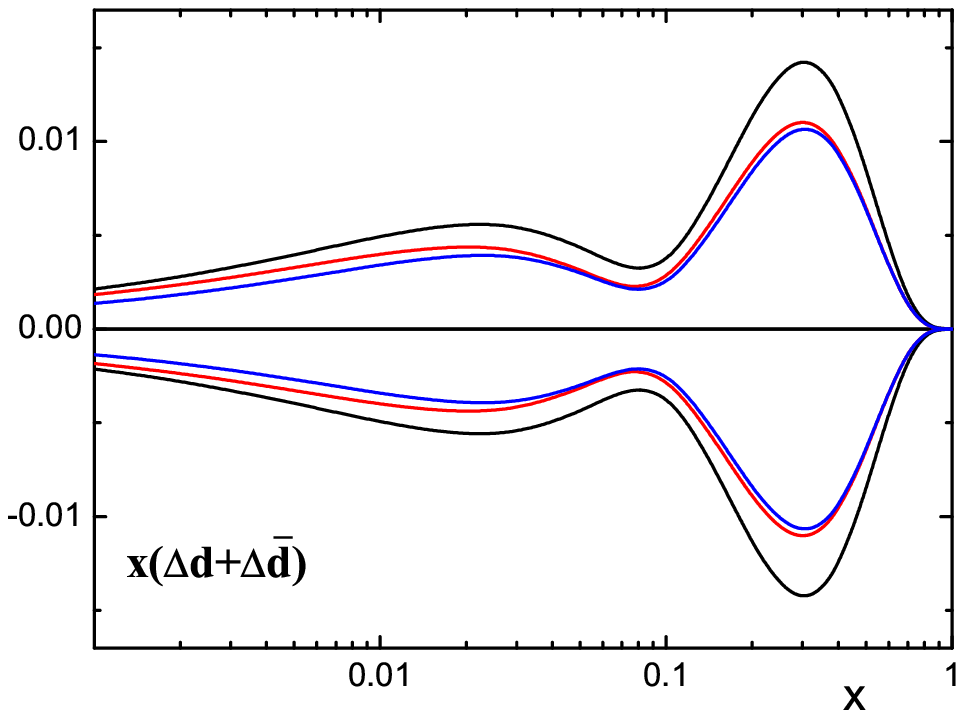, width=\textwidth}
\end{center}
\end{minipage}
\hfill
\vspace{-0.8cm}
\begin{center}
\begin{minipage}[t]{16.5 cm}
\caption{Impact of CLAS'06 and COMPASS'06 data on the uncertainties in $x(\Delta u+\Delta\bar{u})$ and $x(\Delta d+\Delta\bar{d})$ of the LSS'05 analysis. } \label{err-u}
\end{minipage}\end{center}
\end{figure}

\item \textit{The polarized strange quark density.}
This is a highly controversial issue at present. All analyses of
purely DIS data have found negative values for $\Delta s(x) +
\Delta \bar{s}(x)$.
Fig.~\ref{err-s} shows the impact of the CLAS'06 and COMPASS'06 data on the uncertainty
in the LSS'05 determination of $x(\Delta s(x)+ \Delta \bar{s}(x))$.

\begin{figure}[htb!]
\hspace{3.5cm} \begin{minipage}[t]{8 cm}
\epsfig{file=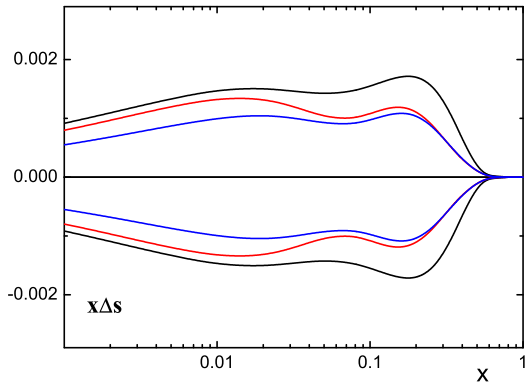,scale=0.7}
\end{minipage} \begin{center}
\begin{minipage}[t]{16.5 cm}
\vspace{-1cm}
\caption{Impact of CLAS'06 and COMPASS'06 data on uncertainties in the LSS'05 determination of $x(\Delta s(x)+ \Delta \bar{s}(x))$. For curve labels see Fig.~\ref{err-u}.}
 \label{err-s}
\end{minipage}
\end{center}
\end{figure}

An important quantity is the first moment
 \beq \label{DeltaS}\Delta S \equiv \int_0^1 \, dx [\Delta s(x) + \Delta
 \bar{s}(x)] .  \eeq
 LSS'06 \cite{Leader:2006xc} give for its value
\beq \label{DeltaSLSS} \Delta S_{\overline{MS}} = -0.126 \pm 0.010
\qquad \textrm{at} \qquad Q^2=1 \, GeV^2 \eeq and in
\cite{Leader:2002az} it was shown that a positive value for the
first moment would imply a huge breaking of $ SU(3)_F $
invariance, far greater than the $\pm 10\% $ breaking estimated by
Ratcliffe \cite{Ratcliffe:1998su}. Nonetheless analyses of  SIDIS
data for kaon production have suggested positive values of $\Delta
s(x) + \Delta \bar{s}(x)$ for $x\geq 0.03$. A hint of this first
emerged in the HERMES analysis of 2004/5 \cite{Airapetian:2004tw}
and was confirmed by the more precise data reported in
\cite{Airapetian:2008qf} where the first moment for the measured
range $0.02 \leq x \leq 0.6 $ was given as
 \beq \label{DeltaSHERMES}  \Delta S = 0.037 \pm 0.019(stat.) \pm 0.027(sys.)
 \eeq
 One could argue that the HERMES results should not be taken too seriously, since
 the  analyses were carried out in LO and relied on derived
  quantities called \emph{purities}, whose accuracy  may have been overestimated. However, the recent combined
 analysis of DIS, SIDIS and $pp\rightarrow \pi^0 \,\textrm{or} \,\textrm{jet} +
 X$ by the DSSV group \cite{deFlorian:2008mr}, which is an $\overline{MS}$ NLO, LT
 analysis, also finds positive values for $\Delta s(x) + \Delta \bar{s}(x)$ for $x\geq 0.03 $, yet
ends up with
a negative first moment $ \Delta S = -0.114$ at $Q^2 = 10 \, GeV^2$.
This is achieved by $\Delta s(x) + \Delta \bar{s}(x)$ becoming
negative below roughly $x=0.02$. But there are essentially no data
in the latter region, which suggests this must be caused by the need
to satisfy the $SU(3)_F$ symmetry condition Eq.~(\ref{eq:a38}).
DSSV state that they do \emph{not} impose $SU(3)_F$ symmetry,
and they multiply the RHS of Eq.~(\ref{eq:a38}) by $(1 +
\epsilon_{SU(3)})$ where $\epsilon_{SU(3)}$ is a free parameter.
However, a $\chi^2$ bias against large values of
$\epsilon_{SU(3)}$ is built into the minimization procedure
\footnote{private communication from Werner Vogelsang}
which explains the turning negative of $\Delta s(x) + \Delta\bar{s}(x)$, 
although this is not stated in the quoted
paper.

The recent AAC'08 analysis \cite{Hirai:2008aj},
based on  the world data on DIS plus the $\pi^0$ production data from RHIC
(incorporated at LO),
also finds  a negative strange quark density.
Fig.~\ref{delsLSSDSSV} compares the LSS'06
version of $\Delta s + \Delta \bar{s}$ with that of DSSV and AAC'08.
Note that the shape  appears somewhat
different in the various DIS analyses. The origin of this has been
traced to the different choices of unpolarized densities used in
the imposition of positivity \cite{Leader:2005kw}, as can be seen
in the figure.

\begin{figure}[htb!]
\begin{center}
\begin{minipage}[t]{10 cm}
\epsfig{file=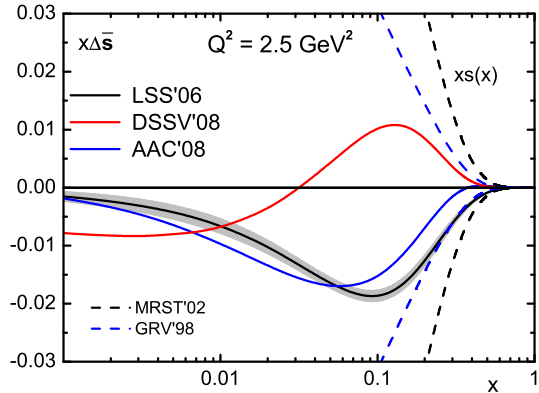,scale=0.8}
\end{minipage}%
\end{center}
 \begin{center}
\begin{minipage}[t]{16.5 cm}
\vspace{-1cm}
\caption{LSS'06, DSSV and AAC'08 ($\overline{MS}$) versions of $x(\Delta s + \Delta \bar{s})$.
} \label{delsLSSDSSV}
\end{minipage}
\end{center}
\end{figure}

It is thus clear that the origin of the positive values for
$\Delta s(x) + \Delta \bar{s}(x)$ lies entirely in the SIDIS data.
 It may be significant
that both the HERMES and DSSV analyses utilize the recent de
Florian, Sassot and Stratmann \cite{de Florian:2007hc} set of
fragmentation functions, which are surprisingly different from
earlier sets of fragmentation functions, especially for kaons. COMPASS are studying
the dependence of $\Delta s(x) + \Delta \bar{s}(x)$ on the choice of fragmentation functions. A preliminary report
can be found in the talk of R.~Windmolders at the SPIN2008 Conference.

Finally, there is an intriguing question as to whether the
standard collinear analysis should be applied to the SIDIS data.
The point is the following.  What we mean by any $q(x, Q^2)$ or
$\Delta q(x, Q^2)$ actually involves an integral over the
intrinsic transverse momentum $k_t$ of the quark up to a maximum
$k_t^2 \approx Q^2$. If, as is the case for some of the SIDIS
data, the magnitude of the transverse momentum of the detected
hadron is less than $Q$, then it is perhaps incorrect to analyze
the data using the conventional collinear densities. Indeed, Bass
\cite{Bass:2002jd} has  shown that for that part of the strange
quark density generated via photon-gluon fusion the \emph{sign} of
$\Delta S$ is sensitive to the upper limit in the  $k_t^2$
integration, but the overall effect is much too small to be relevant.

\item \textit{The flavor singlet first moment $\Delta \Sigma$.} 
All the modern global analyses obtain compatible values for $\Delta \Sigma $. 
In the $\overline{MS}$ scheme, where $a_0(Q^2) =\Delta \Sigma (Q^2)$ 
they find at $Q^2=4$ GeV$^2$:
\begin{table}[!h]
\centering
\begin{tabular}{c c c c}
\hline
\\
LSS'06 & COMPASS'06 & AAC'08 & DSSV \\ [0.5ex]
\hline
\\
$0.24\pm 0.04$ & $0.29\pm 0.01$ & $0.25\pm 0.05$ & 0.24 \\[0.5ex]
\hline
\end{tabular}
\end{table}

For reasons which are not understood these values are somewhat lower than
the values obtained directly from
 $\Gamma_1^d$ as discussed in Section~\ref{subsec:bjork}. 

\item \textit{The polarized gluon density.}
In Sec.~\ref{subsec:QCD} we explained how the ``spin crisis in the parton model"
could be alleviated by generating the experimentally small value of $a_0$ through a cancellation between
relatively large values of $\Delta \Sigma$ and $\Delta G$ (see Eq.~\ref{eq:a0NLO}).
 Present day estimates are $a_0\approx 0.2$. Thus demanding $\Delta \Sigma \approx 0.6$
 requires, for the first moment,
  \beq \label{DeltaGneeded}
 \Delta G \approx 1.7 \qquad \textrm{at} \qquad Q^2=1 \, GeV^2 \eeq
 The question is whether this is compatible with what we know about the polarized gluon density.

 \begin{figure}[htb!]
\vspace{-10cm}
 \hspace{-2.7cm}
 \begin{minipage}[t]{8 cm}
\epsfig{file=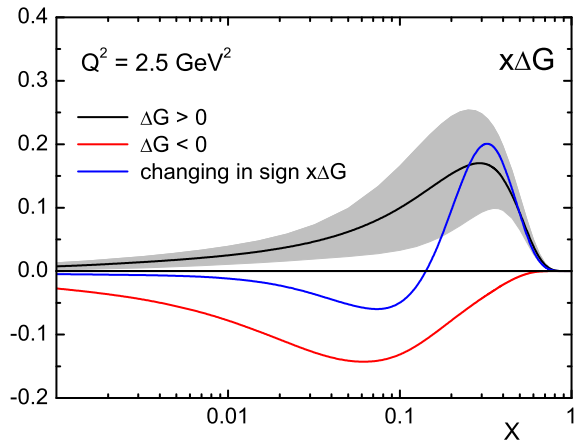,scale=1.1}
\end{minipage}
\vspace{-15.5cm}
\begin{center}
 \begin{minipage}[t]{16.5 cm}
\caption{Comparison of the three types of LSS'06 $\,( \overline{MS}) \,\, x\Delta G(x)$.} \label{LSS3DeltaGs}
\end{minipage} \end{center}
\end{figure}

There are three ways to access $\Delta G(x) $: via polarized DIS, and via the
 measurement of the asymmetry $A_{LL}$ both in SIDIS production
 of charmed quarks or high $p_T$ jets, and in  semi-inclusive polarized $pp$ reactions at RHIC.
 Here we will only discuss DIS. The latter approaches have been treated in 
 Section~\ref{subsec:PDFexp}.

  In determining the parameters of the polarized gluon density
  from fits to the data on $g_1(x, Q^2)$, it should be remembered,
 as already mentioned, that the main role of the gluon is in the
evolution with $Q^2$, but that the range of $Q^2$ is very limited, so
the determination of $\Delta G(x) $ is imprecise.

\begin{figure}[htb!]
\begin{center}
\begin{minipage}[h]{8cm}
\epsfig{file=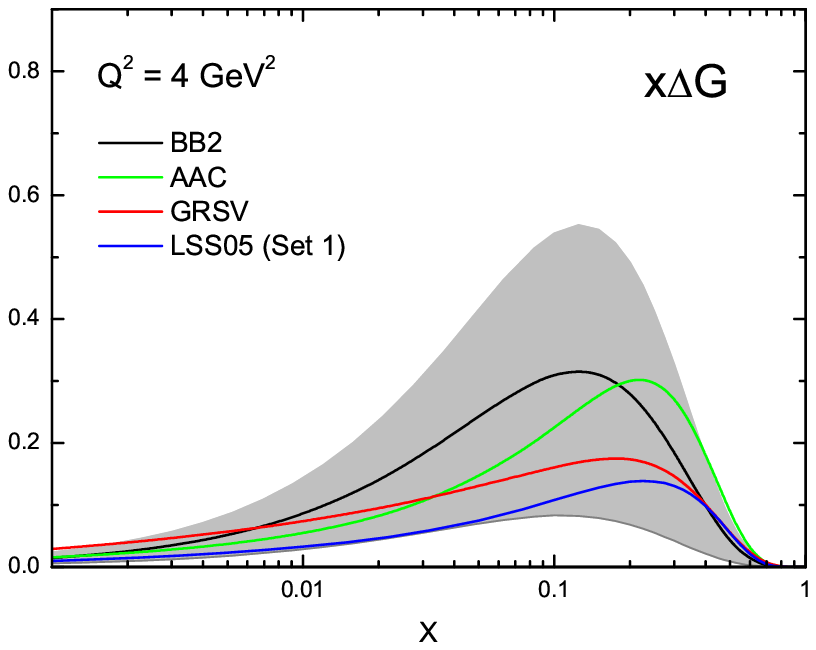, scale=0.8}
\end{minipage}
\begin{minipage}[h]{8cm}
\epsfig{file=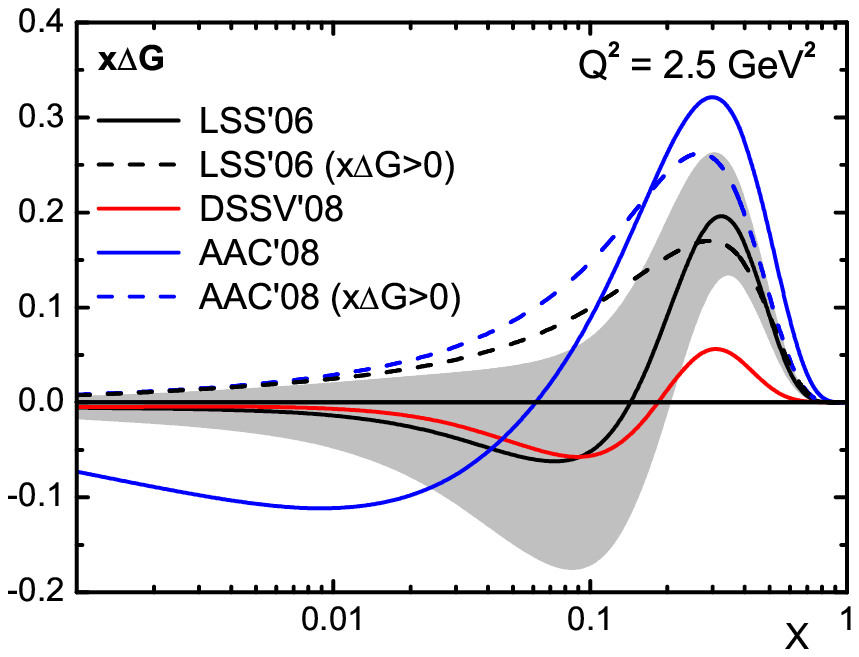,scale=0.8}
\end{minipage}
\end{center}
\begin{center}
\begin{minipage}[t]{16.5 cm}
\vspace{-0.3in}
\caption{Left: a compilation of earlier results for $x\Delta G(x)$; right: present situation. }
\label{DeltaGusual}
\end{minipage} \end{center}
\end{figure}

For a long time all analyses seemed to indicate that $\Delta G(x)
$ was a positive function of $x$. LSS used a very simple
parametrization \beq \label{eq:LSSparamDeltaG}
x\Delta G(x) =
\eta_g A_g x^{a_g}[xG(x)]
 \eeq
 where $A_g$ is fixed so that the first moment of $\Delta G$ is given by $\eta_g$.
In the minimization procedure there was nothing to stop $\eta_g$ from being negative. Yet
the best $\chi^2$ values always corresponded to positive $\Delta
G(x) $.
With the inclusion of recent data,
LSS'06 \cite{Leader:2006xc} find equally good fits with positive,
negative and sign-changing $\Delta G(x) $, \emph{provided} higher
twist terms are included. The latter are particularly demanded by
the CLAS'06 data \cite{Dharmawardane:2006zd}. Fig.~\ref{LSS3DeltaGs} compares the LSS'06 positive, negative and
sign-changing versions of $\Delta G(x) $.

\begin{figure}[htb!]
 \hspace{4cm} \begin{minipage}[t]{8 cm}
\epsfig{file=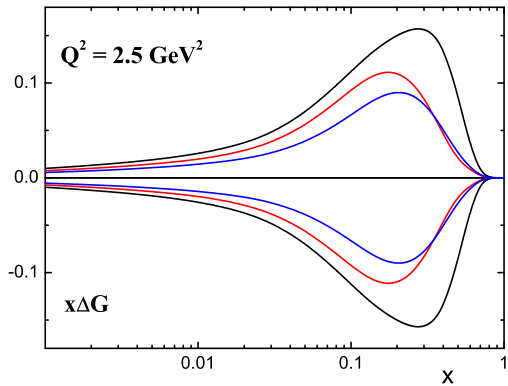,scale=0.7}
\end{minipage} \begin{center} \begin{minipage}[t]{16.5 cm}
\vspace{-1cm}
 \caption{Impact of CLAS'06 and COMPASS'06 data on uncertainties in the LSS'05 determination of $x\Delta G(x)$. For curve labels see Fig.~\ref{err-u}.} \label{err-G}
\end{minipage} \end{center}
\end{figure}

 Also the
 COMPASS'06 analysis \cite{Alexakhin:2006vx}, even though  HT terms are not included, finds
acceptable negative $\Delta G(x) $ fits, but has some
peculiarities (linked to the fact that the flavor combinations
$\Delta q_{3,8}$ and $\Delta \Sigma $ are parametrized, rather
than the individual quark densities), which suggest it may not be very physical.
The DSSV analysis finds a sign-changing $\Delta G(x)$, while the AAC'08 analysis finds both positive and sign-changing densities.
In the left hand panel of Fig.~\ref{DeltaGusual} we show various results for $\Delta
G(x) $ as of a couple of years ago. The right panel shows the present situation.
 In Fig.~\ref{err-G} we show the impact of the CLAS'06 and COMPASS'06 data on the uncertainty
in the LSS'05 determination of $x\Delta G(x)$.

 Regarding the present data, in all fits, and irrespective of the form of the gluon density, the magnitude
 is always found to be very small. Typically one has $|\Delta G| \approx 0.29 \pm 0.32 $,
\emph{much} smaller than the desired $1.7$ !
As discussed in Section~\ref{subsec:PDFexp}, all the present data on SIDIS and $pp$ reactions
are perfectly compatible with this very small magnitude for $\Delta G$ and cannot distinguish
between the various sign possibilities.
\end{itemize}

\subsection{\it The Spin Structure Function $g_2$ \label{subsec:g2}}
The spin structure function $g_2$, unlike $g_1$ and $F_1$, does not have an
intuitive interpretation in the simple quark-parton model. To understand $g_2$ properly,
it is best to start with the operator product expansion (OPE) method (see also
Sections~\ref{subsec:parton},\ref{subsec:delq},\ref{subsec:mom},\ref{subsec:OPE}).
In the OPE, neglecting quark masses, $g_2$ can be cleanly separated into a
twist-2 and a higher twist term:
  \begin{eqnarray}g_2(x,Q^2) = g_2^{WW}(x,Q^2) +g_2^{H.T.}(x,Q^2)~.
  \end{eqnarray}
The leading-twist term can be determined from
$g_1$ as~\cite{Wandzura:1977qf}
  \begin{eqnarray} \label{eq:WW'}
   g_2^{WW}(x,Q^2) = -g_1(x,Q^2) + \int _{x}^1 \frac{g_1(y,Q^2)}{y} dy
  \end{eqnarray}
(this holds also with the inclusion of TM corrections \cite{Blumlein:1998nv}), 
and the higher-twist term $g_2^{H.T.}(x,Q^2)$ arises from quark-gluon correlations.
Therefore, $g_2$ provides a clean way to study higher-twist effects.

Experimentally, $g_2$ can be extracted from combined measurements of both
the longitudinal and the transverse target spin cross section differences 
(Eqs.~\ref{eq:LongXsec},\ref{TransXsec} in Section~\ref{subsec:formalism}), 
or the corresponding
asymmetries $A_{||}, A_{\perp}$. These measurements demand high precision and 
large luminosity, since the factors multiplying $g_2$ tend to be relatively small
in DIS kinematics. Several 
experiments~\cite{Anthony:1996mw,Abe:1997qk,Adams:1994id,Adams:1997hc}
 therefore took only enough statistics on the
transverse target configuration to reduce the systematic error on $g_1$ due to the
unknown contribution from $g_2$ below the statistical error, yielding only crude
information on $g_2$ itself.

Two experiments at SLAC, E143 and E155~\cite{Abe:1998wq,Anthony:2002hy}
took dedicated data to measure $g_2$
and the virtual photon asymmetry $A_2$ (see Eq.~\ref{eq:A_2gi})
with more precision. The most extensive
data set in the DIS region comes from the second run of E155, both for the proton and the deuteron,
with $g_2$ for the neutron extracted from the difference. 
The general trend of these data follows the Wandzura-Wilczek form (Eq.~\ref{eq:WW'}), but
some deviations are seen for the proton, especially at small $x$. The data on $A_2$ show that
this asymmetry is indeed rather small, well below the limit given in Eq.~\ref{eq:A2bound}.

\begin{figure}[htb!]
\begin{center}
\parbox[t]{3.2in}{\epsfig{file=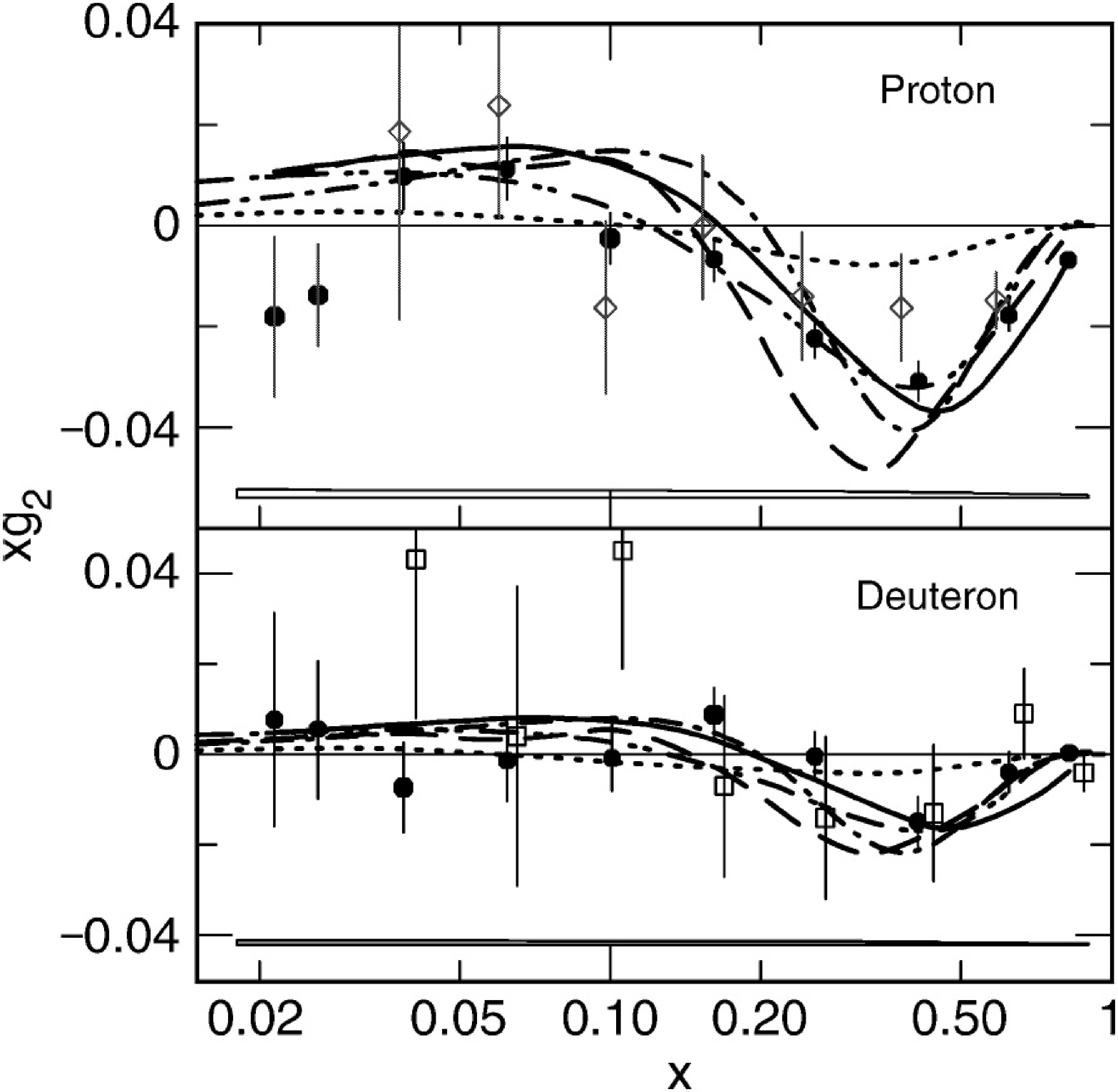,scale=0.2}}
\raisebox{3cm}{\parbox{3.5in}{\epsfig{file=xg2n_vs_x.epsi,scale=0.3,angle=-90}}}
\parbox{3.5in}{\epsfig{file=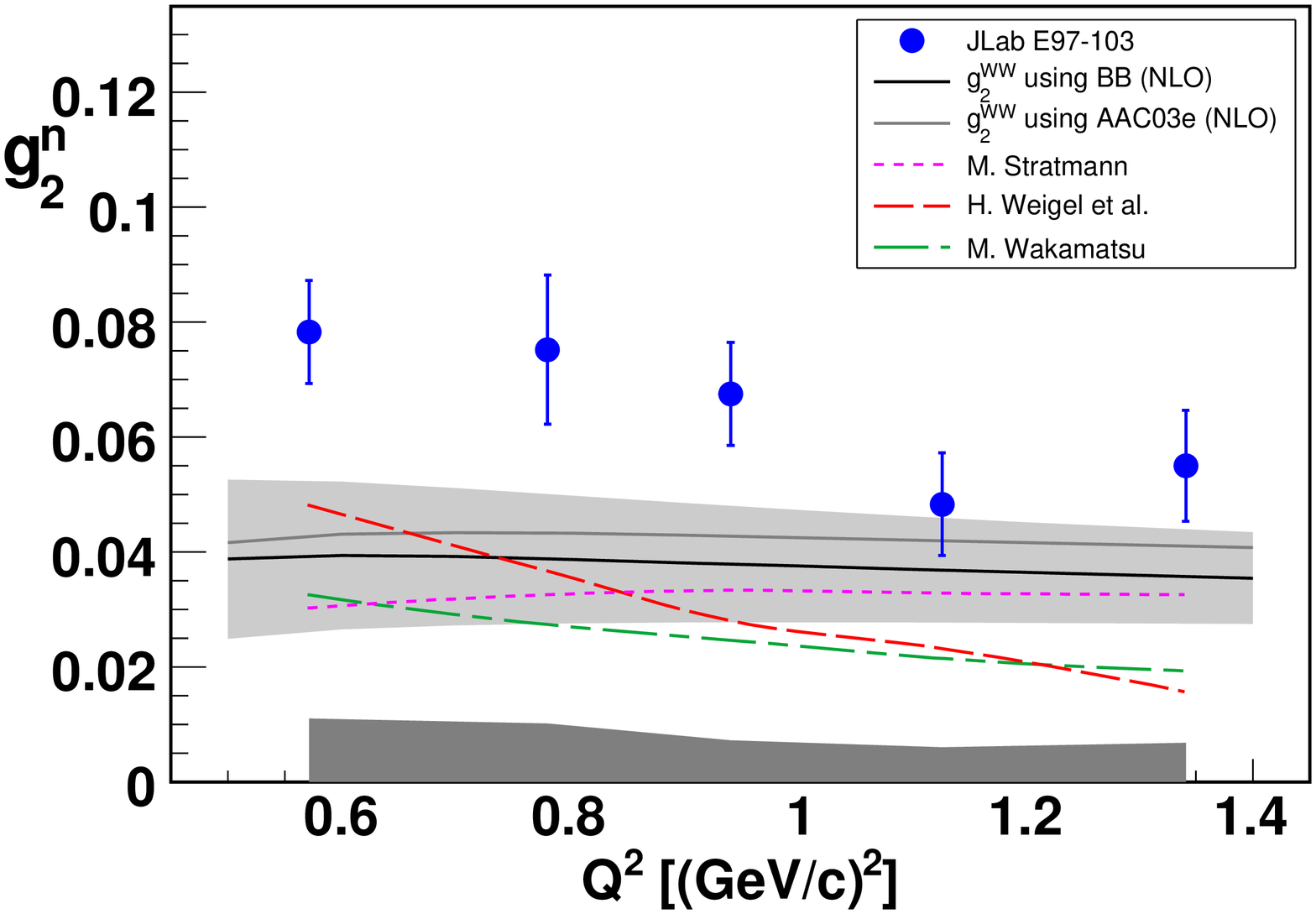,scale=0.3,angle=-90}}
\begin{minipage}[t]{16.5 cm}
\caption{Results for $g_2$ of the proton and the deuteron 
(top-left panel) and the neutron (top-right panel)
as a function of $x$ from SLAC~\protect\cite{Anthony:2002hy} and 
JLab Hall A~\protect\cite{Zheng:2004ce,Kramer:2005qe}, 
in comparison with $g_2^{WW}$ (solid curves).
The left panel also shows
bag model calculations of Stratmann~\protect{\cite{Stratmann:1993aw}}
 (dash-dot-dot) and Song~\protect{\cite{Song:1996ea}}
(dot) and the chiral soliton models of 
 Weigel~\protect{\cite{Weigel:2000gx,Weigel:2003pe}}
(dash dot) and Wakamatsu~\protect{\cite{Wakamatsu:2000ex}} (dash).
The bottom plot shows results for $g_2^n$ from JLab 
Hall A~\protect\cite{Kramer:2005qe} as a function of $Q^2$ 
at an $x$ value around 0.2, in comparison with 
$g_2^{WW}$ and models.
}
\label{fig:g2n}
\end{minipage}
\end{center}
\end{figure}

Figure~\ref{fig:g2n} shows the SLAC results on
the proton, deuteron (top-left panel) and neutron (top-right panel) 
as a function of $x$. 
Also shown on the top-right panel of Fig.~\ref{fig:g2n}
are recent JLab neutron results with 
polarized $^3$He from 
Hall A E99-117~\cite{Zheng:2004ce} and E97-103~\cite{Kramer:2005qe}. 
E97-103 is a dedicated precision measurement of $g_2^n$ in 
the DIS region at low $Q^2$ to study its $Q^2$ dependence. It
covered
five different $Q^2$ values from 0.58 to 1.36 GeV$^2$ at $x \approx 0.2$.
Results on the $Q^2$ dependence of $g_2^n$ are given in the bottom panel of
Fig.~\ref{fig:g2n}. The light-shaded area in the plot
gives the leading-twist contribution, obtained by fitting world 
data~\cite{Bluemlein:2002be} and
evolving to the $Q^2$ values of this experiment. The systematic errors are
shown as the dark-shaded area near the horizontal axis.
The precision reached is more than an order
of magnitude improvement over that of the SLAC data~\cite{Anthony:2002hy}.  The difference
of $g_2$ from the leading twist part ($g_2^{WW}$) is due to
higher-twist effects and is sensitive to quark-gluon correlations.
The measured g$_2^n$ values
are consistently higher than g$_2^{WW}$.
For the first time, there is a clear indication that higher-twist effects
become significantly positive at $Q^2$ below 1 GeV$^2$,
while the bag model~\cite{Stratmann:1993aw} and Chiral Soliton 
model~\cite{Weigel:2003pe,Wakamatsu:2000ex}
predictions of higher-twist effects are negative or close to zero.
The $g_1^n$ data obtained from the same experiment agree with
the leading-twist calculations within the uncertainties.

\subsection{\it Spin Structure in the Resonance Region \label{subsec:res}}
While most measurements of spin structure functions initially focused on the deep--inelastic
region, more recently experiments have also collected a truly massive data
set on $g_1$ and $g_2$ in the region $W < 2$ GeV, the nucleon resonance region, over a
wide range in $Q^2$. 
Because the (resonant) final state in this kinematic region
can have a strong influence on scattering matrix elements, these data are less suitable
for a direct interpretation within the framework of perturbative QCD
(Section~\ref{subsec:QCD}). On the other hand,
they can reveal new and important information on the internal structure of nucleon
resonances and their excitation by the electromagnetic probe. Combining proton, deuteron
and neutron targets, one can separate the various spin-isospin channels that lead to
the excitation of overlapping resonances as well as non-resonant (multi-) meson production.

\begin{figure}[htb!]
\begin{center}
\epsfig{file=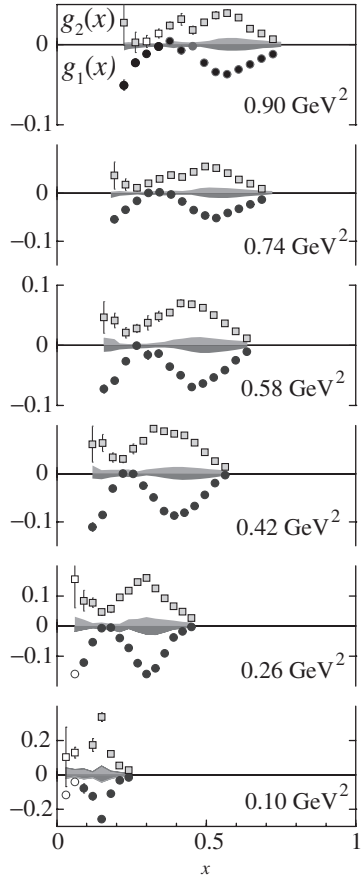,scale=0.5}
\begin{minipage}[t]{16.5 cm}
\vspace{-0.1in}
\caption{Data on the spin structure functions $g_1$ (filled circles) and $g_2$ (open squares)
vs. $x$ in the resonance region, for $^3$He
(approximately a neutron target), from Jefferson Lab's Hall A. 
Statistical errors are shown as error bars, while the systematic errors are indicated
by the shaded bands. The average $Q^2$ for each data set is indicated in the figure.
\label{fig:rra}}
\end{minipage}
\end{center}
\end{figure}

For a given resonance, the (electro- or)
photo-excitation strength can, in general, be given in terms of three 
helicity amplitudes, namely $A_{3/2}(Q^2)$ (transverse photons leading to final state helicity 3/2), 
$A_{1/2}(Q^2)$ (transverse photons leading to final state helicity 1/2) and
$S_{1/2}(Q^2)$ (longitudinal photons). These amplitudes are directly related to the photon
 asymmetries, {\it vz.} 
\bea
A_1 & = & \frac{|A_{1/2}|^2 - |A_{3/2}|^2}{|A_{1/2}|^2 + |A_{3/2}|^2} \\
A_2 & = & \sqrt{2} \frac{Q}{q^*} \frac{S^*_{1/2}A_{1/2}}{|A_{1/2}|^2 + |A_{3/2}|^2} \, . \nonumber
\eea \label{eq:resamps}
Here, $q^*$ is the (virtual) photon three-momentum in the rest frame of the resonance.
As a consequence, electromagnetic excitation of spin-1/2 resonances will have an 
asymmetry $A_1 = 1$ since the amplitude $A_{3/2}$ cannot contribute. Vice versa, for resonances
like the Delta (with spin 3/2) that are predominantly excited via $M1$ transitions, one has
$A_{3/2} \approx \sqrt{3} A_{1/2}$ and therefore $A_1 \approx -0.5$. By studying the 
$Q^2$-dependence of these asymmetries for a given mass range, one can gain information
on the relative strength of overlapping resonances (and non-resonant background) as well
as, for spin-3/2 resonances, the transition from $A_{3/2}$ dominance at low $Q^2$ to the
asymptotic $A_{1/2}$ dominance expected from helicity conservation in collinear pQCD.

\begin{figure}[htb!]
\vspace{-0.2in}
\begin{center}
\begin{minipage}[t]{8.8 cm}
\epsfig{file=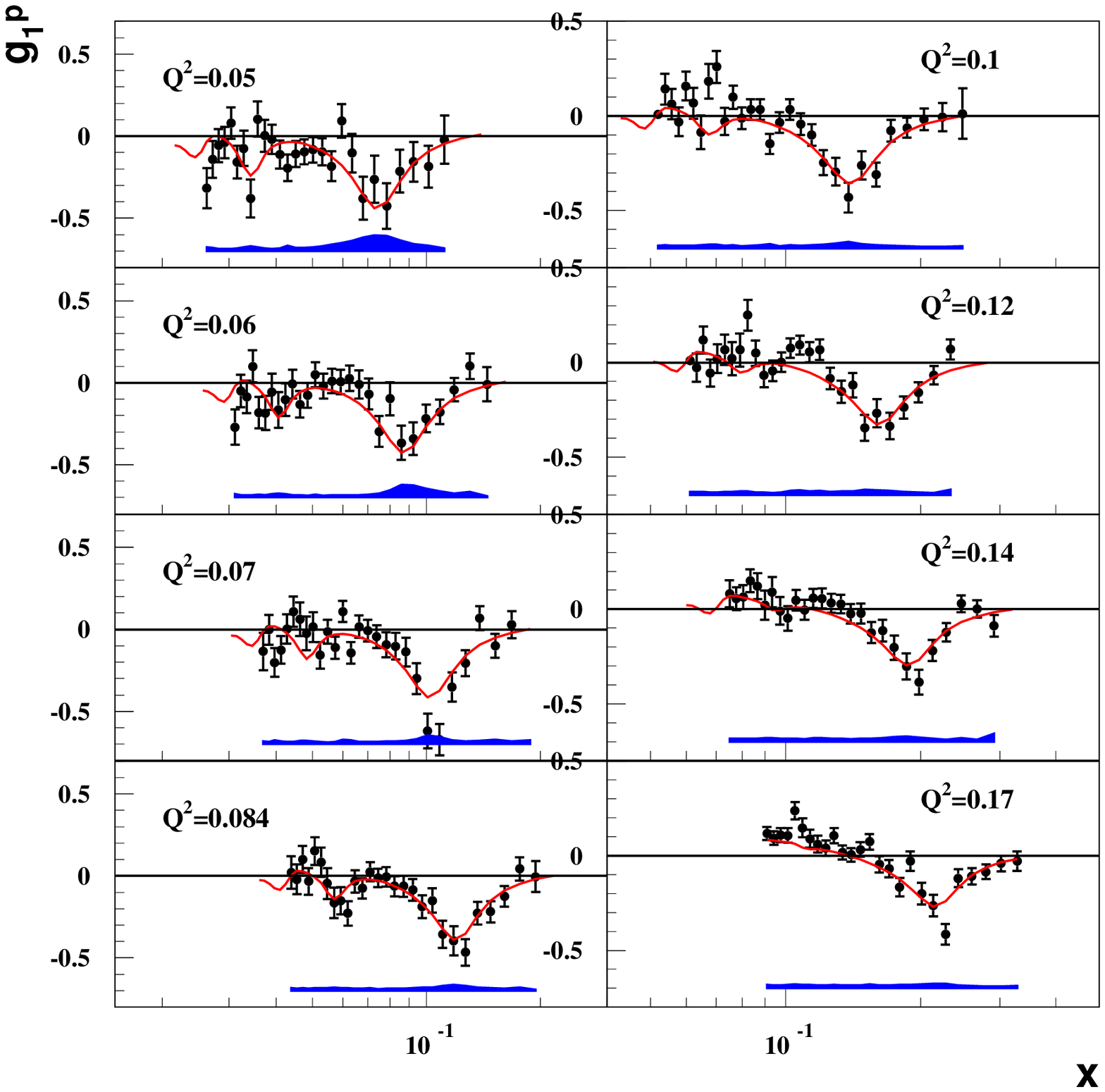, scale=0.48}
\end{minipage}
\begin{minipage}[t]{8.8 cm}
\epsfig{file=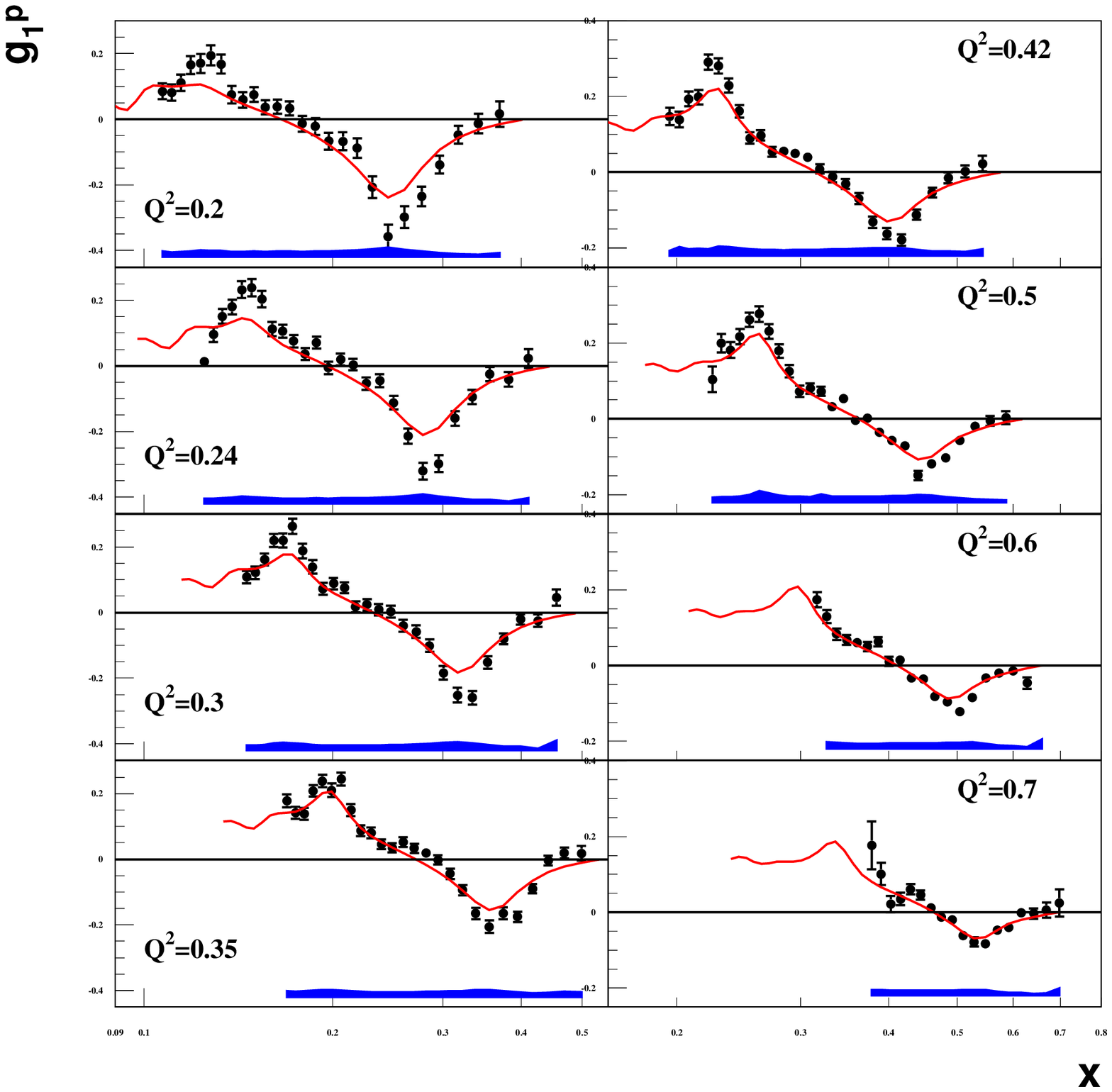, scale=0.48}
\end{minipage}
\end{center}
\begin{center}
\begin{minipage}[t]{9 cm}
\vspace{-0.7in}
\epsfig{file=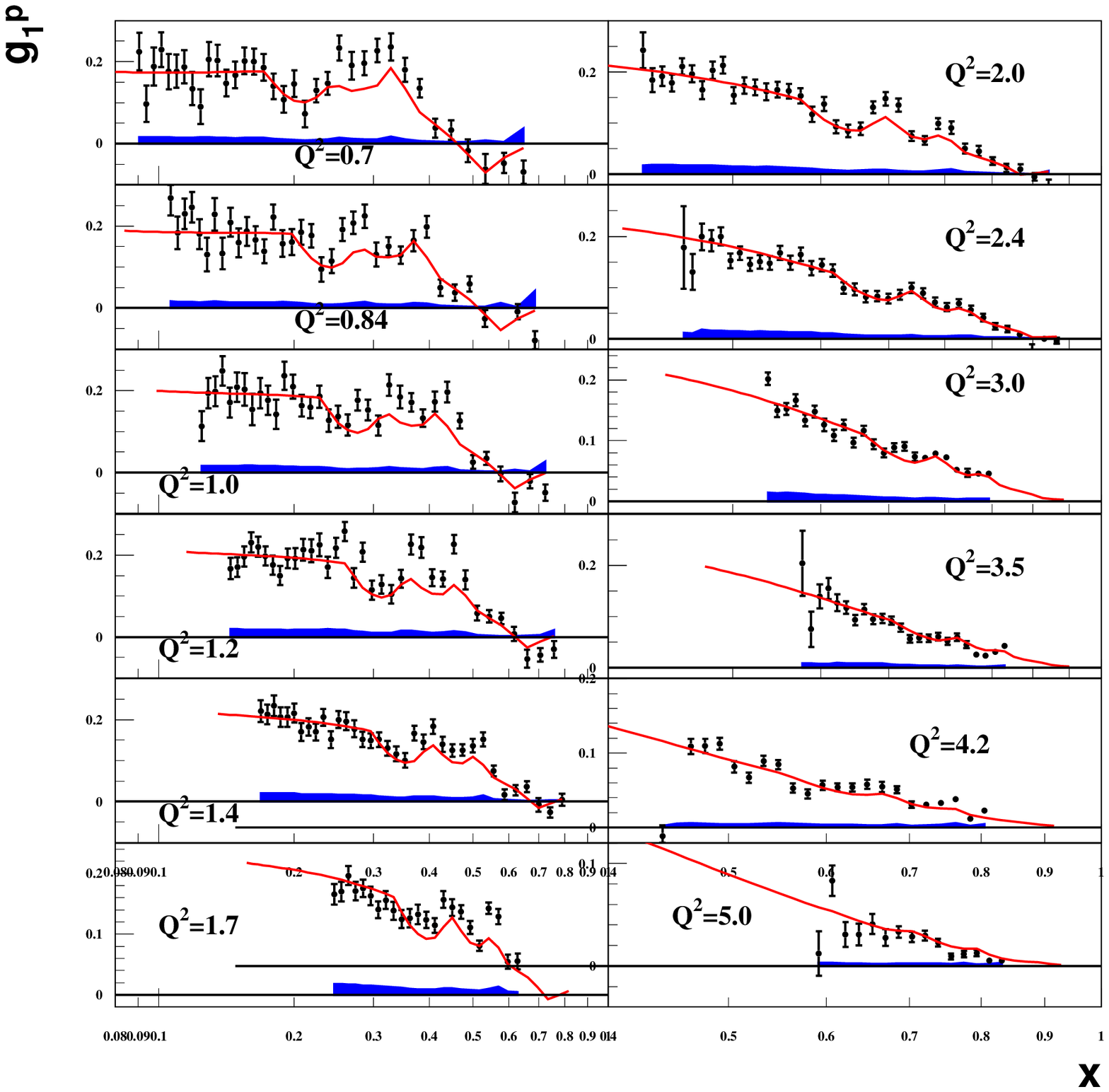, scale=0.52}
\end{minipage}
\begin{minipage}[t]{16.5 cm}
\vspace{-0.08in}
\caption{Data on the spin structure functions $g_1$ of the proton
vs. $x$ in the resonance region and beyond,
from Jefferson Lab's Hall B. 
Statistical errors are shown as error bars, while the systematic errors are indicated
by the shaded bands. The average $Q^2$ for each data set is indicated in the figure.
The curves are from a phenomenological parametrization of the data.
A similarly complete data set exists also for the deuteron.
\label{fig:rrb}}
\end{minipage}
\end{center}
\end{figure}

Measurements of spin structure functions in the resonance region are also required for
the evaluations of moments (Section~\ref{sec:mom}), since in most cases the integrals must
be evaluated up to $x=1$, which includes the resonance and elastic region, particularly
at more moderate $Q^2$. As an example, for $Q^2 = 3$ GeV$^2$, half of the $x$-range, $0.5 < x < 1$,
lies in the resonance region. These measurements are also very important
to study the onset of duality for spin structure functions, see Section~\ref{sec:duality}.
Finally, higher order QED radiative corrections to all measurements
(even in the DIS region)  have contributions
from the resonance region, making good knowledge of spin structure functions at low
$W$ imperative for accurate interpretation.

The first experiment to measure double spin observables at SLAC was also
the first to take data in the resonance region~\cite{Baum:1980mh}. The next dedicated
measurement was by the E143 collaboration~\cite{Abe:1996ag}, who used a lower
beam energy of 9.7 GeV to cover the resonance region at $Q^2 = 0.5$ GeV$^2$ and 1.2 GeV$^2$.
These data showed that the asymmetry $A_1$
is indeed negative in the $\Delta$ region at moderate $Q^2$, while it rises rapidly to
rather large values in the second resonance region (dominated by the S$_{11}$ and
D$_{13}$ resonances). 

\begin{figure}[htb!]
\begin{center}
\begin{minipage}[t]{9 cm}
\epsfig{file=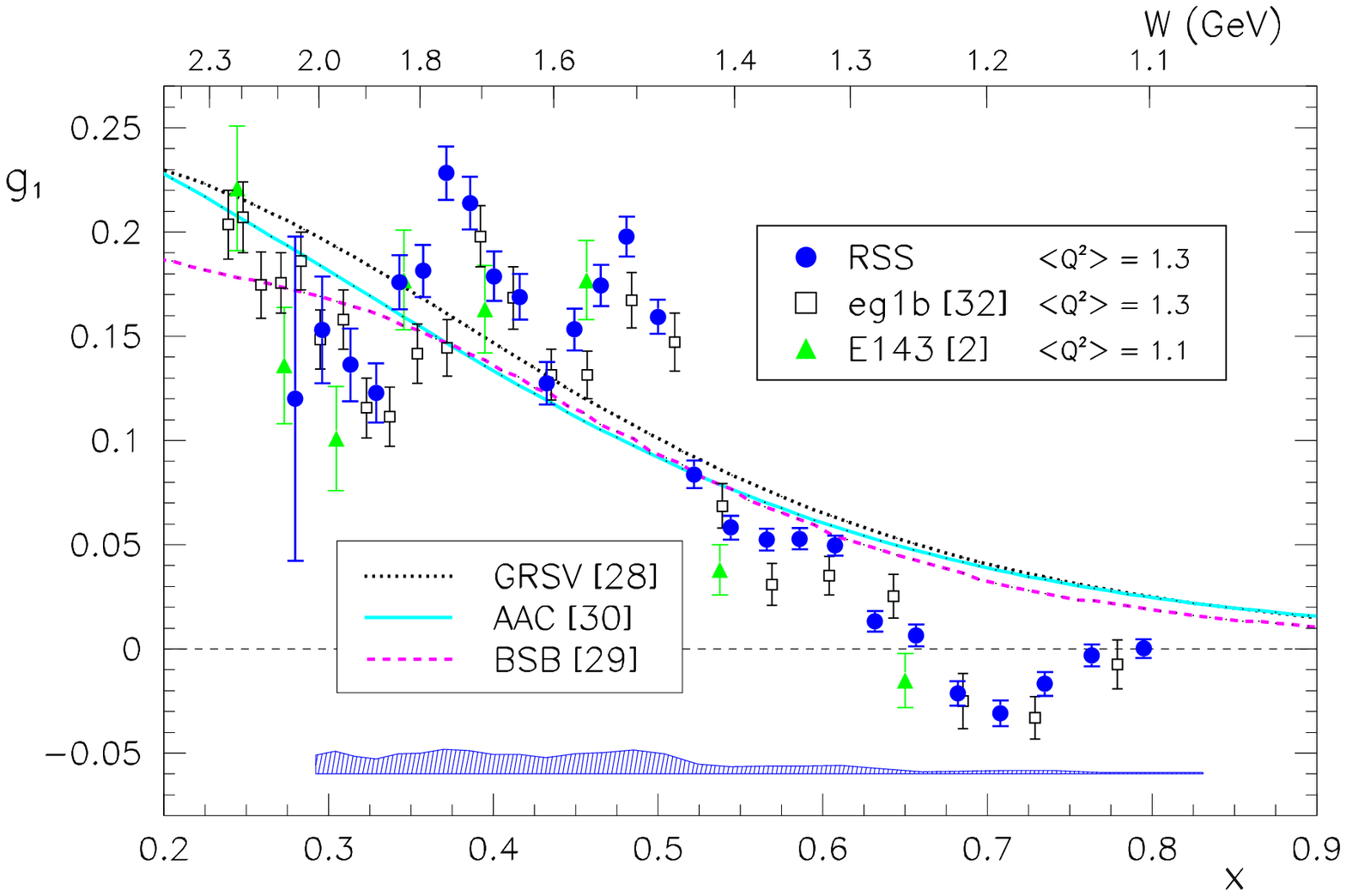,scale=0.45}
\end{minipage}
\begin{minipage}[t]{8.8 cm}
\epsfig{file=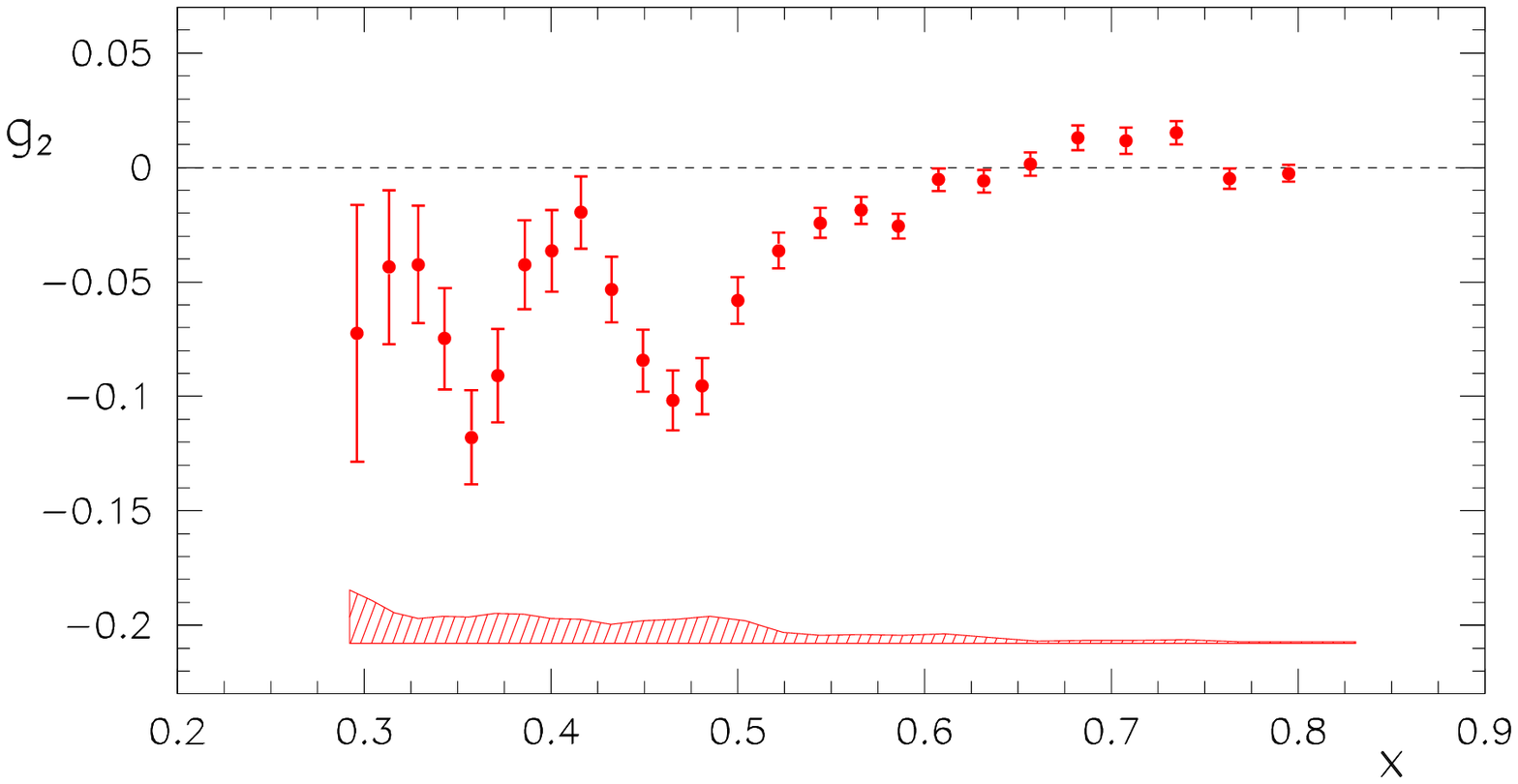,scale=0.45}
\end{minipage}
\begin{minipage}[t]{16.5 cm}
\caption{Data on the spin structure functions $g_1$  and $g_2$ of the proton
versus $x$ in the resonance region, for an average $Q^2$ of 1.3 GeV$^2$,
from Jefferson Lab's Hall C  (dark filled symbols) and other experiments (as indicated). 
Statistical errors are shown as error bars, while the systematic errors are indicated
by the shaded bands. 
\label{fig:rrc}}
\end{minipage}
\end{center}
\end{figure}

Since then, experiments in all three halls at Jefferson Lab have collected a vast
data set on all inclusive spin structure functions throughout the resonance region,
for proton, deuteron and $^3$He targets, spanning several orders of magnitude in
$Q^2$ (from 0.015 to 5 GeV$^2$). Figures~\ref{fig:rra} through~\ref{fig:rrc} show
some samples from this data set. 
Double spin asymmetries in the resonance region were also
measured at MIT Bates with the BLAST detector~\cite{Filoti:2007} 
on the proton and the deuteron at 0.85 GeV beam energy.
The kinematic region covered ranges from $Q^2 = 0.08 - 0.38$  GeV$^2$ and
from the elastic peak to $W = 1.33$ GeV.
More data in the region of the $\Delta(1232)$ resonance were also obtained at NIKHEF for
$Q^2 = 0.11$ GeV$^2$~\cite{vanBuuren:2002im}.

The data in Fig.~\ref{fig:rra} are from experiment E94-010 in Jefferson Lab's Hall 
A~\cite{Amarian:2003jy}. Both spin structure functions $g_1$ and $g_2$ were
extracted directly from cross section differences for polarized
electrons scattering off a $^3$He target polarized both along and transverse
to the beam (see Eqs.~\ref{eq:LongXsec},\ref{TransXsec}). These data
approximate the spin structure functions for a free neutron, although nuclear corrections
were not applied. 
One can clearly see the strongly negative response of $g_1$
in the $\Delta(1232)$ resonance region (high $x$). 
The structure function $g_2$ is close to the negative
of $g_1$, which implies that $A_2$, which is proportional to $g_1 + g_2$, is close to zero.

Figure~\ref{fig:rrb} shows a sample from the vast data set collected with CLAS during
the EG1 experiment~\cite{Dharmawardane:2006zd}
 on the spin structure function $g_1$ for the proton and the deuteron.
$g_1$ was extracted from measurements of the asymmetry $A_\|$ (see
Eq.~\ref{eq:Alongg1A2}) over a range in $x$ covering both the resonance and
deep inelastic regions, for over 25 bins in $Q^2$. 
As $Q^2$ increases over 2 orders in magnitude, the asymmetry (and therefore $g_1$)
changes from a predominantly negative value ($A_{3/2}$-dominance) to the strongly positive
DIS limit for
nearly all $x$ except in the region of the $\Delta(1232)$ resonance.

Finally, we show in Fig.~\ref{fig:rrc} the data from the ``RSS'' experiment in Hall 
C~\cite{Wesselmann:2006mw}.
This experiment measured $g_1$ and $g_2$ in the resonance region
for one bin in $Q^2$, centered around
1.3 GeV$^2$, for both proton and deuteron, using transversely and longitudinally
polarized targets.
The $g_1$ data  ``oscillate'' around the deep inelastic curves (from various NLO fits),
while $g_2$ is in rough agreement with the Wandzura-Wilczek form (not shown).

\section{Sum Rules and Spin Polarizabilities} \label{sec:mom}

\subsection{\it Moments of Spin Structure Functions \label{subsec:mom}}
Moments of structure functions are a powerful tool to study fundamental properties
of nucleon structure, like the total momentum fraction carried by quarks or the total
contribution of quark helicities to the spin of the nucleon. While a complete description
of structure functions based on fundamental QCD principles may be unattainable
for now, moments of structure functions can be directly compared to rigorous
theoretical results, like sum rules, lattice QCD calculations and chiral perturbation theory. The original ``spin crisis''
was directly due to a discrepancy between the data and an approximate ``sum rule'' by
Ellis and Jaffe~\cite{Ellis:1973kp}. On the other hand, a detailed measurement of the
$Q^2$--evolution of the Bjorken sum rule~\cite{Bjorken:1968dy} provides a significant
test of pQCD in the spin sector.


As mentioned in Section~\ref{subsec:parton}, via the Operator Product Expansion (OPE) the moments of $g_{1,2}$ can be related to
hadronic matrix elements of current operators. The moments are given as a sum, ordered according to the
\emph{twist} $\tau = (\textrm{dimension} - \textrm{spin})$ of the current operators, beginning with the lowest twist
$\tau =2$. Each additional unit of $\tau$ produces a factor of order $\frac{\Lambda_{QCD}}{Q}$, and is thus less
important in the high $Q^2$ regime. Strictly speaking, the higher twist terms are mixed up with correction
terms of order $\frac{M^2}{Q^2}$, of purely kinematic origin (target mass corrections), which can be calculated exactly,
so as to expose the HT terms of dynamic origin. Some of the twist 2  terms are directly \textit{measurable}
in \textit{other} processes. Twist 3 and higher terms can sometimes be determined from combinations
of measured quantities.\footnote{The reader is warned that the notation in the literature on this subject is in a state of anarchy, with identical symbols often being used for significantly different quantities in different papers. Thus $G_{1,2}$ sometimes means $M^3$ times our $G_{1,2}$, but perhaps more bizarre is the use of the symbol $\nu$ for $ P\cdot q $ in \cite{Ji:1993mv,Ji:1993sv}}

After a long series of manipulations one finds expressions for the
$n^{th}$ moment $\int_0^1 \, dx\,x^{n-1}\, g_1(x,Q^2)$ for $n=1,3,5......$ of $g_1$ and  the $n^{th}$ moment $\int_0^1 \, dx\,x^{n-1}\, g_2(x,Q^2)$ for $n=3,5,7......$  of $g_2$, in terms of hadronic matrix
elements  of certain local operators. (For details see  Section 22.2 of \cite{Leader:1996hm}).

The most important case  is the first moment, $n=1$, of $g_1$, where the operators involved are the octet of well known axial-vector currents which control neutron and hyperon
$\beta$-decay:
\beq \label{eq:AVcurrent}
J^i_{5\mu} = {\bar{\bm{\psi}}} \gamma_{\mu}\gamma_5
\Biggl(\frac{{\bm{\lambda}}_i }{2}\Biggl) {\bm{\psi}}\qquad
(i=1,2,...,8),
\eeq
where the ${\bm{\lambda}}_j$ are the usual Gell-Mann matrices and
${\bm{\psi}}$ is a column vector in flavor space
\begin{equation}\label{eq:psi}
{\bm{\psi}} =
\left(\begin{array}{c}\psi_u\\\psi_d\\\psi_s\end{array} \right)\, .
\end{equation}
The flavor singlet current, which does not play a role in $\beta$-decay, is
\begin{equation}\label{eq:AVsinglet}
J^0_{5\mu} = {\bar{\bm{\psi}}} \gamma_\mu \gamma_5 {\bm{\psi}}\,. 
\end{equation}

The relation of the hadronic matrix elements to the first  moments $ a_3 \equiv g_A, \, a_8, \,a_0 $ of the flavor combinations of quark densities introduced in Eqs.~\ref{eq:a3,8,0} is
\beqy \label{eq:defain}
\langle P,S|J^3_{5\mu}|P,S\rangle  = Ma_3S_\mu  \nn \\
\langle P,S|J^8_{5\mu}|P,S\rangle  = \frac{M}{\sqrt{3}}\,\,a_8S_\mu  \nn \\
\langle P,S|J^0_{5\mu}|P,S\rangle = 2M a_0(Q^2) S_\mu . \eeqy
The $Q^2$ dependence of $a_0$ arises because the singlet current has to be renormalized and it is customary and convenient to choose $ Q^2 $ as the renormalization scale.

At leading twist the expression for $\Gamma_1^{p,n}$, valid for $Q^2\gg M^2$, can be written
\beq \label{eq:GammaTotal}
\Gamma_1^{p,n} = \frac{1}{12}\big[ \big(\pm a_3 + \frac{1}{3}\,a_8 \big)E_{NS}(Q^2) + \frac{4}{3}\,a_0(Q^2) E_S(Q^2)\big] , \eeq
where the  coefficient functions are given, for 3 active flavors, by
\beq \label{ENS}
E_{NS}(Q^2) = 1- \big(\frac{\alpha_s}{\pi}\big)-3.58\,\big(\frac{\alpha_s}{\pi})^2 - 20.22 \big(\frac{\alpha_s}{\pi}\big)^3 ... \eeq
and in the $\overline{MS}$ scheme
\beq \label{ES}
E_{S}(Q^2) = 1- \big(\frac{\alpha_s}{\pi}\big)-1.096\,\big(\frac{\alpha_s}{\pi})^2 ...\, . \eeq
Note that the singlet coefficient here multiplies $a_0(Q^2)$ and not the invariant quantity referred to in various papers as $\hat{a}_0$,\, \, $ g_A^{(0)}|_{inv}$ and $\Delta \Sigma|_{invariant} $, which are the value of $a_0(Q^2)$ as $Q^2\rightarrow \infty$.

There exist important sum rules for the moments of $g_{1,2}$ which we now discuss.

\begin{itemize}
\item \textit{The Bjorken sum rule}. Using only isospin invariance Bjorken \cite{Bjorken:1968dy} showed that as $Q^2\rightarrow \infty $,
\beq \label{eq:Bjsum}
\int_0^1 dx\,[g_1^p(x,Q^2) - g_1^n(x,Q^2)] = \frac{g_A}{6}\, \big[ 1 - \big(\frac{\alpha_s}{\pi}\big) - 3.58 \big(\frac{\alpha_s}{\pi}\big)^2 - 20.22 \big(\frac{\alpha_s}{\pi}\big)^3 \cdot \cdot \cdot \big] , \eeq
where the square bracket on the RHS contains the perturbative 
corrections Eq.~\ref{ENS}, for 3 active flavors, to Bjorken's original result $g_A/6$. The sum rule  is quite rigorous, involving only the assumption of isospin invariance, 
and inspired Feynman to say that its failure would signal the demise of QCD.

\item \textit{The Efremov, Leader, Teryaev sum rule }. The ELT sum rule \cite{Efremov:1996hd} is also a rigorous result, following from the OPE,  and states that
\beq \label{eq:ELTsum}
\int_0^1 dx\,x\,[g_{1,f}^V(x) + 2 g_{2,f}^V(x) ] =0 , \eeq
where $V$ implies \emph{valence} contribution and the result holds for each flavor $f$.
It was originally thought reasonable to assume that the sea-quark densities are the same in protons and neutrons, in which case  Eq.~\ref{eq:ELTsum} can be written as a kind of analogue to the Bjorken sum rule
\beq \label{eq:ELTBj}
\int_0^1 dx\,x\,[g_2^p(x) - g_2^n(x) ]=\frac{1}{2}\int_0^1 dx\,x\,[g_1^n(x) - g_1^p(x) ] . \eeq
However, the above assumption is tantamount to assuming $\Delta \bar{u} =\Delta \bar{d}$ 
in a proton, which, at least for the {\em unpolarized} antiquark densities, is not a good approximation.
For further interesting sum rules of this type see \cite{Blumlein:1998nv, Blumlein:1996vs}.

\item \textit{The Burkhardt-Cottingham sum rule.} The BC sum rule \cite{Burkhardt:1970ti}
\beq \label{eq:BCsum}
\int_0^1 dx\, g_2(x) = 0  \eeq
does {\em not} follow from the OPE, although it looks as if it does, because the $n^{th}$ moment of $g_2$ is proportional to $(n-1)$; but, as mentioned, the OPE result for $g_2$ only holds for $n\geq 3$.

The BC sum rule is based on an assumption about the asymptotic behavior of a particular virtual Compton amplitude (see discussion below). The argument is not watertight and it may be that
\beq \label{eq:g2diverge}
g_2(x) \rightarrow \frac{1}{x^2}  \qquad \textrm{as} \qquad x\rightarrow 0 , \eeq
thus making the integral in Eq.~\ref{eq:BCsum} diverge. For a detailed discussion of this issue, see Section~5.2 of \cite{Anselmino:1994gn}.
Note that the Wandzura-Wilczek approximation to $g_2$, Eq.~\ref{eq:WW} automatically satisfies the BC sum rule
\beq \label{eq:WWBCrule}
\int_0^1 dx\, g_2^{WW}(x) = 0  . \eeq

\item \textit{Higher twist corrections.}

At lower values of $Q^2$ higher twist contributions become important. Of particular interest is $ \Gamma_{1}(Q^2)$, which can be expanded, for $Q^2> \Lambda_{QCD}^2$, in inverse powers of $Q^2$ (twist expansion)
\beq \label{eq:twistexp}
\Gamma_{1}(Q^2)= \mu_2(Q^2) + \,\frac{\mu_4(Q^2)}{Q^2} + \, \frac{\mu_6(Q^2)}{Q^4} + \,......  \eeq
with $\mu_2$, the leading twist term, given by Eq.~(\ref{eq:GammaTotal}).

The evolution in $Q^2$ of the $\mu_n(Q^2)$  can, in principle, be calculated in perturbative
QCD, but results are  known only for $\mu_2$ , and it is usual, therefore, 
to ignore the $Q^2$ dependence of 
$\mu_{4,6}$. The functions $\mu_n$ are related to matrix elements of operators of twist $ \tau\leq n $;
 the occurrence of operators with twist lower than $n$ is a kinematic effect, 
 a consequence of target mass corrections.

We consider now the structure of the higher twist 
(i.e., $1/Q^2$) corrections to $\Gamma_1(Q^2)$ as defined in Eq.~\ref{eq:twistexp}. 
As mentioned  $\mu_4(Q^2)$ contains both target mass (TM) corrections and 
genuine dynamical higher twist contributions. The TM corrections for $g_{1,2}(x,Q^2)$ 
were given in \cite{Piccione:1997zh}. For the first moment they yield
\beq \label{eq:mu4TM}
\mu_4^{TM}(Q^2)= \frac{2M^2}{9}\int_0^1 \, dx \, x^2\,[5g_1(x,Q^2) + 6g_2(x,Q^2)] . \eeq

$g_2$, as discussed in Section~\ref{subsec:g2}, contains a twist-2 (and 
kinematic twist-3) part, $g_2^{WW}$, 
given in Eq.~\ref{eq:WW'}, and a dynamic twist-3 part given by $g_2-g_2^{WW}$. Writing
\beq \label{eq:g2split}
g_2= g_2^{WW} + (g_2 - g_2^{WW}) \eeq
one finds, after a little algebra, that
\beq \label{eq:mu4g1g2}
\mu_4^{TM}(Q^2)= \frac{2M^2}{9}\Big\{\int_0^1 \, dx \, x^2\,g_1(x,Q^2) + 6 \int_0^1 \, dx \, x^2\,[g_2(x,Q^2)- g_2^{WW}(x,Q^2)]\Big\} . \eeq
The twist-2 combination of moments is usually referred to as  $a_2$ and the twist-3 as $d_2$. More precisely
\beq \label{eq:defa2d2}
a_2(Q^2)\equiv 2\int_0^1 \, dx \, x^2\,g_1(x,Q^2) \qquad \quad d_2(Q^2)\equiv 3\int_0^1 \, dx \, x^2\,[g_2(x,Q^2)- g_2^{WW}(x,Q^2)] . \eeq

There is also a genuine twist-4 contribution to $\mu_4$, written as $\frac{4M^2}{9}f_2$, so that finally
\beq \label{eq:mu4final}
\mu_4(Q^2) = \frac{M^2}{9}[ a_2(Q^2) + 4 d_2(Q^2) + 4 f_2(Q^2)] . \eeq

The twist-4 part $f_2$ cannot be written in terms of moments of the standard scaling functions. In terms of operator matrix elements it is defined by
\beq \label{eq:deff2}
f_2(Q^2) M^3 S^{\mu} = \frac{1}{2}\sum_{flavors} e_f^2 \langle P , S\,|\, g \bar{\psi}_f \,\tilde{G}^{\mu\nu}\, \gamma_\nu \,\psi_f\, | P, S\,\rangle \eeq
where $\tilde{G}^{\mu\nu} = (1/2)\epsilon^{\mu\nu\alpha\beta} G_{\alpha\beta}$ with the sign convention $\epsilon^{0123} = +1$. ( Note that in some papers the normalization of the covariant spin vector is taken to be $S^2=-M^2$, in which case $M^2$ will appear on the LHS of Eq.~\ref{eq:deff2}.)

The first correct expressions for $d_2$ and $f_2$ were given by Ji and Unrau \cite{Ji:1993sv}  and by Ehrnsperger, Mankiewicz and Sch\"{a}fer \cite{Ehrnsperger:1993hh}. ( Note that in Ji's paper the twist-4 term in Eq.~\ref{eq:mu4final} appears with a minus sign. This is because his coupling $g$ in the operator in Eq.~\ref{eq:deff2} has the  opposite sign to the usual convention i.e. he uses for the covariant derivative acting on quark fields  $D^\mu = \partial^\mu + igA^\mu $.)

Since $\mu_4$, $a_2$ and $d_2$ can be measured, Eq.~\ref{eq:mu4final} yields a measurement of the twist-4 matrix element $f_2$. Both $d_2$ and $f_2$ can then be compared to non-perturbative QCD calculations.

Finally, it is of some interest to consider the color electric and color magnetic polarizabilities of the nucleon \cite{Chen:2005tda}. They are defined by
\beq \label{eq:pols}
\chi_E 2M^3 \bm{S} = \langle \, N, \bm{S} \,| \, \bm{j}_a \times \bm{E}_a \,| N, \bm{S}\,\rangle \qquad \chi_B 2M^3 \bm{S} = \langle \, N, \bm{S}\, | \, j^0_a  \bm{B}_a \,| N, \bm{S}\,\rangle  \eeq
where $\bm{S} $ is the rest frame spin vector of the nucleon and $j^\mu_a$ is the quark current. $\bm{E}_a$ and $\bm{B}_a$ are the color electric and magnetic fields respectively. The color polarizabilities can be expressed in terms of $d_2$ and $f_2$ as
\beq \label{eq:chis}
\chi_E = \frac{2}{3}(2d_2 + f_2) \qquad \chi_M = \frac{1}{3}(4d_2 - f_2) . \eeq

\item \textit{Extrapolation to low $Q^2$: the Gerasimov, Drell, Hearn sum rule.} The twist expansion does not converge for very small $Q^2$. Thus to study 
polarized structure functions in the low $Q^2$ region it is necessary to make contact with soft, non-perturbative physics i.e. to study the relation to Compton scattering with \emph{real}, or almost real  photons. In this region care must be exercised in defining moments of structure functions. As already mentioned in Section~\ref{subsec:parton}  relations between moments of structure functions and
matrix elements of operators are only valid if the moments include the elastic contributions located at $x=1$. In the \emph{deep} inelastic region the elastic contributions are negligible and are not included in experimental estimates of the moments, but at low $Q^2$, in the resonance region, the distinction is important. Thus moments in the latter region, \emph{without} an elastic contribution will be labeled $\bar{\Gamma}$.

The GDH sum rule \cite{Gerasimov:1965et, Drell:1966jv} concerns Compton scattering with \emph{real} photons,
 i.e., with $Q^2=0$, and follows from the fact that the  value of the forward spin-flip amplitude, $f_2(\nu)$, at $\nu=0$, calculated to order $e^2$, is given by low energy theorems, and from the assumption that the dispersion relation for it does not require subtractions. This leads to
\beq \label{eq:GDH}
\int_0^\infty \frac{d\nu}{\nu} \, [\sigma_A(\nu) -\sigma_P(\nu)] = - \frac{2\pi^2\alpha}{M^2}\, \kappa^2 \eeq
where $\kappa$ is the anomalous magnetic moment of the  nucleon. The $\sigma_{A,P}$ are the total cross-sections for the absorption of a circularly polarized photon by a proton polarized with spin antiparallel/parallel to the photon spin.

Since, to order $e^2$, the cross-sections are zero below pion production threshold $\nu_0$
(see below), and following the convention used by experimentalists, we write Eq.~\ref{eq:GDH} in the form
\beq \label{eq:GDH'}
   I(0)\equiv \int_{\nu_0}^\infty \frac{d\nu}{\nu} \, [\sigma_A(\nu) -\sigma_P(\nu)] =
   -\frac{2 \pi^2 \alpha}{M^2}\kappa^2 . \eeq

In the language of DIS the above cross-section difference is referred to as $\sigma_{TT}$ and Eq.~\ref{eq:GDH'}, generalized to non-zero $Q^2$,  is usually written
\beq \label{eq:GDHDIS}
\lim_{Q^2\rightarrow 0}\, \bar{I}_{TT}(Q^2)= - \kappa^2/4 \eeq
where
\beqy \label{eq:ITTdef}
\bar{I}_{TT}(Q^2)& = &M^2 \int_{\nu_0 (Q^2)}^{\infty}  \frac{d\nu}{\nu^2} [ \nu M G_1(\nu, Q^2) - Q^2 G_2(\nu, Q^2)]\\
&=& \frac{2M^2}{Q^2}\bar{\Gamma}_{TT}(Q^2)  \label{eq:ITTdef'}\eeqy
with
\beq \label{eq:nu0Q}
\nu_0 (Q^2)= \frac{Q^2 + m_\pi ( 2M + m_\pi)}{2M}  \eeq
and where $ \bar{\Gamma}_{TT}(Q^2)$ is the \emph{inelastic} portion of the first moment
\beq \label{eq:gammabar}
\bar{\Gamma}_{TT}(Q^2)\equiv \int_0^{x_0(Q^2)} dx \, \big[g_1(x,Q^2) - \frac{4M^2x^2}{Q^2}g_2(x,Q^2)\big] \eeq
and $x_0(Q^2)$ is the threshold for pion production,
\beq \label{eq:pionthresh}
 x_0=\frac{Q^2}{Q^2 + m_{\pi}(2M + m_{\pi})} . \eeq

 Note that the generalization of $I(0)$ in Eq.~\ref{eq:GDH'} to arbitrary $Q^2$ is then given by
\beq \label{eq:IrelITT}
 I(Q^2)=\frac{8 \pi^2\alpha}{M^2}\, \bar{I}_{TT}(Q^2) . \eeq

 The relevance of Eqs.~\ref{eq:GDHDIS} and \ref{eq:ITTdef'} to deep inelastic scattering was first pointed out by Anselmino, Ioffe and Leader \cite{Anselmino:1988hn}, who noted that the first moment $\bar{\Gamma}^p_{TT}(Q^2)$, which is a slowly (logarithmically) varying, positive function in the DIS region, would have to change drastically in order to satisfy Eq.~\ref{eq:GDHDIS} at small $Q^2$. However \cite{Anselmino:1988hn} failed to distinguish between the full moment $\Gamma^p_{TT}(Q^2)$ and its inelastic version $\bar{\Gamma}^p_{TT}(Q^2)$, and Ji \cite{Ji:1993mv} pointed out that there is a significant difference between the extrapolation to $Q^2=0$ of the full and the inelastic moments . The reason is a non-uniformity of the limits $\nu \rightarrow 0$ and $Q^2 \rightarrow 0$ in the generalization of  $f_2(\nu)$ to $ Q^2\neq 0 $, namely,
 \beq \label{eq:nonunif}
 \lim_{Q^2\rightarrow 0}\, \, \lim_{\nu \rightarrow 0}\, f_2(\nu, Q^2) \neq \lim_{\nu \rightarrow 0} \,\,  \lim_{Q^2\rightarrow 0}\, f_2(\nu, Q^2) . \eeq
 
Thus whereas the RH limit does not inherit an elastic contribution from the Born terms in Compton Scattering the LH one does. And since the integrals involved in the moments above correspond to $\nu = 0$ in $f_2(\nu, Q^2)$, it is the LH limit which must be used in the extrapolation to $Q^2=0$. Thus for the \emph{full} moment 
Eq.~\ref{eq:GDHDIS} is altered to
\beq \label{eq:GDHDISfull}
 \lim_{Q^2\rightarrow 0}\, I_{TT}(Q^2)= \frac{M^2}{Q^2} [ F^{el}_1(Q^2) + F^{el}_2(Q^2)]^2 - \kappa^2/4   \eeq
where $F^{el}_{1,2}$ are the Dirac and Pauli \emph{elastic} from factors normalized to
\beq \label{eq:FFs}
F^{el}_{1p}(0) = 1 \qquad F^{el}_{1n}(0) = 0 \qquad F^{el}_{2p,n}(0)= \kappa_{p,n} . \eeq

In a similar way one can generalize to arbitrary $Q^2$ \cite{Ji:1993mv} the spin-dependent Compton amplitudes $S_{1,2}$, normalized so that
\beq \label{eq:S12}
Im S_{1,2}(\nu, Q^2) = 2\pi \, G_{1,2}(\nu, Q^2) .  \eeq

For $S_1$ one obtains the dispersion relation
\beq \label{eq:S1def}
S_1(\nu, Q^2) = 4 \int_{\nu_0(Q^2)}^\infty \frac{G_1(\nu',Q^2) \nu' d\nu'}{\nu'^2 - \nu^2} . \eeq
Then
\beqy \label{eq:I1def}
\frac{M^3}{4}S_1(0,Q^2)=\bar{I}_1(Q^2)& \equiv &M^2 \int_{\nu_0(Q^2)}^{\infty}  \frac{d\nu}{\nu}   M G_1(\nu, Q^2)\\
&=& \frac{2M^2}{Q^2}\bar{\Gamma}_1(Q^2)  \label{eq:I1def'}\eeqy
where $ \bar{\Gamma}_1(Q^2)$ is the \emph{inelastic} portion of the first moment $\Gamma_1(Q^2)$
\beq \label{eq:gamma1bar}
\bar{\Gamma}_1(Q^2)\equiv \int_0^{x_0(Q^2)} dx \, g_1(x,Q^2)  \eeq
and
\beq \label{eq:GDHDIS1}
 \lim_{Q^2\rightarrow 0}\, \bar{I}_1(Q^2)= - \kappa^2/4 . \eeq

For the full moment,
\beq \label{eq:GDHD1ISfull}
 \lim_{Q^2\rightarrow 0}\, I_1(Q^2)= \frac{M^2}{Q^2} F^{el}_1(Q^2) \,[ F^{el}_1(Q^2) + F^{el}_2(Q^2)] - \kappa^2/4 .     \eeq

 In the DIS region the elastic contribution to the moments is totally negligible. However in the extrapolation down through the resonance region towards $Q^2=0$ it is vitally important to distinguish between the two cases.

 We turn now to the Compton amplitude $S_2(\nu)$ which underlies the BC sum rule. If one assumes that
  $S_2(\nu, Q^2)$ satisfies a superconvergence relation, i.e., that it vanishes as $\nu \rightarrow \infty $ fast enough so that both $S_2$ and $\nu S_2$ satisfy unsubtracted dispersion relations, then one can  show that
 \beq \label{eq:BCgen}
 \int_0^\infty Im S_2(\nu, Q^2) d\nu = 2 \pi\int_0^\infty G_2(\nu, Q^2) d\nu =0  \eeq
 which leads to Eq.~\ref{eq:BCsum} for the full moment. From a knowledge of the elastic terms in $S_2(0,Q^2)$ one obtains an expression for the inelastic integral
 \beq \label{eq:I2}
 \bar{I}_2(Q^2) \equiv\frac{2M}{Q^2}\int_0^{x_0(Q^2)} g_2(x,Q^2) dx = \frac{1}{4}F_2^{el}(Q^2)[F_1^{el}(Q^2) + F_2^{el}(Q^2)]  . \eeq

For a discussion of the convergence properties of the amplitudes see Section 2.10 of \cite{Ioffe:1985ep}.

 \item \textit{Generalization of the GDH approach.}

 The above type of analysis, based on the dispersion relations for $f_2(\nu, Q^2), S_{1,2}(\nu Q^2)$, has been generalized to all the amplitudes in virtual Compton scattering by Drechsel, Pasquini and Vanderhaeghen \cite{Drechsel:2002ar}. For  very small energies, $\nu < \nu_0(Q^2) $, the amplitudes can be expanded in powers of $\nu^2$. The coefficients of the next-to-leading terms are called  \emph{generalized forward spin polarizabilities} of the nucleon. They can be expressed in terms of moments of the structure functions and can thus be measured. They are important because they reflect soft, non-perturbative aspects of nucleon structure, and can be calculated, at least approximately, using various forms of chiral perturbation theory and lattice methods. They thus provide benchmark tests for these theories. Here we shall simply give the expressions for experimentally relevant polarizabilities. For details of the amplitudes etc, see \cite{Drechsel:2002ar}.

 \beqy \label{eq:gamma0}
 \gamma_0(Q^2) \equiv \gamma_{TT}(Q^2) &=& 2\alpha \int_{\nu_0}^{\infty}  \frac{d\nu}{\nu^4}\, \big[ \nu M G_1(\nu, Q^2) - Q^2 G_2(\nu, Q^2)\big] \\
 &=& \frac{16\alpha M^2}{Q^6}\int_0^{x_0} x^2 \, \big[g_1(x,Q^2) - \frac{4M^2x^2}{Q^2}g_2(x,Q^2)\big]\,dx \eeqy

 \beqy \label{eq:delta0}
 \delta_0(Q^2) \equiv \delta_{LT}(Q^2) &=& 2\alpha \int_{\nu_0}^\infty \frac{d\nu}{\nu^3}\,\big[MG_1(\nu,Q^2) + \nu G_2(\nu, Q^2)\big] \\
 &=& \frac{16\alpha M^2}{Q^6}\int_0^{x_0} x^2 \big[g_1(x,Q^2) + g_2(x,Q^2)\big]\, dx \eeqy

 The comparison between theory and experiment is discussed in the next few sections.

\end{itemize}

\subsection{\it First Moment of $g_1$ at High $Q^2$ and Sum Rules \label{subsec:bjork}}
After the EMC announced their surprising result~\cite{Ashman:1987hv},
the next round of experiments 
at SLAC, CERN and DESY was focused on the goal
 to measure the first moments of $g_1$ for the proton, deuteron and
neutron ($^3$He) in the DIS region with increasing precision.
The main interest was to gain information on the
contribution of quark helicities to the nucleon spin
and to test the
Bjorken sum rule~\cite{Bjorken:1968dy}.

Experimentally, these integrals can only be measured over some 
finite range in $x$, since the
limit $x \rightarrow 0$ for finite $Q^2$ requires infinitely high energy transfers $\nu$.
In practice, all existing experiments accessing $g_1$ are fixed target experiments, with
a lower limit of $x$ of the order $x = 0.004 - 0.01$ (if $Q^2 \geq 1$ GeV$^2$). 
Experimental results for the moments $\Gamma_1$
therefore include some extrapolation to $x=0$, typically based on Regge 
phenomenology~\cite{Bass:2006dq,Bianchi:1999qs,Simula:2001iy}, NLO fits or
some other assumption about the functional form of the underlying 
quark densities.
This extrapolation to the lower integral limit is the largest source of 
theoretical uncertainty for the
results obtained so far. Most experiments also use an extrapolation at high $x$, since 
small statistics and limited momentum resolution prevent an accurate measurement in this
kinematic region. However, at sufficiently large $Q^2$, the contribution from
this extrapolation is very small and
well defined. 

Table~\ref{tab:gamma1} lists the final results for the first moments measured
 at $Q^2 \geq 1$ and evolved to a fixed $Q^2$ (as indicated) for the
 more recent experiments. 
 The quoted uncertainties include both statistical and systematic errors added in
 quadrature. The results from different experiments are in agreement within 
 uncertainties, after taking $Q^2$-evolution into account. 
 The possible exception is the deuteron result of SMC~\cite{Adeva:1998vv}
 which is significantly lower than all others; this is due to a strongly negative
 contribution from the unmeasured low-$x$ region which since has been
 ruled out by COMPASS data.
 In particular, the Bjorken sum rule is confirmed to within 7-15\% of
 its value $0.182 \pm 0.002$ (evolved to the average $Q^2 = 5$ GeV$^2$
 of the data listed in Table~\ref{tab:gamma1}); by far the largest remaining
 uncertainty stems from the low-$x$ convergence of the integral.
 This agreement can be considered a successful test of $Q^2$-evolution and the value of
 the strong coupling constant $\alpha_s$.
 
\begin{table}
\begin{center}
\begin{minipage}[htb!]{16.5 cm}
\caption{Results for the first moments of the spin structure functions $g_1$
from different experiments. Each experiment evolved its data
to a fixed value of $Q^2$ which is indicated.
The results for the deuteron are {\em not} corrected for its D-state,
 but are ``per nucleon''.}
\label{tab:gamma1}

\begin{tabular}{cccccl}
& & & & & \\[-2mm]
\hline
& & & & & \\[-2mm]
$Q^2$ [GeV$^2$] & $\Gamma_1^p$ & $\Gamma_1^n$ & $\Gamma_1^d$ & $\Gamma_1^p - \Gamma_1^n$ & Ref.\\
& & & & &  \\[-2mm]
\hline
& & & & &   \\[-2mm]
10 & 0.120 $\pm$ 0.016 & - & 0.019 $\pm$ 0.015 & 0.198 $\pm$ 0.023 & SMC~\cite{Adeva:1998vv}\\
3 & - & -0.033 $\pm$ 0.011 & - & - & E142~\cite{Anthony:1996mw}\\
3 & 0.133 $\pm$ 0.010 & -0.032 $\pm$ 0.018 &0.047 $\pm$ 0.007 & 0.164 $\pm$ 0.023 & E143~\cite{Abe:1998wq}\\
5 & - & -0.056 $\pm$ 0.009 & - & 0.168 $\pm$ 0.010 & 
E154\footnotemark[1]~\cite{Abe:1997dp}\\
5 & 0.118 $\pm$ 0.008 & -0.058 $\pm$ 0.009 & - & 0.176 $\pm$ 0.008 & 
E155\footnotemark[1]~\cite{Anthony:2000fn}\\
2.5 & 0.120 $\pm$ 0.009 & -0.028 $\pm$ 0.009 & 0.043 $\pm$ 0.004 & 0.148 $\pm$ 0.017 &
 HERMES\footnotemark[2]~\cite{Airapetian:2007mh}\\
3 & - & - & 0.046 $\pm$ 0.006 & - & COMPASS~\cite{Alexakhin:2006vx}\\[+2mm]
\hline
& & & &  \\[-2mm]
5 & Bjorken Sum Rule & & & 0.182  $\pm$ 0.002& \\
& & & &  \\[-2mm]
\hline
\end{tabular}
\addtocounter{footnote}{-1}
\footnotetext{$^1$ From a NLO analysis}
\footnotetext{$^2$ Over measured region $x>0.021$ only}
\end{minipage}
\end{center}
\end{table} 

 COMPASS and HERMES used their most recent result on the deuteron
to extract an ``experimental estimate'' of the singlet axial charge $a_0$
(which equals $\Delta \Sigma$ in the $\overline{MS}$ scheme), yielding
 $\Delta \Sigma \approx 0.35 \pm 0.06$ for COMPASS (at $Q^2 = 3$ GeV$^2$) and 
 $\Delta \Sigma \approx 0.33 \pm 0.04$
 for HERMES (evaluated at $Q^2 = 5$ GeV$^2$). These numbers are 
 somewhat higher than most recent 
 NLO analyses (see Section~\ref{subsec:delq}), but in agreement with
 similar analyses by other experiments. All results on $\Gamma_1$ point
 towards a negative contribution to the integral from strange quarks and antiquarks
 of order $\approx -0.1$,
 while semi-inclusive results from HERMES~\cite{Airapetian:2004zf,Airapetian:2008qf} 
 are consistent with 
 $\Delta s \geq 0$ in their measured range, see Sections~\ref{subsec:PDFexp},\ref{subsec:delq}.

\subsection{\it The BC and ELT Sum Rules \label{subsec:gamma2}}
 While the first moment of $g_1$ depends on $Q^2$ and changes significantly from
the DIS region to the real photon point, the Burkhardt-Cottingham sum rule for
the first moment of $g_2$ should apply at all values of $Q^2 > 0$, as long as the
elastic contribution is included in the integral. It can therefore be tested both by
DIS data and at lower $Q^2$, where a significant contribution to the integral comes from the 
nucleon resonance region.

The SLAC E155 data discussed in Section~\ref{subsec:g2} 
yielded a first test of the sum rule for the proton and the deuteron. The data were integrated
over the measured region, $0.02 < x < 0.8$ at $Q^2 = 5$ GeV$^2$. The contribution from the
unmeasured large-$x$ region is negligible. For the low-$x$ expansion, one can use the
assumption that $g_2$ follows the Wandzura-Wilczek form (Eq.~\ref{eq:WW'}) which yields
a result independent of our knowledge of $g_1$ below $x = 0.02$. Under these assumptions,
E155 found the integral for the proton to be $-0.022 \pm 0.008$ and that for the 
deuteron as $-0.002 \pm 0.011$, after averaging with the data from E143. While the proton
result appears to be inconsistent with the BC sum rule at the $ 2.75 \sigma$ level,
any firm conclusion depends strongly on the behavior of $g_2$ at small $x$ which 
is not known with enough precision and may
not follow the Wandzura-Wilczek  form.

The same data were also used to estimate the value for the ELT integral (Eq.~\ref{eq:ELTsum})
using the approximation given in Eq.~\ref{eq:ELTBj}, with the neutron structure functions inferred from
the measured proton and deuteron ones. The integral over the measured region is consistent
with the expected value of zero, within errors of $\pm 0.008$. Here, the extrapolation to
small $x$ is less critical because of the extra factor of $x$ in the integral.

\begin{figure}[htb!]
\begin{center}
\begin{minipage}[t]{8 cm}
\vspace {0.1cm}
\epsfig{file=BC_He3.eps,scale=0.4}
\end{minipage}
\begin{minipage}[t]{8 cm}
\epsfig{file=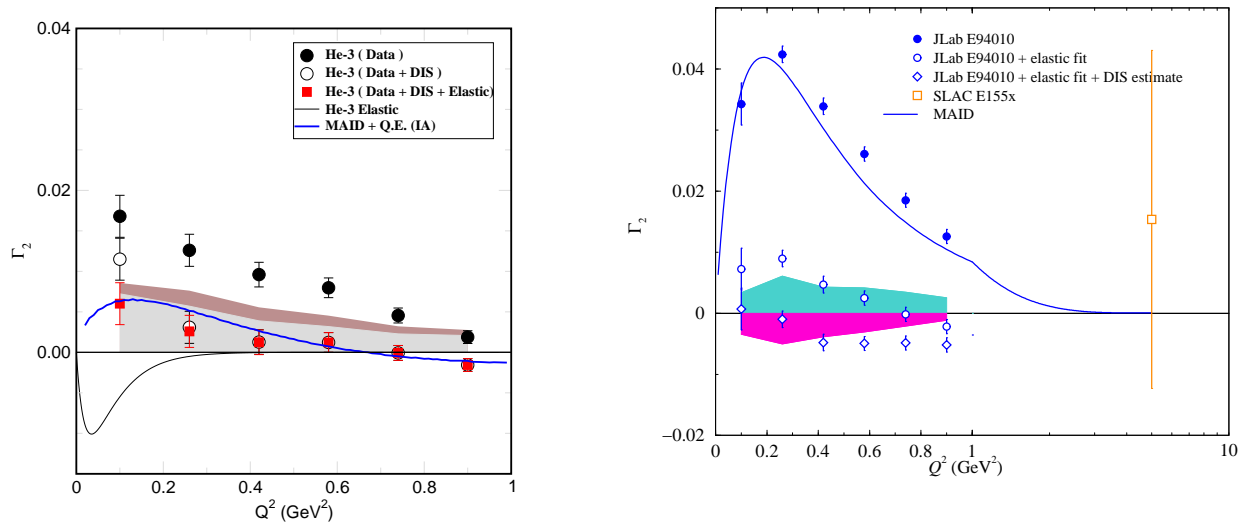,scale=0.4,angle=-90}
\end{minipage}
\begin{minipage}[t]{16.5 cm}
\caption{Results for the BC sum $\Gamma_2 (Q^2)$ for $^3$He~\protect\cite{Slifer:2008re}
and the neutron~\protect\cite{Amarian:2003jy}. The integrals
over the measured region (filled circles) are compared with MAID model calculations. Light (dark) shaded bands indicate experimental (extrapolation) systematic
errors.}
\label{fig:BC}
\end{minipage}
\end{center}
\end{figure}

The most extensive measurement of the BC sum rule at smaller $Q^2$ comes
from an experiment using a longitudinally and transversely
polarized $^3$He target in Hall A.
The integral of $\Gamma_2^{^3He}=\int g_2^{^3He}(Q^2) dx$
is plotted in the left panel of Fig.~\ref{fig:BC} for the measured region (from break-up
threshold on up and covering
the resonance region; solid circles).
For comparison, we also show a calculation using the MAID code~\cite{Drechsel:1998hk} 
which is a unitary isobar model describing the nucleon resonance region, with parameters
fit to meson photo- and electro-production data. Added to the MAID result is
 an estimate of the $pd$ and $ppn$
breakup contributions via a calculation of quasi-elastic single-proton knockout
in the impulse approximation.
The open circles show
the experimental results after adding an estimated
DIS contribution.
The solid squares
correspond to the results obtained after adding the elastic contributions for $^3$He
and are in good agreement with the expected value of zero within the
systematic errors (indicated by the error bands for the experimental
systematic error and the uncertainty 
on the low-$x$ extrapolation).

After applying some nuclear corrections, results for the neutron can be extracted
from the $^3$He data.
The result for $\Gamma_2^n$ 
is plotted in the right panel of 
Fig.~\ref{fig:BC} in the measured region (solid circles), while the
open circles include the elastic contribution and the open diamonds
correspond to the results obtained after
adding an estimated DIS 
contribution assuming $g_2 = g_2^{WW}$ for the neutron. 
Taking the difference between proton and deuteron data from E155, one can estimate the result for the
neutron at high $Q^2$ (open square), which is consistent with zero but with a rather 
large error bar. 
The results from the Hall-A experiment for the neutron  integral are the most precise
data on the BC sum rule and
are consistent with the expectation of zero, within systematic errors. 
The MAID parametrization agrees with the general trend but is slightly lower than the 
 data in the measured (resonance) region, presumably because multi-pion final
 states are not included in the MAID code. 
 
The RSS experiment in Hall C~\cite{Wesselmann:2006mw}
 also took data on $g_2$ for the proton and the deuteron,
at an average $Q^2$ of about 1.3 GeV$^2$. For the proton, they find a negative integral
over the measured region ($1$ GeV $<W<2$ GeV), but after adding the elastic contribution and an
estimate for the small-$x$ part, they find a 
(preliminary) result for the BC sum very close to zero,
with (small) statistical and (dominant) systematical errors of about $\pm 0.005$.
This result is in disagreement with the SLAC result, but the unmeasured region
($x<0.29$) is rather large and $Q^2$ is significantly lower. 
Preliminary results on the deuteron are slightly negative but consistent with zero
within an error of about $\pm 0.015$.

\subsection{\it Higher Moments and Higher Twist \label{subsec:OPE}}
 
While the moments of the structure function $g_1$ provide information
on the parton densities, the moments 
of  $g_2$ are related to higher twist
effects and provide information on quark-gluon interactions (correlations).
The $x^2$ moment of the spin-structure functions given in Eq.~\ref{eq:defa2d2}, 
\begin{equation}
d_2 = 3 \int _{0}^{1} x^2 [g_2(x)-g_2^{WW}(x)] dx,
\end{equation}
is of special interest: 
at high $Q^2$ it is a twist-3 matrix element and is related to the color 
polarizabilities~\cite{Ji:1997gs}, see Section~\ref{subsec:mom}. It can be calculated in 
lattice QCD~\cite{Gockeler:2000ja} and a number
of theoretical models~\cite{Stratmann:1993aw,Weigel:2003pe,Wakamatsu:2000ex}. 

\begin{figure}[ht!]
\centering \scalebox{0.5} {\includegraphics {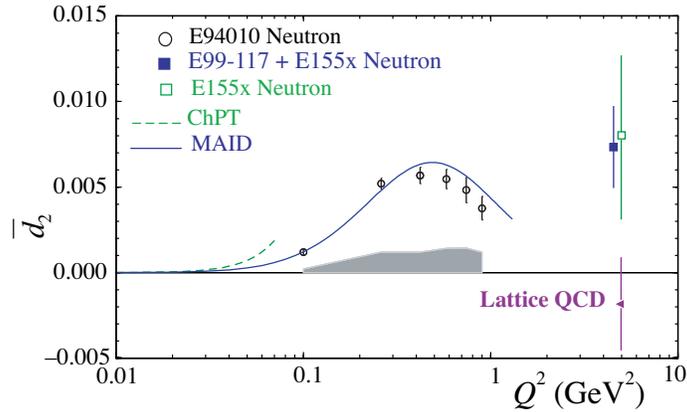}}
\vspace{-1.0in}
\begin{center}
\begin{minipage}[t]{16.5 cm}
\vspace{-0.2in}
\caption{Results for $\bar d_2^n (Q^2)$ from JLab Hall A~\protect\cite{Amarian:2003jy,Zheng:2004ce}
and SLAC~\protect\cite{Anthony:2002hy}, together with 
Lattice QCD calculations~\protect\cite{Gockeler:2000ja}.}
\label{fig:d2n}
\end{minipage}
\end{center}
\end{figure}

Experimentally, due to the $x^2$ weighting, the contributions are dominated by the high-$x$ region and the problem of low-$x$ extrapolation is avoided. The
SLAC E155 data~\cite{Anthony:2002hy}
allowed the first extraction of $\bar{d}_2^p$ and 
$\bar{d}_2^n$ (excluding the elastic contribution). 
Combining E155 results with earlier SLAC data, at an average $Q^2$ 
of 5 GeV$^2$, $\bar{d}_2^p=0.0032 \pm 0.0017$ and $\bar{d}_2^n=0.0079 \pm 0.0048$.
Combining these data with the JLab $g_2$ from Hall A E99-117~\cite{Zheng:2004ce}
a new value for
the third moment $\bar{d}_2^n$
was extracted at an average $Q^2$ of 5 GeV$^2$.
Compared to the previously published SLAC result~\cite{Anthony:2002hy}, 
the uncertainty on $\bar{d}_2^n$ has 
been improved by about a factor of 2 (see Fig.~\ref{fig:d2n}). 
 While a negative or near-zero value was 
predicted by Lattice QCD and most models, the new result for $\bar{d}_2^n$ 
is positive. Also shown in Fig.~\ref{fig:d2n} are the low $Q^2$ (0.1-1 GeV$^2$) results 
for $\bar{d}_2^n$ from another Hall A experiment,
 E94-010~\cite{Amarian:2003jy}, which were
compared to a Chiral Perturbation Theory calculation~\cite{Ji:1999pd} and a 
prediction based on MAID~\cite{Drechsel:2000ct}. The elastic part, while negligible at high
$Q^2$, is significant at low $Q^2$ ($<1$ GeV$^2$)
and was subtracted from both theoretical calculations. The MAID calculation
represents the low-$Q^2$ data rather well, probably due to the $x^2$-weighting
which suppresses the contribution of higher mass final states.

A new precision experiment to measure $d_2^n$ at an average $Q^2$ of 
3 GeV$^2$ is planned in Hall A in early 2009~\cite{d2n6GeV}. Further 
measurements~\cite{d2n12GeV} of $d_2^n$ at constant 
$Q^2$ ranging from 3 to 5 GeV$^2$ are planned at JLab after the 12 GeV energy upgrade. 

The Hall C RSS experiment~\cite{Wesselmann:2006mw} measured $g_2$ on the proton 
and the deuteron 
and extracted $\bar{d}_2^p = 0.0072 \pm 0.0017$ at a $Q^2$ value of 1.3 GeV$^2$. 
A more comprehensive measurement of $g_2^p$ and $d_2^p$ is scheduled in Hall C~\cite{SANE}. It will cover a wide $Q^2$ range from 2.5 to 6.5 GeV$^2$.

%
The higher-twist contributions to $\Gamma_1$ can be obtained
by a fit with an OPE series, Eq.~(\ref{eq:twistexp}), truncated to an order
appropriate for the precision of the data.
The goal is to determine the twist-4 matrix element $f_2$.
Once $\mu_4$ is obtained, $f_2$ is extracted 
by subtracting the leading-twist contributions of $a_2$ and $d_2$ 
following Eq.~(\ref{eq:mu4final}). To have an idea how the higher-twist terms
(twist-6 and above) affect the twist-4 term extraction,
it is necessary to study the convergence of the expansion 
and to choose the $Q^2$ range in a region where $\mu_8$ term is not
significant. This study is made possible only with the 
availability of the new low-$Q^2$ data from JLab.

Higher-twist analyses 
have been performed on the proton~\cite{Deur:2005jt,Osipenko:2004xg,Osipenko:2005nx}, 
the neutron~\cite{Meziani:2004ne} and the Bjorken sum 
($p-n$)~\cite{Deur:2004ti,Deur:2008ej}. 
$\Gamma_1$ at moderate $Q^2$ was obtained from the JLab $g_1$ data with details
described in 
Section~\ref{subsec:chiral}.
For consistency, the unmeasured low-$x$ parts of the JLab  and 
 the world data on $\Gamma_1$ were re-evaluated using the same prescription~\cite{Meziani:2004ne}. 
The elastic contribution, negligible above 
$Q^2$ of 2 GeV$^2$ but significant (especially for the proton) at lower values 
of $Q^2$, was added using the parametrization of Ref.~\cite{Mergell:1995bf}.
The leading-twist term $\mu_2$ was determined by fitting the 
data at $Q^2 \ge 5$ GeV$^2$ assuming that higher twists
in this $Q^2$ region are negligible. 
Using the proton (neutron) data alone, and with input of 
$a_3 (=g_A)$ from the neutron beta decay and $a_8$ from hyperon 
decay (assuming SU(3) flavor symmetry), 
$\Delta\Sigma=0.15 \pm 0.07$ for the proton analysis~\cite{Deur:2005jt} and
$\Delta\Sigma=0.35\pm 0.08$ for the neutron analysis~\cite{Meziani:2004ne} were obtained.  
Note that there is a difference of nearly two standard deviations between these two 
results; this difference presumably comes from the different data sets used
for the two analyses.


The fit results using an expansion up to ($1/Q^6$) in determining $\mu_4$ are 
summarized in Table~\ref{tab:mu2}. 
The extraction on $p-n$ was performed 
twice: first~\cite{Deur:2004ti} with the combined Hall A E94-010 neutron and Hall B EG1A proton
data and then the second time~\cite{Deur:2008ej} with the EG1B proton and deuteron data. 
In order to extract $f_2$, shown in Table~\ref{tab:f2},
the target-mass corrections $a_2$ were evaluated using the Bl\"umlein-B\"ottcher NLO fit to the world data~\cite{Bluemlein:2002be} for the proton and a JLab fit to the world neutron data, which includes the recent
high precision neutron results at large $x$~\cite{Zheng:2003un,Zheng:2004ce}. The $d_2$ values 
used are from SLAC E155~\cite{Anthony:2002hy} (proton) and JLab E99-117~\cite{Zheng:2004ce}
(neutron). 
\begin{table}[ht]
\begin{center}
\begin{minipage}[t]{16.5 cm}
\caption{Results of $\mu_4$, $\mu_6$ and 
$\mu_8$ at $Q^2$ = 1 GeV$^2$ for proton, neutron and $p-n$. The uncertainties are
first statistical then systematic.} 
\label{tab:mu2}
\end{minipage}
{\begin{tabular}{|c|c|c|c|c|}
\hline 
Target& $Q^{2}$ (GeV$^{2}$)&
$\mu_4/M^2$&
$\mu_6/M^4$&
$\mu_8/M^6$\tabularnewline
\hline
\hline 
proton~\cite{Deur:2005jt} & 0.6-10.0&
-0.065$\pm 0.012 \pm 0.048$&
0.143$\pm 0.021 \pm 0.056$&
-0.026$\pm 0.008 \pm 0.016$\tabularnewline
neutron~\cite{Meziani:2004ne}&0.5-10.0&
0.019$\pm 0.002 \pm 0.024$&
-0.019$\pm 0.002 \pm 0.017$&
0.00$\pm 0.00 \pm 0.03$\tabularnewline
$p-n$~\cite{Deur:2004ti}&0.5-10.0&
-0.060$\pm 0.045 \pm 0.018$&
0.086$\pm 0.077 \pm 0.032$&
0.011$\pm 0.031 \pm 0.019$\tabularnewline
$p-n$~\cite{Deur:2008ej}&0.66-10.0&
-0.039$\pm 0.010 \pm 0.026$&
0.084$\pm 0.011 \pm 0.024$&
0.047$\pm 0.026$\tabularnewline
\hline
\end{tabular}}
\end{center}
\end{table}

\begin{table}[ht]
\begin{center}
\begin{minipage}[t]{16.5 cm}
\caption{
Results of $f_2$, $\chi_E$ and 
$\chi_B$ at $Q^2$ = 1 GeV$^2$ for proton, neutron and $p-n$.
The uncertainties are
first statistical then systematic.} 
\label{tab:f2}
\end{minipage}
{\begin{tabular}{|c|c|c|c|}
\hline 
Target & $f_2$ & $\chi_E$ & $\chi_B$ \tabularnewline
\hline
\hline 
$p$~\cite{Deur:2005jt} & -0.160 $\pm 0.028 \pm 0.109$ & -0.082 $\pm 0.016 \pm 0.071$ &
0.056 $\pm 0.008 \pm 0.036$\tabularnewline
$n$~\cite{Meziani:2004ne} & 0.034 $\pm 0.005 \pm 0.043$ & 0.031 $\pm 0.005 \pm$ 0.028 & 
-0.003 $\pm$ 0.004 $\pm$ 0.014 \tabularnewline
$p-n$~\cite{Deur:2004ti} & $-0.136 \pm 0.102 \pm 0.039$ & $-0.100 \pm 0.068 \pm 0.028 $
& $0.036 \pm 0.034 \pm 0.017 $ \tabularnewline
$p-n$~\cite{Deur:2008ej} & $-0.101 \pm 0.027 \pm 0.067$ & $-0.077 \pm 0.050 $
& $0.024 \pm 0.028 $ \tabularnewline
\hline
\end{tabular}}
\end{center}
\end{table}

The fits were repeated varying the minimum $Q^2$ threshold 
to study the convergence of the OPE series. 
The extracted quantities have large uncertainties (dominated by the 
systematic uncertainty) but are stable with respect to the minimal $Q^2$ 
threshold when it is below 1 GeV$^2$. 
The results do not vary significantly when the 
$\mu_8$ term is added, which justifies \emph{a posteriori} the use of the 
truncated OPE series in the chosen $Q^2$ range. In the proton case, the 
elastic contribution makes a significant contribution to the $\mu_6$ term at 
low $Q^2$ but this does not invalidate \emph{a priori} the validity of the
series since the elastic contribution affects mainly $\mu_6$ and $\mu_8$ remains
small compared to $\mu_4$. We notice the alternation of signs between the 
coefficients. This leads to a partial suppression of the higher-twist 
effects and may be a reason for quark-hadron duality in
the spin sector (see Section~\ref{sec:duality}). We also note that the sign 
alternation is opposite for the proton and neutron. 
Following Eq.~(\ref{eq:chis}), the electric and magnetic color polarizabilities were determined,
see 
Table~\ref{tab:f2}.
 We observe a sign change 
in the electric color polarizability between the proton and the neutron. 
We also expect a sign change in the color magnetic polarizability. 
However, with the large uncertainty and the 
small negative value of the neutron $\chi_B$, it is difficult to confirm 
this expectation.

Additional data from JLab experiment EG1 are presently under analysis and will
significantly
improve the precision of the first moments $\Gamma_1^p$ and $\Gamma_1^d$
around $Q^2 = 1$ GeV$^2$, which is the crucial region for 
the extraction of higher-twist contributions.
These data should therefore allow a more precise extraction of quantities like
$f_2$ and the color polarizabilities.

\subsection{\it GDH Sum Rule and Chiral Expansion \label{subsec:chiral}}
In recent years
the Gerasimov, Drell and Hearn (GDH) sum rule~\cite{Gerasimov:1965et,Drell:1966jv} at $Q^2=0$
has attracted much experimental and theoretical~\cite{Chen:2005td,Drechsel:2004ki} 
effort that has provided us with rich information. 
Experimental measurements of the GDH sum on the proton and deuteron were performed
at Mainz~\cite{Ahrens:2001qt} and Bonn~\cite{Dutz:2003mm,Dutz:2005ns}, and the
contribution from the first resonance region and below was also measured at 
LEGS~\cite{Hoblit:2008iy}.
The results show that the GDH sum rule is
satisfied for the proton, assuming a model~\cite{Drechsel:2004ki} for the unmeasured regions. 
A review on the overall status of the GDH sum rule for real photons has been published
recently in this journal~\cite{Helbing:2006zp}.

A generalized sum rule~\cite{Ji:1999mr} connects the GDH sum rule with the 
Bjorken sum 
rule and provides a clean way to compare theories with experimental data over the
entire $Q^2$ range. The (generalized) GDH sum rule and similar
spin sum rules relate the moments of the spin-structure functions to the 
nucleon's static properties (e.g., its anomalous magnetic moment) and to
real or virtual Compton amplitudes
(see Eqs.~\ref{eq:S12}-\ref{eq:GDHD1ISfull},\ref{eq:gamma0}-\ref{eq:delta0}),
 which can be calculated theoretically.
Several papers~\cite{Chen:2005td,Drechsel:2002ar,Drechsel:2004ki} 
provide comprehensive reviews on this subject.

\begin{figure}[htb!]
\begin{center}
\begin{minipage}[t]{8 cm}
\vspace {-0.1cm}
\epsfig{file=GDH_He3.eps,scale=0.45}
\end{minipage}
\begin{minipage}[t]{8 cm}
\epsfig{file=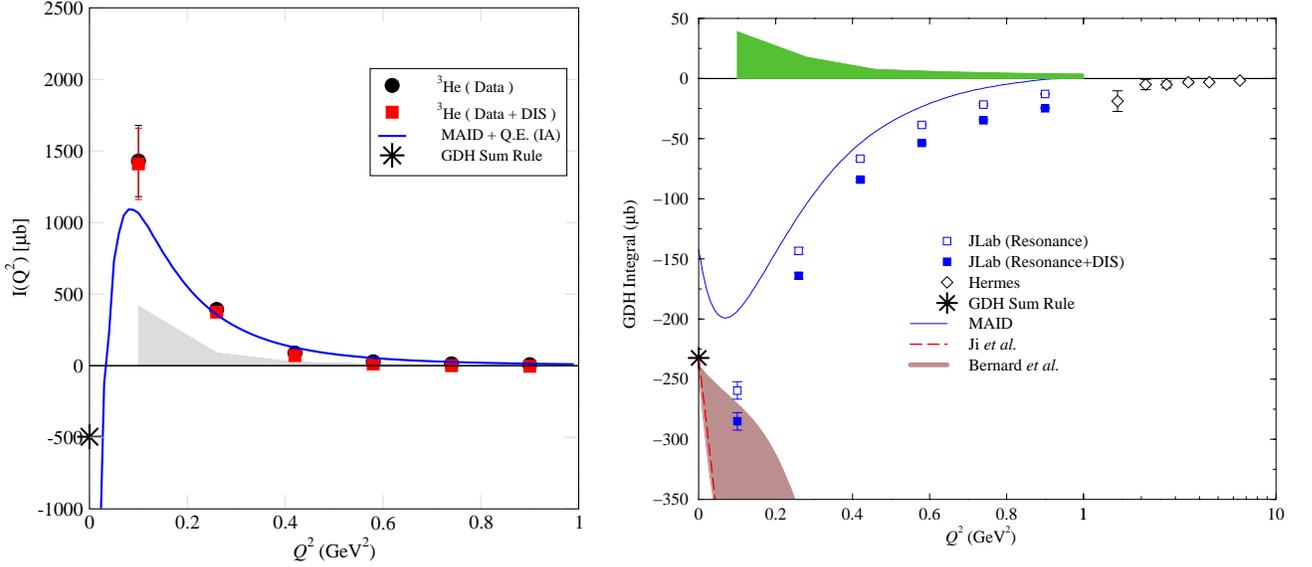,scale=0.45,angle=-90}
\end{minipage}
\begin{minipage}[t]{16.5 cm}
\caption{Results of the GDH sum $I(Q^2)$ for $^3$He~\protect\cite{Slifer:2008re} (left)
 and for the neutron~\protect\cite{Amarian:2002ar} (right). The $^3$He GDH results are compared 
 with the MAID model plus a contribution from quasi-elastic nucleon knock-out. 
 The neutron GDH results are compared with $\chi$PT calculations of 
Ji~\protect\cite{Ji:1999pd} (dashed line) and Bernard {\it et al.}~\protect\cite{Bernard:2002pw} 
(shaded area).  
The MAID model calculation~\protect\cite{Drechsel:2000ct}
is represented by a solid line.  Data from HERMES~\protect\cite{Airapetian:2002wd} are also shown.}
\label{fig:GDH}
\end{minipage}
\end{center}
\end{figure}

%
Fig.~\ref{fig:GDH} shows the extended GDH integrals  
$I(Q^2)=\int_{\nu_0}^\infty [\sigma_{1/2}(Q^2)-\sigma_{3/2}(Q^2)] d\nu/\nu$ 
for $^3$He (left) and  for the neutron (right),
which were extracted from Hall A experiment E94-010~\cite{Slifer:2008re,Amarian:2002ar}, 
from break-up threshold for $^3$He and from pion threshold for the neutron to $W=2$ GeV.
The uncertainties, when visible, represent statistics only; the systematics are shown by the shaded bands.   
The solid squares include an estimate of the unmeasured high-energy part.
The corresponding
uncertainty is included in the systematic uncertainty band.


The $^3$He results rise with decreasing $Q^2$. Since the GDH sum 
rule at $Q^2=0$ predicts a large negative value, a drastic turn around 
should happen at $Q^2$ lower than 0.1 GeV$^2$. A simple model using MAID~\cite{Drechsel:2000ct} 
plus quasi-elastic contributions estimated from a PWIA model~\cite{CiofidegliAtti:1994cm} 
indeed shows the expected turn
around. The data at low $Q^2$ should be a good testing
ground for few-body Chiral Perturbation Theory calculations.

\begin{figure}[htb!]
\centering{\scalebox{0.38}{\includegraphics {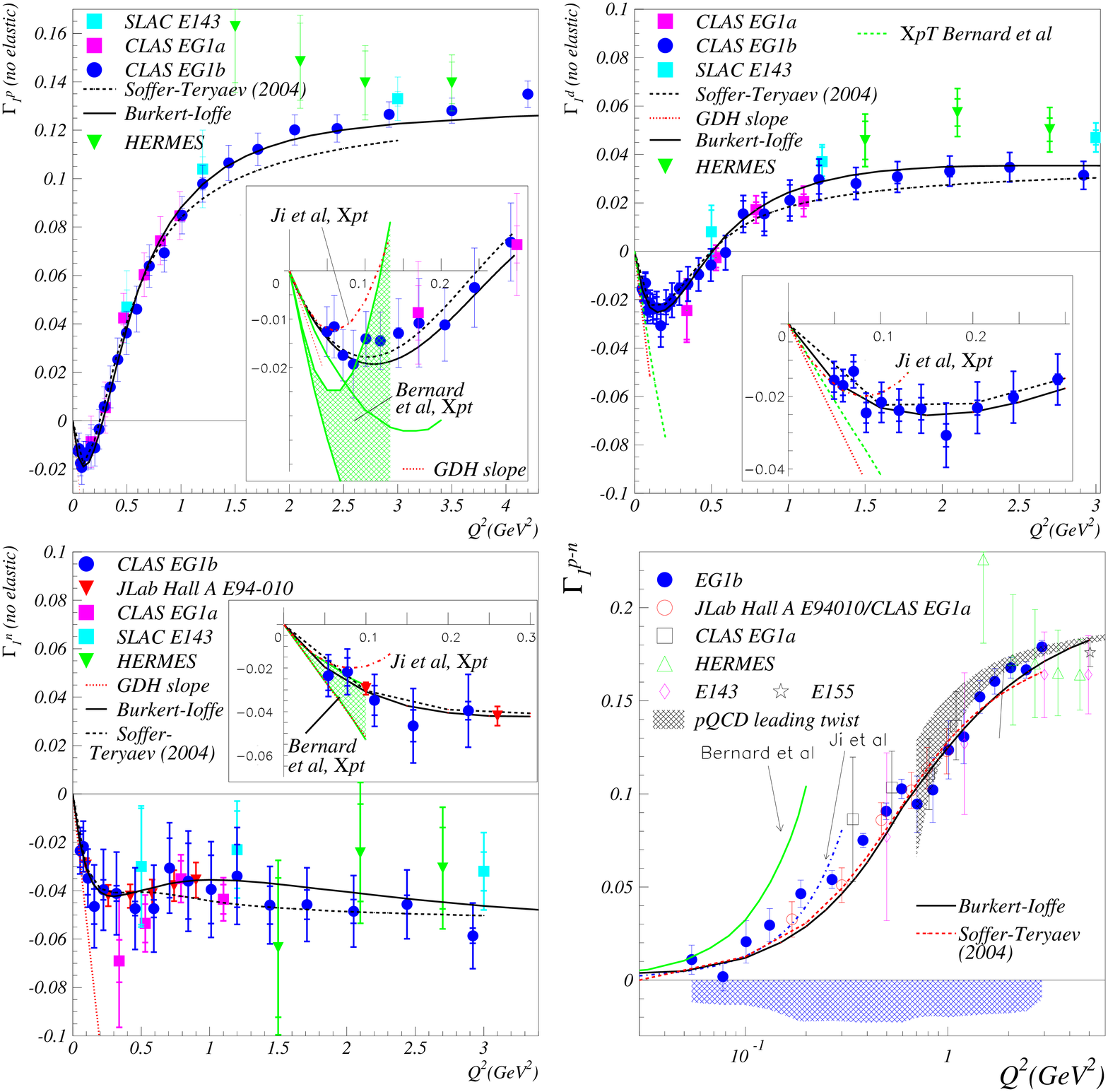}}}
\begin{center}
\begin{minipage}[t]{16.5 cm}
\vspace{-0.2in}
\caption{Results on $\overline{\Gamma}_1 (Q^2)$ for p, d, n, and p-n from 
JLab Hall A~\protect\cite{Amarian:2003jy} and 
CLAS EG1a~\protect\cite{Fatemi:2003yh} and EG1b~\protect\cite{Prok:2008ev},
together with data from SLAC and HERMES.
The slopes at $Q^2=0$ predicted by the GDH sum rule 
are given by the red dotted lines. The
 solid (short-dashed) lines are the predictions from the Burkert-Ioffe (Soffer-Teryaev) 
 parametrizations~\protect\cite{Burkert:1992tg,Soffer:2004ip}.
The leading twist $Q^2$-evolution of the p-n moment is given by the grey band. 
 The insets show comparisons with $\chi$PT calculations. The dot-dashed
lines (green lines and bands) at low $Q^2$ are the next-to-leading order $\chi$PT 
predictions by Ji \emph{et al.}~\protect\cite{Ji:1999pd} 
(Bernard \emph{et al.}~\protect\cite{Bernard:2002pw}). } 
\label{fig:gamma1pn}
\vspace {-4mm}
\end{minipage}
\end{center}
\end{figure}

The neutron results indicate a smooth variation of $I(Q^2)$  to increasingly 
negative values as $Q^2$ varies from $0.9\,{\rm GeV^2}$ 
towards zero.
The data are more negative than the MAID model calculation\cite{Drechsel:2000ct},
again presumably due to the contribution from multi-pion final states. 
(Since the calculation only includes contributions to $I(Q^2)$ for $W \le 2\,{\rm GeV}$, 
it should be compared with the
open circles.) The GDH sum rule 
prediction, $I(0)=-232.8\,\mu{\rm b}$, is indicated in Fig.~\ref{fig:GDH}, along with  
extensions to $Q^2>0$ using two next-to-leading order $\chi$PT
calculations, one using the Heavy Baryon approximation (HB$\chi$PT)~\cite{Ji:1999pd} 
(dashed line) and the other
based on Relativistic Baryon $\chi$PT (RB$\chi$PT)~\cite{Bernard:2002pw}. 
The shaded band shows the RB$\chi$PT calculation
including resonance effects~\cite{Bernard:2002pw}, 
which have an associated
large uncertainty due to the resonance parameters used. 
%
%

In Fig.~\ref{fig:gamma1pn} we present the existing data set on the
related quantity $\bar \Gamma_1(Q^2)$ (Eq.~\ref{eq:gamma1bar}) at low to moderate $Q^2$.
The upper two panels show the data on the proton and the deuteron
 from EG1a~\cite{Fatemi:2003yh} together with
new results from the Hall B EG1b experiment~\cite{Prok:2008ev}.
The bottom row shows
published results for the neutron from Hall A~\cite{Amarian:2003jy} and 
for the proton-neutron difference obtained by combining all of these
data sets.
Also shown are data from 
SLAC E143~\cite{Abe:1998wq} and E155~\cite{Anthony:2000fn} 
as well as from HERMES~\cite{Airapetian:2007mh}.
The inner error bars
indicate the statistical uncertainties while the outer ones are the quadratic sums of the 
statistical and systematic uncertainties, except for the lower-right panel where
the systematic errors are indicated by a shaded band.

As $Q^2 \rightarrow 0$, the slopes of  $\bar{\Gamma}_1$ 
(shown as dotted lines in Fig.~\ref{fig:gamma1pn}) and 
of $\bar{\Gamma}_{TT}$ are the same, both given by the GDH sum rule 
(see Eqs.~\ref{eq:GDHDIS}-\ref{eq:gammabar},\ref{eq:gamma1bar},\ref{eq:GDHDIS1}).
The behavior at low $Q^2$ can be calculated with $\chi$PT. 
We show results of calculations by Ji {\it et al.}~\cite{Ji:1999pd} 
using HB$\chi$PT and by Bernard {\it et al.}~\cite{Bernard:2002pw} without and with the 
inclusion of vector 
mesons and $\Delta$ degrees of freedom. 
The calculations are in reasonable agreement with the data at the lowest $Q^2$
values of 0.05 - 0.1 GeV$^2$.
At moderate and large $Q^2$ data are compared with two phenomenological model 
calculations~\cite{Soffer:2004ip,Burkert:1992tg}. 
The model by Soffer and Teryaev~\cite{Soffer:2004ip} assumes a smooth
variation with $Q^2$  of the integral $\int{[g_1(x) + g_2(x)]dx}$ 
(which is constrained by DIS data at large $Q^2$ and by sum rules at the photon point)
and subtract the contribution from $\bar{\Gamma}_2$ (using the BC sum rule) to extract
$\bar{\Gamma}_1(Q^2)$. The curve by Burkert and Ioffe~\cite{Burkert:1992tg}
uses a parametrization of the resonance region (similar to MAID) and adds a
Vector Meson Dominance-inspired ansatz to connect the GDH sum rule at the photon point
with the DIS data for $\bar{\Gamma}_1$.
Both models describe the data well.  

The lower-right panel in Figure \ref{fig:gamma1pn} shows the moment of $g_1^p-g_1^n$, 
the Bjorken integral. This is of special interest because it contains 
contributions only from the 
flavor non-singlet (or isovector) part,
and the contribution from the $\Delta (1232)$ resonance cancels~\cite{Burkert:2000qm}. 
The data at high $Q^2$ were used
to test the Bjorken sum rule as described in Section~\ref{subsec:bjork}.
They were also used to extract a value of the strong coupling constant, $\alpha_s$,
assuming the validity of the sum rule.
The leading-twist pQCD evolution of the Bjorken sum rule
 is shown by the grey band. It tracks the data down to
surprisingly low $Q^2$, which indicates an overall suppression of higher-twist
effects (see Section~\ref{subsec:OPE}). 
The new JLab data at low $Q^2$ provide interesting information in the low energy region, 
where the strong interaction is truly strong and non-perturbative.

\begin{figure}[htb!]
\centering {\scalebox{0.40}{\includegraphics {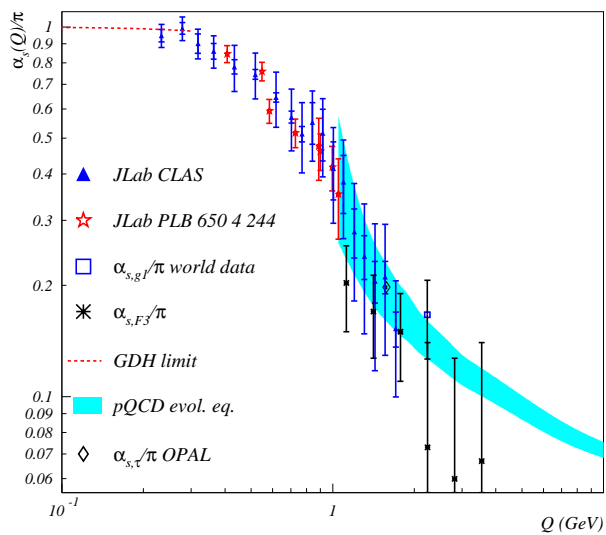}}}
\begin{center}
\begin{minipage}[t]{16.5 cm}
\vspace{-0.3in}
\caption{Results for $\alpha_{s,g_1}$~\protect\cite{Deur:2008rf}
 extracted from the generalized Bjorken sum, together with other world data on 
 effective strong couplings.}
\label{fig:alphas}
\end{minipage}\end{center}
\end{figure}

A new attempt~\cite{Deur:2008rf,Deur:2005cf} was made to extract an effective strong coupling, 
$\alpha_{s,g_1}$ in the low $Q^2$ region (Figure \ref{fig:alphas}). The extracted $\alpha_{s,g_1}$,
which by definition must converge towards a finite value as  $Q^2 \rightarrow 0$,
shows a clear trend of weakening $Q^2$-dependence with decreasing $Q^2$. 
With the GDH sum 
rule as a constraint at $Q^2=0$, a model fit to the extracted $\alpha_{s,g_1}$ 
shows a small $Q^2$-slope at the origin. 
This is consistent 
with a conformal behavior, which may be important for any attempt to apply
AdS/CFT~\cite{Brodsky:2006uqa} for the strong interaction in the low-energy region.


The generalized spin polarizabilities provide additional benchmark tests
of $\chi$PT calculations, and their measurement is
an important step in understanding the dynamics of
QCD in the low $Q^2$ region. 
Since they have an extra $1/\nu^2$ 
weighting compared to the first moments, these 
integrals have less contributions from the large-$\nu$ 
region and converge much faster, which minimizes the uncertainty due to
the unmeasured region at large $\nu$. 

Generalized spin
polarizabilities have been evaluated with next-to-leading order 
$\chi$PT 
calculations~\cite{Kao:2002cp,Bernard:2002pw}.
One issue in the $\chi$PT calculations is how to properly
include the nucleon resonance contributions, especially the $\Delta(1232)$ resonance.
As was pointed out in~\cite{Kao:2002cp,Bernard:2002pw} , while $\gamma_0$ is sensitive to 
resonances, $\delta_{LT}$ is insensitive to the $\Delta$ 
resonance.

\begin{figure}[!htb]
\centering{\scalebox{0.4}{\includegraphics{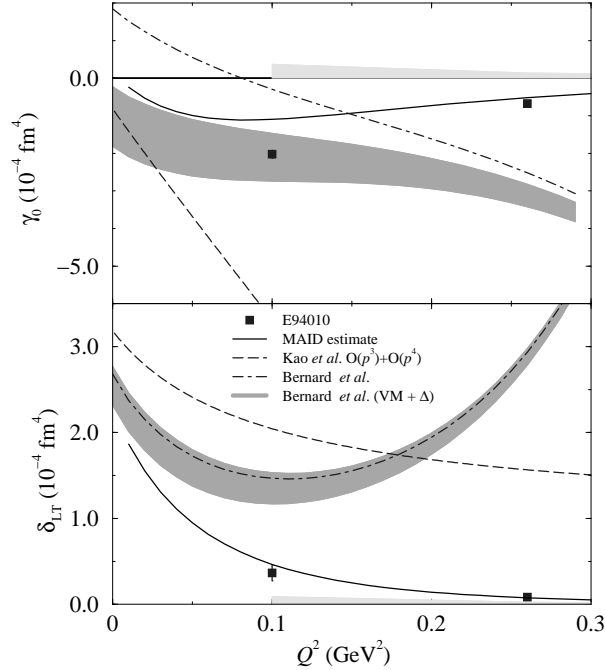}}}
\begin{center}
\begin{minipage}[t]{16.5 cm}
\vspace{-0.3in}
\caption{Results for the neutron spin polarizabilities 
$\gamma_0$ (top panel) and $\delta_{LT}$
(bottom panel). Solid squares represent the results with statistical 
uncertainties. The light bands represent the systematic uncertainties. 
The dashed curves represent the heavy baryon $\chi $PT 
calculation~\protect\cite{Kao:2002cp}. 
The dot-dashed curves and the dark bands represent the relativistic baryon
$\chi $PT calculation 
without and with~\protect\cite{Bernard:2002pw} the $\Delta $ and vector meson 
contributions, respectively.
Solid curves represent the MAID model~\protect\cite{Drechsel:2000ct}.
}
\label{fig:fig3}
\end{minipage}\end{center}
\end{figure}

The first results for the neutron generalized forward
spin polarizabilities $\gamma_0(Q^2)$ and $\delta_{LT}(Q^2)$
were obtained at Jefferson Lab Hall A~\cite{Amarian:2004yf}.
The results for $\gamma_0^n(Q^2)$ are shown
in the top panel of Fig.~\ref{fig:fig3}. 
The statistical uncertainties are smaller than the
size of the symbols. 
The data are compared with 
a next-to-leading order ($O(p^4)$) HB$\chi$PT 
calculation~\cite{Kao:2002cp}, a next-to-leading order RB$\chi$PT
calculation and the same calculation explicitly including 
both the $\Delta$ resonance and vector meson contributions~\cite{Bernard:2002pw}.
Predictions from the MAID model~\cite{Drechsel:2000ct} are also shown.
At the lowest $Q^2$ point,
the RB$\chi$PT
calculation including the resonance contributions
is in good agreement with the experimental result.
For the HB$\chi$PT calculation without explicit resonance contributions, 
discrepancies are large even at $Q^2 = 0.1$ GeV$^2$. 
This might indicate the significance of the resonance contributions or a
problem with the heavy baryon approximation at this $Q^2$.
The higher $Q^2$ data point is in good agreement with the MAID
prediction,
but the lowest data point at $Q^2 = 0.1 $ GeV$^2$ is significantly lower.

Results for $\gamma_0$ on the proton has been recently submitted for publication~\cite{Prok:2008ev}.  
They show significant disagreement with both $\chi$PT 
calculations~\cite{Kao:2002cp,Bernard:2002pw}. 
 An isospin separation of  $\gamma_0$ was performed and discussed in Ref.~\cite{Deur:2008ej}.
 The isoscalar ($\gamma_0^{p+n}$) and isovector ($\gamma_0^{p-n}$) combinations  
  also disagree with $\chi$PT, in spite of the fact that for the latter the
 contribution from the  $\Delta(1232)$ resonance cancels.

Since $\delta_{LT}$ is 
insensitive to this resonance contribution for each nucleon individually,
it was believed that $\delta_{LT}$ should be
more suitable than $\gamma_0$ to serve as a testing ground for the chiral 
dynamics of QCD~\cite{Kao:2002cp,Bernard:2002pw}.
The bottom panel of Fig.~\ref{fig:fig3} shows $\delta_{LT}$ 
compared to
$\chi$PT calculations and the MAID predictions. While the MAID predictions are in good 
agreement with the results, it is surprising to see
that 
the data are in significant disagreement with the $\chi$PT calculations 
even at the lowest $Q^2$, 0.1 GeV$^2$. 
This disagreement (``$\delta_{LT}$ puzzle'') presents a significant challenge to the present 
Chiral Perturbation Theory.

New experimental data have been taken at very low $Q^2$, down to 0.02 GeV$^2$
for the neutron ($^3$He)~\cite{E97110} for both longitudinal and transverse target polarizations, 
but only for longitudinal target polarization for the proton and the
deuteron~\cite{E03006}.
Preliminary results just became available for the neutron. 
Analysis is underway for the proton and deuteron data.
These results will provide benchmark tests of
$\chi$PT calculations at the kinematics where they are expected to work.
A new proposal~\cite{E08027} was recently approved to measure $g_2^p$ with a transversely 
polarized proton target in the low $Q^2$ region. It will provide an isospin separation of the spin 
polarizabilities to shed light on the ``$\delta_{LT}$'' puzzle.

 \section{Quark-Hadron Duality in Spin Structure \label{sec:duality}}
In the previous sections, we have interpreted spin structure function data and their integrals
either in terms of quark and gluon degrees of freedom (Sections~\ref{subsec:parton}--\ref{subsec:g2}, 
\ref{subsec:mom}--\ref{subsec:OPE}) or in terms of low-energy effective theories 
(Sections~\ref{subsec:res}, \ref{subsec:chiral}).
While quarks and gluons are the fundamental degrees of freedom for the theory of strong
interactions, QCD, a description in terms of their hadronic composites (nucleon, nucleon resonances
and light mesons) is often more economical at lower resolution $Q^2$, where perturbative
 QCD is not applicable and calculations are extremely difficult.
At intermediate resolution, models like the constituent quark model may be able
to describe some of the observed phenomena.

Both from a theoretical and a practical point of view, it is interesting to investigate where and
how these different pictures of the nucleon connect and overlap. In particular, if we can find
a kinematic region where both the quark-parton description and the hadronic description
are found to be valid to some degree, we can potentially expand the kinematic range of
experiments that attempt to extract information on quarks in the nucleon. At the same time,
we could gain new insights on the transition from quasi-free quarks at short length scales to 
their confinement at larger distances.

Such ``dual'' descriptions of experimental data have been successful in cases including
$e^+e^-$ annihilation, semi-leptonic heavy meson decays and lepton scattering; for a review 
see~\cite{Melnitchouk:2005zr}. 
In particular, in the 1970's Bloom and Gilman~\cite{Bloom:1970xb} found that
the unpolarized structure function $F_2(x,Q^2)$ in the nucleon resonance region ($W<2$ GeV)
at moderate $Q^2$, when averaged over suitable intervals in $x$, was quite close to the
deep inelastic, scaling structure function 
$F_2(x)$ measured at much higher $Q^2$ and $W$ but the same average $x$. 
This finding has been confirmed and
expanded in much greater detail by measurements at Jefferson Lab~\cite{Niculescu:2000tk}.
The agreement between resonance region and high-$Q^2$ data is typically improved if
one accounts for target mass effects and the logarithmic $Q^2$-evolution of deep inelastic 
structure functions. 
Nachtmann~\cite{Nachtmann:1973mr} has shown how to handle 
the finite mass of the nucleon target exactly;
an approximate method simply employs the modified 
Nachtmann scaling variable 
$\xi = 2 x / (1 + \sqrt{1 + 4M^2 x^2 / Q^2})$.
This has been extended to the polarized case by
Piccione and Ridolfi~\cite{Piccione:1997zh}.

In the framework of the OPE, a 
quark-parton description of structure function {\em moments} at intermediate $Q^2$
should in principle be always possible if one includes contributions of sufficiently high
twists. This so-called
 {\em global} duality then becomes simply the observation that these higher-twist contributions
are either small or cancel in integrated observables~\cite{DeRujula:1976ke}. Experimentally,
this seems to be the case for unpolarized structure functions of the proton; however, it
is non-trivial whether the same is true for polarized structure functions and for neutron
structure functions. In particular, the polarized structure function $g_1$ might not necessarily
exhibit {\em local} duality (for a restricted range in $x$), 
since at low $Q^2$ and $W$ it is dominated by the $\Delta(1232)$ resonance, which
should have a negative spin asymmetry $A_1 \approx - 0.5$ (see Section~\ref{subsec:res}), 
while at large $Q^2$ and $x$ one expects this
asymmetry to approach unity (see Section~\ref{subsec:val}).
Therefore, a detailed study of the limits of validity of duality for polarized structure functions
is of great interest.

First measurements of spin structure functions in the resonance 
region~\cite{Baum:1980mh,Abe:1996ag} found that, with the exception of the $\Delta(1232)$
resonance, duality seemed to hold approximately within errors, especially
for the higher $Q^2$ data of E143~\cite{Abe:1996ag}. 
Early data from the Jefferson Lab EG1 program with CLAS~\cite{Yun:2002td,Fatemi:2003yh}
confirmed that local duality does not work well at low $Q^2$, with
significantly higher statistics. The paper by
Yun et al.~\cite{Yun:2002td} explicitly looked at the approach of the structure function
$g_1^d(\xi,Q^2)$ towards the scaling limit for increasing $Q^2$ and found that this limit
had not been reached yet up to $Q^2 = 1$ GeV$^2$. The HERMES 
collaboration~\cite{Airapetian:2002rw} averaged their 
measured asymmetry $A_1^p$ over the
resonance region 1 GeV $<W<$ 2 GeV for 5 different bins in $x$
and $Q^2$ and found the results
in reasonable agreement with DIS results, within their relatively large errors.

\begin{figure}[htb!]
\begin{center}
\begin{minipage}[t]{10 cm}
\epsfig{file=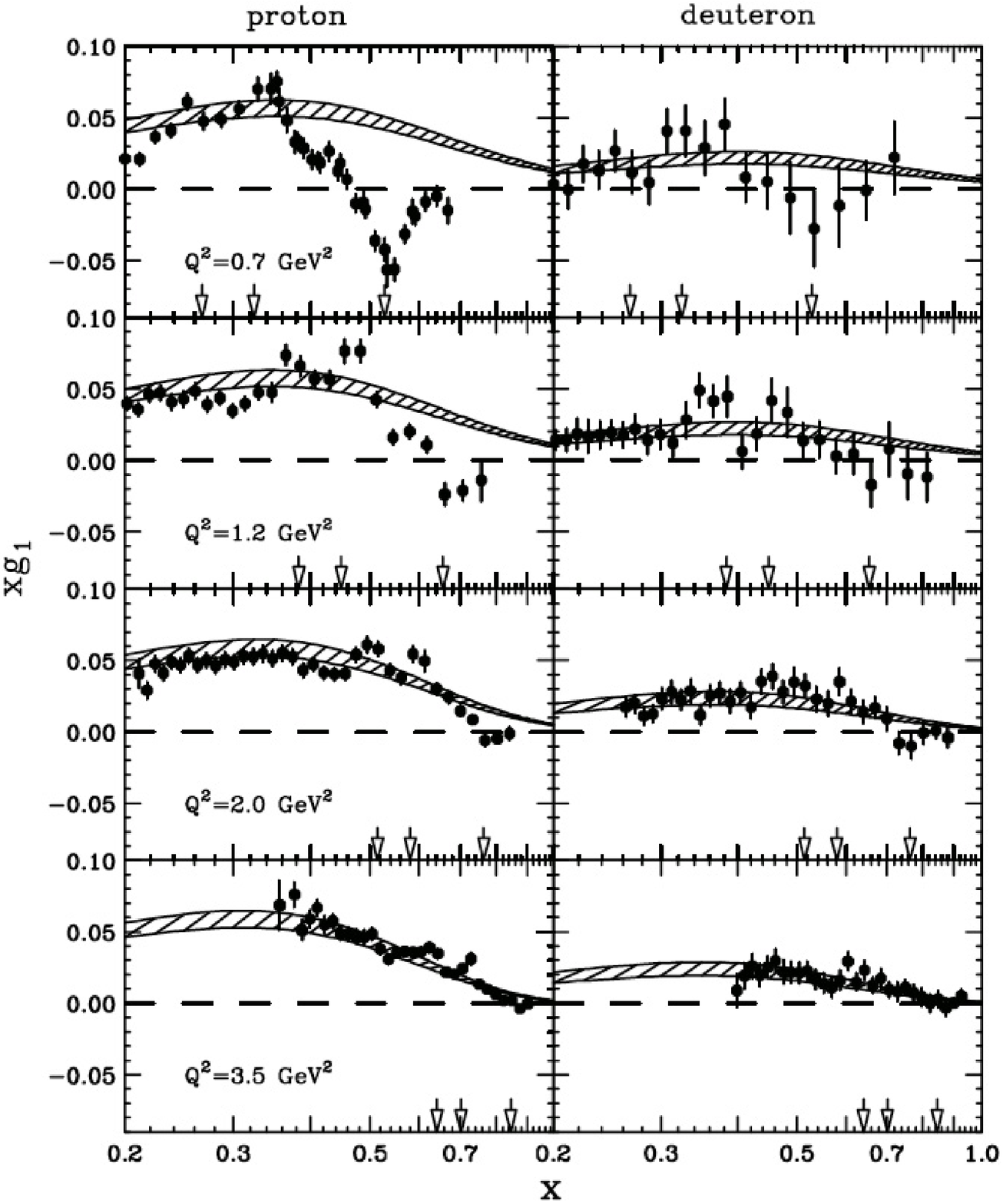, scale=0.35}
\end{minipage}
\begin{minipage}[t]{6 cm}
\epsfig{file=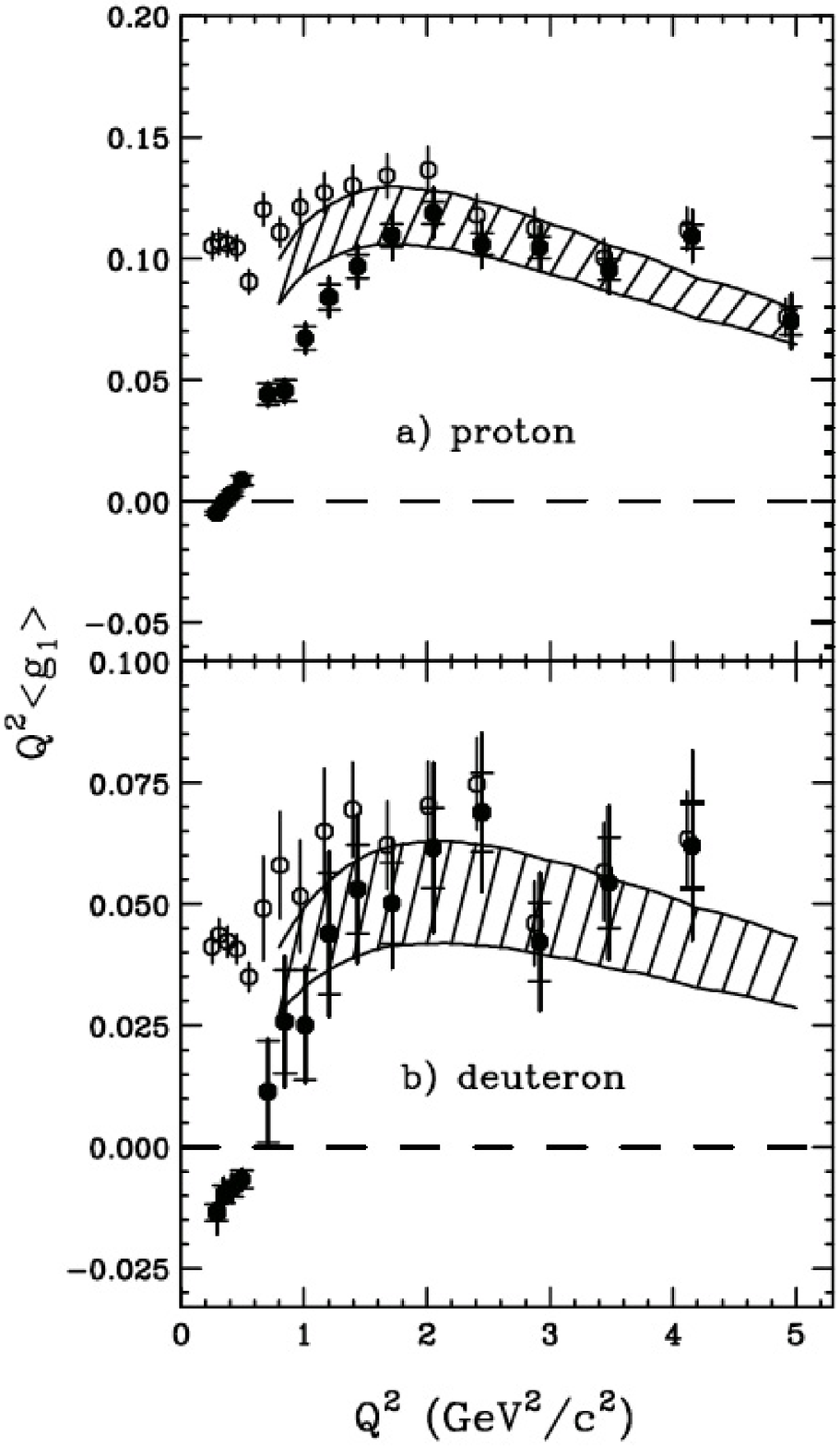, scale=0.375}
\end{minipage}
\end{center}
\begin{center}
\begin{minipage}[t]{16.5 cm}
\vspace{-0.2in}
\caption{Data on the spin structure functions $g_1(x,Q^2)$ of the proton and the deuteron
from Jefferson Lab's Hall B (left panel) and its average over the resonance region (right panel).
Prominent resonances are indicated by arrows in the left panel. The hatched curves
represent the range of extrapolated DIS results from modern NLO fits (GRSV and AAC, see 
Section~\ref{subsec:delq}), evolved to the $Q^2$ of the data and corrected for
target mass effects. The open circles in the right panel include the elastic contribution,
while the filled circles are only integrated over $W > 1.08$ GeV.
\label{fig:dualeg1}}
\end{minipage}
\end{center}
\end{figure}

By far the most detailed study of duality in the spin structure function $g_1$ has been
published by the EG1 collaboration~\cite{Bosted:2006gp}. As shown in Fig.~\ref{fig:dualeg1},
one observes a clear trend of strong, resonant deviations from the scaling curve at
lower $Q^2$, towards a pretty good agreement at intermediate $Q^2$. The integral
of $g_1$ over the whole resonance region begins to agree with the NLO results above
$Q^2 \approx 1.7$ GeV$^2$. 
At lower $Q^2$, that integral rises above the NLO extrapolation if one includes the
elastic contribution. This shows that the elastic peak somewhat overcompensates
the negative contribution from the $\Delta(1232)$ resonance; however, leaving out the
elastic part of the integral leads to integral values much
below the NLO curves.
The results on the proton and deuteron from EG1~\cite{Bosted:2006gp} thus
indicate a much slower approach to ``global'' duality for the polarized structure
function $g_1$ than has been observed for unpolarized structure functions. 

A more detailed examination shows that $g_1$ averaged over the first resonance region
($1.08 < W < 1.38$ GeV), which includes the $\Delta(1232)$ resonance, is substantially below the NLO
expectations even to the highest $Q^2$ measured by EG1 (5 GeV$^2$). On the other hand,
the three resonances (P$_{11}$(1440), S$_{11}$(1535), D$_{13}$(1520)) in the second
resonance region ($1.38 < W < 1.58$ GeV) all have dominant $A_{1/2}$ transition
amplitudes already at rather modest $Q^2$, so that the averaged $g_1$ in this region tends
to overshoot the NLO limit. The cancellation of these two opposite effects, together with
a rather quick approach to ``local duality'' for the higher resonance region ($W > 1.56$ GeV)
leads to the approach to global duality seen in Fig.~\ref{fig:dualeg1}. 

\begin{figure}[htb!]
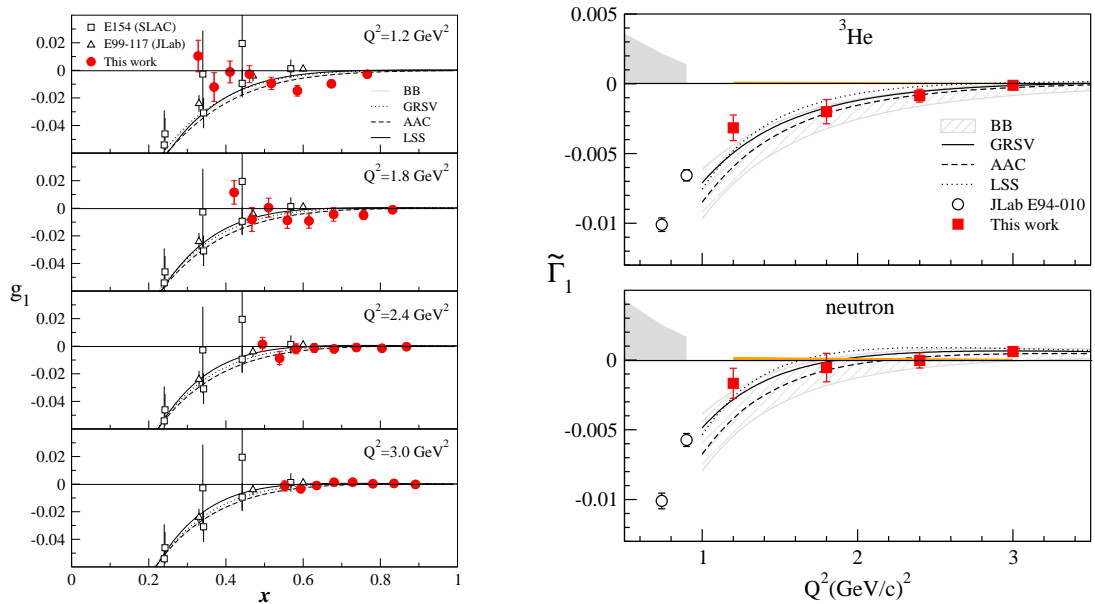

\begin{center}
\begin{minipage}[t]{7 cm}
\epsfig{file=g1_dual.eps, scale=0.4}
\end{minipage}
\begin{minipage}[t]{9 cm}
\epsfig{file=gamma1_dual.eps, scale=0.5}
\end{minipage}
\end{center}
\begin{center}
\begin{minipage}[t]{16.5 cm}
\vspace{-0.2in}
\caption{New data on the spin structure functions $g_1(x,Q^2)$ of $^3$He 
in the resonance region
from Jefferson Lab's Hall A (solid circles, left panel), together with NLO curves for a
combination of proton and neutron $g_1$ with proper polarization factors to mimic
the nuclear spin structure function and evolved to the $Q^2$ of the data.
The right panel show these data (solid squares)
averaged over the resonance region for $^3$He (top half)
 and corrected for nuclear effects
to get the partial integrals for the neutron (bottom half).
The NLO curves shown
represent the range of extrapolated DIS results from modern NLO fits 
(Bl\"umlein and B\"ottcher, GRSV, LSS and AAC, see 
Section~\ref{subsec:delq}), evolved to the $Q^2$ of the data and corrected for
target mass effects.
\label{fig:dualhalla}}
\end{minipage}
\end{center}
\end{figure}

The data taken
by the RSS collaboration in Hall C (see Section~\ref{subsec:res}) corroborate these
observations and add more precise data points for $Q^2 \approx 1.3$ GeV$^2$.
They find that the integral over their measured region, 1.09 GeV $< W < 1.91$ GeV,
is about $17\% \pm 8\%$ lower than the same integral over NLO evaluations of
$g_1^p$ evolved to $Q^2 =1.3$ GeV$^2$, even after applying target mass corrections.

Very recently, new results have become available from a duality study in Hall A,
using a $^3$He target~\cite{Solvignon:2008hk}. By combining NLO fits for
the spin structure functions $g_1^p$ and $g_1^n$ with the proper polarization
factors, one can directly compare these fits with the data on the nucleus $^3$He
(left panel of Fig.~\ref{fig:dualhalla}). Once again, at the lowest $Q^2$ point
the data oscillate around the NLO curves and show a strong deviation in the
region of the $\Delta(1232)$ resonance at least out to $Q^2=1.8$ GeV$^2$. At higher
$Q^2$, both the data and the NLO fits are very close to zero, in contrast to
$g_1^p$ which is markedly positive for the extrapolated NLO fits.
The partial integrals (up to $W = 1.91$ GeV) of these data are shown in the
right panel of Fig.~\ref{fig:dualhalla}). Except for the lowest $Q^2$ point, 
good agreement with the integrals of NLO fits (both for $^3$He and directly for the neutron)
is seen, confirming the onset of duality around $Q^2 = 1.7$ GeV$^2$ for both protons and neutrons.

\section{Summary and Outlook \label{sec:sum}}

After 30 years of dedicated experiments, we are beginning to accumulate a rather detailed
picture of a fundamental aspect of the nucleon, namely its spin structure. 
In spite of early worries (after the initial 
``spin crisis''), it appears that quark helicities {\em do} contribute a small but significant fraction
of the (longitudinal) spin of the nucleon, up to one third. Most importantly, the fundamental
Bjorken sum rule~\cite{Bjorken:1968dy} and its perturbative QCD evolution
seem to be at least consistent with the data. In fact, none of the existing
experiments show any features that would contradict pQCD in its realm of
applicability. 

Most 
pQCD-based analyses of the data agree fairly well on the contribution of various
quark flavors to the proton spin, although there is still some controversy on the
role played by strange sea quarks. We also don't know yet whether 
polarized up and down
anti-quark densities show the same difference as the unpolarized ones.
On the other hand, new high-precision data from 
Jefferson Lab 
constrain very nicely the polarized densities of valence up and down quarks at higher $x$. 
These data indicate that relativistic
quark models, with one-gluon exchange corrections~\cite{Isgur:1998yb}, 
and augmented by pion cloud contributions~\cite{Myhrer:2007cf},
yield a rather good description
in the valence region where sea-quarks contribute little. The emergent picture
(also from recent NLO analyses, see Section~\ref{subsec:delq})
is one where valence quarks 
($\Delta q_V\equiv \Delta q - \Delta \bar{q} $)
carry roughly the expected fraction $(\simeq 60\%)$ of the
nucleon spin, while the (on average) negative helicity of sea quarks reduces
this to about 30-35\%. Future 
experiments~\cite{E12-06-109,E12-06-110,E12-06-122} 
at Jefferson Lab, after the approved
energy upgrade to 11-12 GeV, will extend our knowledge of polarized quark densities
out to $x>0.8$ and will decisively test predictions from pQCD and QCD-inspired models.
These same experiments, as well as presently scheduled 6 GeV 
ones~\cite{E-05-113,E07-011},
will further improve the precision with which we can study scaling violations and
therefore extract polarized gluon and sea-quark densities.
Additional information on individual quark and antiquark flavor contributions
to the nucleon spin will come from future experiments at RHIC (in particular
from direct $W^{\pm}$ production at 500 GeV center-of-mass energy) and
the new FAIR facility in Darmstadt, Germany~\cite{Seitz:2007zz} as well
as semi-inclusive measurements with COMPASS and at Jefferson Lab.
Much additional information on the small-$x$ behavior of quark- and antiquark
polarization densities could be gotten from the electron-ion collider 
(EIC) that has recently been proposed~\cite{Aidala:2008ic}, 
see below.

In addition to logarithmic violations of scaling, expected from pQCD, we also see
some evidence for non-zero twist-3 and twist-4 matrix elements; however,
the overall higher twist corrections to moments of spin structure
functions seem rather modest even at fairly low $Q^2$, indicating a partial cancellation
of contributions from different orders in the twist expansion. 
At these lower values of $Q^2$, we see a transition from partonic (quark and gluon)
degrees of freedom to hadronic ones, without any dramatic break in the rather smooth
transition region. Thanks to the vast data set in the nucleon resonance region,
we can better constrain transition amplitudes and test this quark-hadron duality
in detail. Averages over the resonance region seem to agree
fairly well with extrapolations of deep inelastic data over the same range in $x$,
albeit not to quite as low $Q^2$ as in the unpolarized case. 
Future measurements at Jefferson Lab with 
6 GeV~\cite{SANE,d2n6GeV,E08-015} 
and 12 GeV~\cite{d2n12GeV} beams will
add to our knowledge of higher-twist matrix elements, and of 
the second spin structure function
$g_2$ and its moments.

At the lower end of the $Q^2$ scale,
 the Gerasimov-Drell-Hearn sum rule at the real photon point seems to
be well confirmed. However, it is too early to state with
certainty whether the effective theory of QCD at low energies, $\chi$PT, yields a good
description of the data very close to that point. Some tentative successes in describing the first moment 
$\Gamma_1$ of $g_1$ will be tested very soon, when the new data on 
the neutron ($^3$He), the proton and the deuteron collected in Halls A and B
at Jefferson Lab become available~\cite{E97110,E03006}. 
For the polarizabilities like $\gamma_0$ and $\delta_{LT}$, 
which  are given by higher moments of $g_{1,2}$, 
the agreement at present is unsatisfactory; it may be necessary to
go to a higher order in $\chi$PT before agreement is reached. New data on these
quantities will also be collected at Jefferson Lab~\cite{E08027}.

\begin{figure}[htb!]
 \hfill
\begin{minipage}[t]{.45\textwidth}
\begin{center}
\epsfig{file=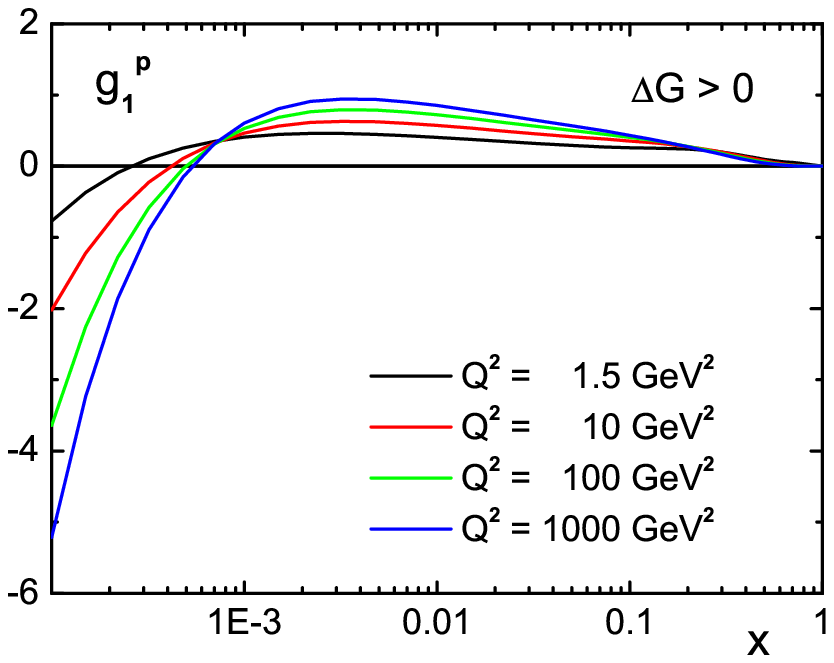, width=\textwidth}
\end{center}
\end{minipage}
\hfill
\begin{minipage}[t]{.45\textwidth}
\begin{center}
\epsfig{file=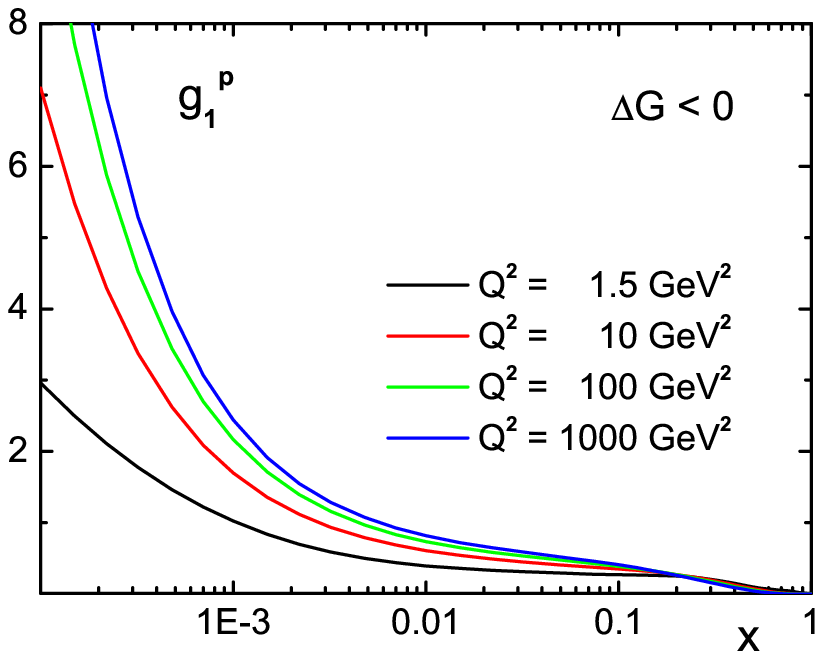, width=\textwidth}
\end{center}
\end{minipage}
\hfill
\vspace{-0.6cm}
\caption{$g_1(x,Q^2)$ at various values of $Q^2$ attainable at EIC  for positive and negative 
$\Delta G$ \protect{\cite{LSSunpub}}.} \label{EIC}
\end{figure}

One of the most important remaining open questions is the provenance of the remaining
2/3 of the nucleon spin that is not carried by quark helicities. We know now that
the gluon contribution cannot be very large, but its precise magnitude
and shape are not known yet, and it could still be an important fraction of the total.
Improved statistics from the direct measurements
at RHIC and COMPASS will help, especially with expanded kinematics
(higher and lower beam energies, different rapidity and $p_T$ ranges,
and new final state channels) which can begin to separate the
low-$x$ and high-$x$ behavior of $\Delta G$.

Additional information will come from future experiments of the spin
structure functions at both extremes of the $Q^2$ scale,
including the upcoming Jefferson Lab experiments
 with the future 11 GeV beam~\cite{E12-06-109},
and further COMPASS data. 
In this context, it is an interesting
question whether future DIS measurements alone can distinguish between
positive and negative $\Delta G(x)$. They can, indeed, but this
 requires a very high energy lepton-nucleon collider like the  EIC proposal~\cite{Aidala:2008ic}.
This can be seen in  Fig.~\ref{EIC} which shows  $g_1^p(x) $ at very small
$x$ for $1\, <\, Q^2 \,< 1,000 \,GeV^2 $ for the two signs of $\Delta G(x) $. Clearly the behavior of
$g_1^p(x) $ at small $x$ and large
$Q^2$ is quite different in the two cases.

\begin{figure}[htb!*]
 \hfill
\begin{minipage}[t]{.45\textwidth}
\begin{center}
\epsfig{file=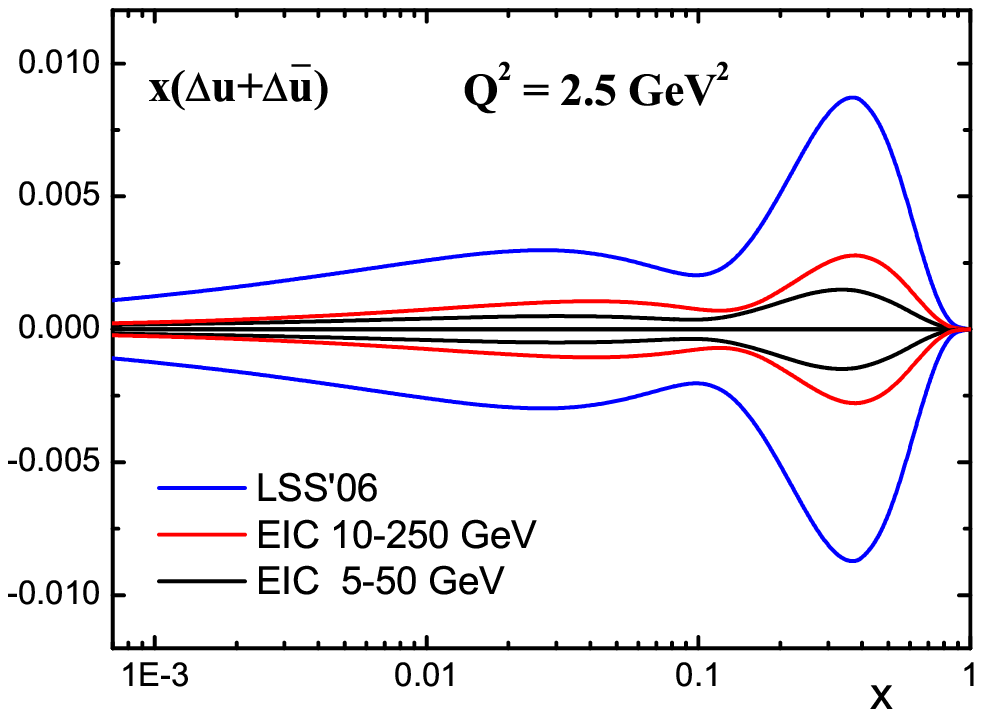, width=\textwidth}
\end{center}
\end{minipage}
\hfill
\begin{minipage}[t]{.45\textwidth}
\begin{center}
\epsfig{file=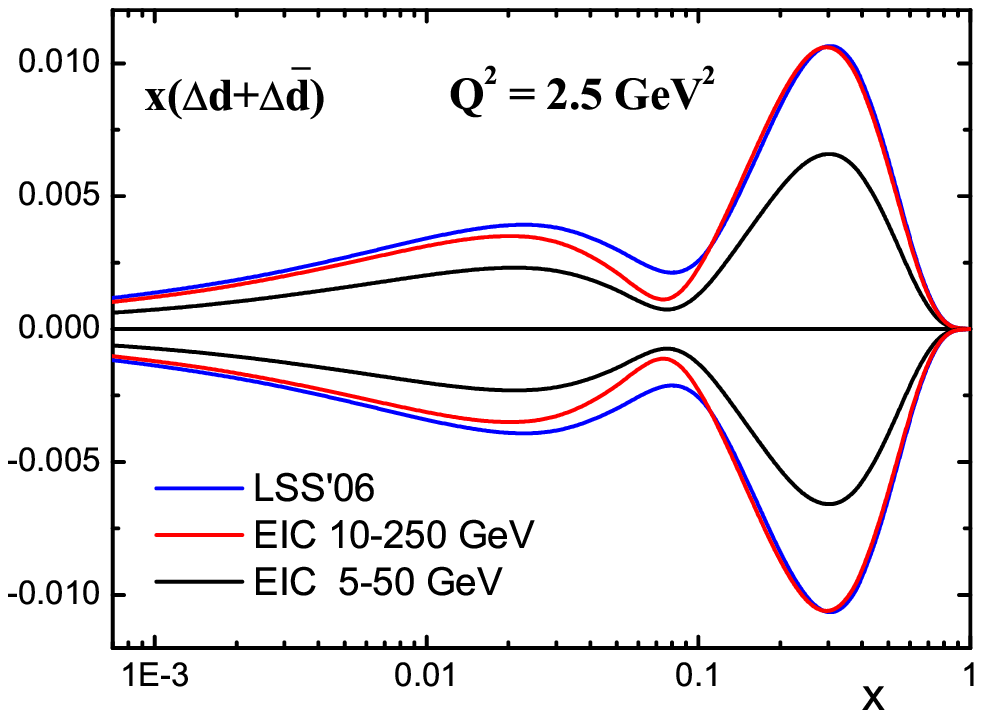, width=\textwidth}
\end{center}
\end{minipage}
\hfill
\vspace{-0.5cm}
\caption{Impact of two versions of EIC on the $u$ and $d$ uncertainties
 \protect{\cite{LSSunpub}}.}\label{err_EICud}
\end{figure}

\begin{figure}[htb!*]
 \hfill
\begin{minipage}[t]{.45\textwidth}
\begin{center}
\epsfig{file=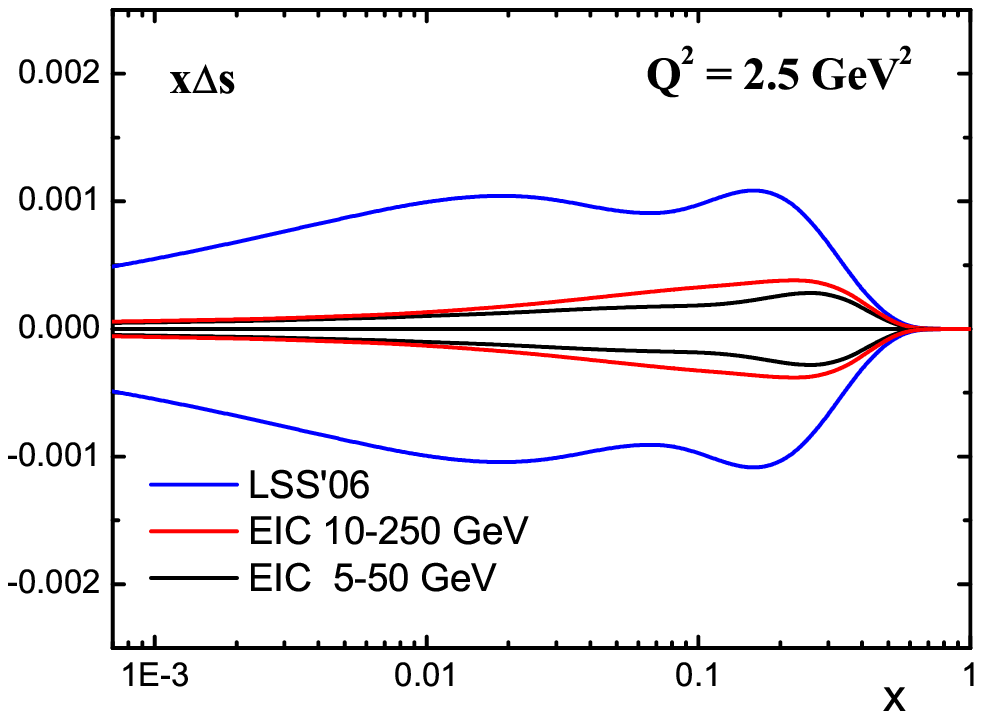, width=\textwidth}
\end{center}
\end{minipage}
\hfill
\begin{minipage}[t]{.45\textwidth}
\begin{center}
\epsfig{file=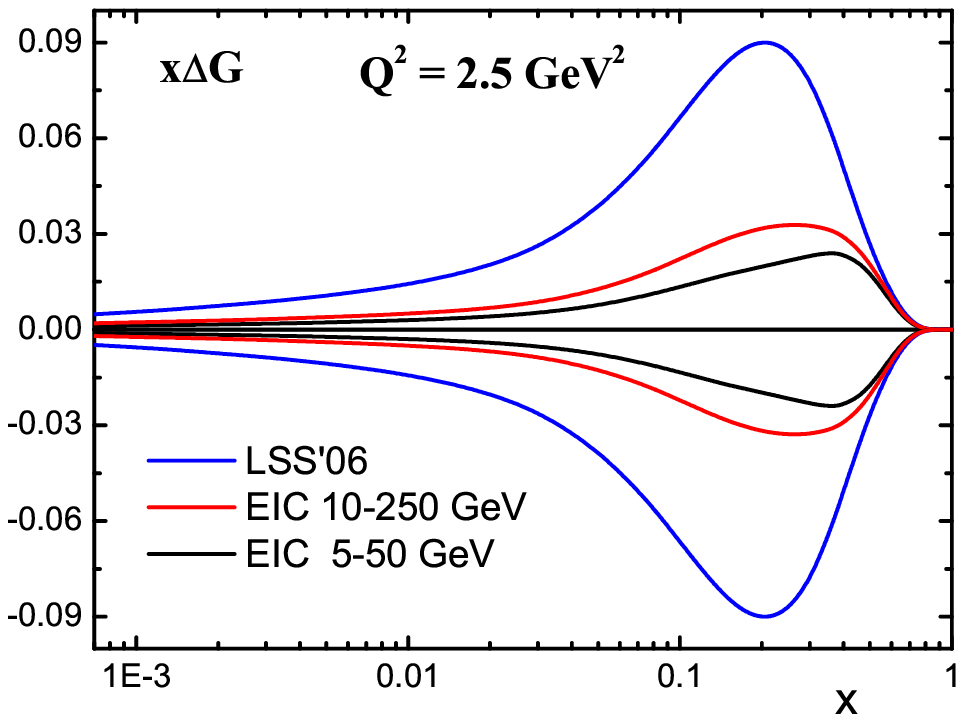, width=\textwidth}
\end{center}
\end{minipage}
\hfill
\vspace{-0.5cm}
\caption{Impact of two versions of EIC on the $s$ and $G$ uncertainties
 \protect{\cite{LSSunpub}}.}\label{err_EICsG}
\end{figure}

A collider of the EIC type would also have a dramatic effect in
reducing the uncertainties in the polarized parton densities. This is illustrated in
Figs.~\ref{err_EICud},\ref{err_EICsG}
where it can be seen that the improvement 
even at moderate to large $x$, especially for $\Delta s$ and
$\Delta G$, is remarkable.

Finally, the contribution from quark angular momentum is also an important
ingredient in the total spin balance of the nucleon. While a direct measurement
is not available, one can learn much about the transverse distribution and motion
of quarks from semi-inclusive measurements of single spin asymmetries. This
is a fairly new field, with a very rich potential and a rapidly growing body of
experimental data, but lies outside the scope of this article. It will be 
addressed by an upcoming review in this journal.

A second avenue towards a full accounting of the nucleon spin lies in measurements
that are sensitive to Generalized Parton Distributions (GPDs), in particular
Deeply Virtual Compton Scattering (DVCS). Moments of certain combinations
of GPDs can be related to the total angular momentum (spin and orbital) carried
by various quark flavors, as expressed in Ji's sum rule~\cite{Ji:2003qt,Ji:1996ek}.
Again, seminal experiments in this area have
already collected data that indicate that the necessary assumption of
factorization (and the validity of the handbag diagram) 
seems to be fulfilled, and even give some
model-dependent indication of the value of u- and d-quark angular 
momentum~\cite{Mazouz:2007vj}.
Future measurements at COMPASS and an extensive program at the energy-upgraded
Jefferson Lab will add substantially to this database. Ultimately, an
electron-ion collider could map out GPDs in detail at significantly lower $x$ than
fixed-target experiments.

In summary, the interest in the nucleon spin structure continues unabated, with many
novel experimental and theoretical approaches towards the ultimate understanding
of all contributions to the nucleon's spin. An extensive and rich experimental program
lies already ahead of us at existing facilities like CERN (COMPASS), RHIC and
Jefferson Lab (both during the remaining years of 6 GeV running and in the
new era of 11-12 GeV beams). Ultimately, this field would benefit tremendously
from a new collider of both polarized ions and polarized leptons, which is now
in the planning stage.

\section{Acknowledgments}

S.K. acknowledges the support from the United States Department of Energy (DOE) under
grant DE-FG02\_96ER40960. 
E.L. is grateful to M.~Anselmino, A.V.~Efremov, A.V.~Sidorov and D.B.~Stamenov 
for helpful discussions, and to Jefferson Lab for its hospitality.
Jefferson Science Associates operates
the Thomas Jefferson National Accelerator Facility for the
DOE under contract DE-AC05-84ER-40150.

\end{document}